\documentclass[12pt,a4paper,roman]{aa}  
\usepackage{rotating} 
\usepackage[english]{babel}
\usepackage{setspace}
\usepackage{siunitx}
\usepackage[utf8]{inputenc}
\usepackage{polski}
\usepackage[T1]{fontenc}
\usepackage{graphicx}
\usepackage{wrapfig}
\usepackage{makecell}
\usepackage{booktabs}
\usepackage{txfonts}
\usepackage{amsmath}
\usepackage{float}
\usepackage{natbib}
\usepackage{graphicx}
\usepackage{enumitem}
\usepackage{dblfloatfix}
\usepackage{txfonts}
\usepackage{subcaption}
\usepackage{multirow}
\usepackage{caption}

\usepackage{amsmath}
\usepackage{float}
 \usepackage[table,xcdraw]{xcolor}
\usepackage[hidelinks,colorlinks=true,linkcolor=blue,citecolor=blue]{hyperref}
\DeclareUnicodeCharacter{2212}{-}
\usepackage{xcolor}

\def \magperarcsec{mag arcsec$^{-2}$}
\def \mue{$\overline{\mu}_{\mathrm{eff},g}$}
\def \re{$r_{\mathrm{eff},g}$}
\def \msun{$\,M_{\odot}\,$}
\def \muo{$\mu_{0}$}
\begin{document}

     \title{DES to HSC: Detecting low surface brightness galaxies in the Abell 194 cluster using transfer learning}

   \author{H. Thuruthipilly \inst{1}\and
          Junais\inst{2,3}\and
          J. Koda \inst{4}
          A. Pollo\inst{1,5}\and
          M. Yagi \inst{6}\and
          H. Yamanoi \inst{7}   \and
          Y. Komiyama \inst{6,8,9}   \and
          M. Romano \inst{10,11,1} \and 
          K. Małek \inst{1,12}\and
         D. Donevski \inst{1}}

   \institute{National Centre for Nuclear Research, Pasteura 7, PL-02-093 Warsaw, Poland. \\
             \email{hareesh.thuruthipilly@ncbj.gov.pl},\\
             \email{junais@ncbj.gov.pl, jin.koda@stonybrook.edu}
             \and Instituto de Astrof\'{i}sica de Canarias, V\'{i}a L\'{a}ctea S/N, E-38205 La Laguna, Spain 
            \and Departamento de Astrof\'{i}sica, Universidad de La Laguna, E-38206 La Laguna, Spain
            \and Department of Physics and Astronomy, Stony Brook University, Stony Brook, NY 11794-3800, USA. 
            \and Astronomical Observatory of Jagiellonian University, Orla 171, 30-244 Krakow, Poland
            \and National Astronomical Observatory of Japan, Mitaka, Tokyo, 181-8588, Japan.
            \and Faculty of Science and Technology, Seikei University, 3-3-1 Kichijoji-Kitamachi, Musashino, Tokyo 180-8633, Japan.
            \and Department of Advanced Sciences, Faculty of Science and Engineering, Hosei University, 3-7-2 Kajino-cho, Koganei, Tokyo 184-8584, Japan.
            \and Graduate University for Advanced Studies (SOKENDAI), Mitaka, Tokyo 181-8588, Japan. 
            \and Max-Planck-Institut für Radioastronomie, Auf dem Hügel 69, 53121 Bonn, Germany.
            \and INAF - Osservatorio Astronomico di Padova, Vicolo dell'Osservatorio 5, I-35122, Padova, Italy.
            \and Aix-Marseille Univ., CNRS, CNES, LAM, Marseille, France.}

   \date{Received XXX / Accepted YYY}
  \abstract
   {Low surface brightness galaxies (LSBGs) are important for understanding galaxy evolution and cosmological models. Nevertheless, the physical properties of these objects remain unknown, as even the detection of LSBGs can be challenging. The upcoming large-scale surveys are expected to uncover a large number of LSBGs, which would require accurate automated or machine learning-based methods for their detection.}
   {We study the scope of transfer learning for the identification of LSBGs. We use transformer models trained on Dark Energy Survey (DES) data to identify LSBGs from dedicated Hyper Suprime-Cam (HSC) observations of the Abell 194 cluster, which are two magnitudes deeper than DES. A new sample of LSBGs and ultra-diffuse-galaxies (UDGs) around the Abell 194 is compiled, and their properties are investigated.}
   {We use eight models, divided into two categories: LSBG Detection Transformer (LSBG DETR) and LSBG Vision Transformer (LSBG ViT). The data from DES and HSC were standardized based on the pixel-level surface brightness. We used an ensemble of four LSBG DETR models and another ensemble of four LSBG ViT models to detect LSBGs. This was followed by a single-component S\'ersic model fit and a final visual inspection to filter out potential false positives and improve sample purity.}
   {We present a sample of 171 low surface brightness galaxies (LSBGs) in the Abell 194 cluster using HSC data, including 87 new discoveries. Of these, 159 were identified using transformer models, and 12 additional LSBGs were found through visual inspection. The transformer model achieved a true positive rate (TPR) of 93\% in HSC data, without any fine-tuning. Among the LSBGs, 28 were classified as ultra-diffuse galaxies (UDGs). The number of UDGs and the radial UDG number density suggests a linear relationship between UDG numbers and cluster mass on a log scale.
   UDGs share similar S\'ersic parameters with dwarf galaxies and occupy the extended end of the $R_{\mathrm{eff}}-M_g$ plane, suggesting they might be an extended subpopulation of dwarf galaxies. We also found that LSBGs and UDGs near the cluster centre are brighter and redder than those in outer regions. 
   }
   {We have demonstrated that transformer models trained on shallower surveys can be successfully applied to deeper surveys with appropriate data normalization. This approach allows us to use existing data and apply the knowledge to upcoming and ongoing surveys, such as the Rubin Observatory Legacy Survey of Space and Time (LSST) and Euclid.}

   \keywords{Low Surface Brightness Galaxies; Galaxy Surveys; Machine Learning}
\titlerunning{DES to HSC with transfer learning.} 
   \maketitle
%
%

\section{Introduction}\label{introduction}
Low surface brightness galaxies (LSBGs) are usually defined as galaxies that are fainter than the night sky \citep{Bothun}. Both simulations (e.g. \citealt{Martin}) and observations (e.g. \citealt{Dalcanton, Neil}) indicate that the bulk of the galaxy population resides in the low surface brightness regime. The LSBGs are estimated to account for a significant fraction $(30\%\sim60\%)$ of the total number density of galaxies \citep{McGaugh, Bothun, Neil, Haberzettl, Martin}. In addition, LSBGs have been found to contribute as much as $15\%$ of the dynamical mass content of the Universe \citep{Driver, Minchin}. Hence, LSBGs play a crucial role in understanding galaxy evolution \citep{Bullock,de_Blok, Sales} and offer important observational constraints for cosmological models \citep{Moore_cosmology, Bullock_Boylan, Laudato}.

In the literature, galaxies with \textit{B}-band central surface brightness $\mu_0(B)$ above a specific threshold are classified as LSBGs. This threshold varies across studies, ranging from $\mu_0(B) \geq 23.0$ mag arcsec$^{-2}$ \citep{Bothun} to $\mu_0(B) \geq 22.0$ mag arcsec$^{-2}$ \citep{Burkholder}.
LSBGs could be classified into several sub-classes based on their physical size, surface brightness, and gas content. For instance, ultra-diffuse galaxies (UDGs) are extended LSBGs with effective radii \re{} $> 1.5$ kpc and central surface brightness $\mu(g, 0) > 24$ \magperarcsec{}. Although the term `UDG' was coined by \citet{Dokkum}, such galaxies had been already identified in several earlier studies \citep{Sandage, McGaugh_bothun, Dalcanton, Conselice_udgs}. Similarly, giant LSBGs (GLSBGs) form another sub-class of LSBGs that are extremely gas-rich (M$_{\rm HI} > 10^{10}$ \msun{}), faint and extended \citep{Sprayberry1995, Saburova2023, Junais_malin2}. 
The formation and evolution of extreme sub-classes of LSBGs, such as UDGs and GLSBGs, are still debated, providing a robust platform to test models of galaxy evolution and cosmology\citep{Amorisco, Cintio, Saburova2021, Benavides, Laudato, Montes}.

The new era of large-scale surveys such as the Hyper Suprime-Cam Subaru Strategic Program (HSC-SSP; \citealt{Aihara}), Euclid \citep{laureijs2011euclid},
and Rubin Observatory’s Legacy Survey of Space and Time (LSST; \citealt{Ivezi__2019}) are expected to uncover more than $10^5$ LSBGs \citep{Greco, Tanoglidis1, haressh_lsbs}.
One of the major obstacles to overcome in detecting LSBGs in large-scale surveys is the separation of contaminants from those faint galaxies. As noted by \citet{Tanoglidis1}, contaminants primarily consist of diffuse light from nearby bright objects, galactic cirrus, star-forming tails of spiral arms, and tidal streams. These contaminants often pass the simple selection criteria meant to select LSBGs and dominate the LSBG candidate sample. Their removal is often achieved through semi-automated methods with a low success rate, followed by time-consuming visual inspection.

Recently, advancements in deep learning (DL) have opened up numerous opportunities, with convolutional neural networks (CNNs) and transformers effectively analyzing astronomical data \citep{2017ApJ...836...97C, 2019PASP..131j8002P, Pearson, Hareesh, Huang_tranformer, Jia, Hwang_VIT, Hareesh2, Grespan}. Training DL models typically requires large datasets. Recent surveys like SDSS \citep{Zhong_SDSS_lsbs}, HSC-SSP \citep{Greco}, and DES \citep{Tanoglidis1} have provided sufficient LSBGs to build such training sets \citep{Tanoglidis2, Yi_2022, Xing2023, haressh_lsbs, Su_LSBGNET}. Consequently, CNNs \citep{Tanoglidis2, Su_LSBGNET} and transformers \citep{haressh_lsbs} have been also employed to identify LSBGs from large-scale datasets. However, previous studies trained and tested these DL models using data from the same survey. 

In this context, a key question arises: to what extent can knowledge gained from one survey be transferred to another? Additionally, how effective are these DL models in cross-survey implementations? For example, can a DL model trained on DES data be effectively used to detect LSBGs in deeper datasets from HSC-SSP, LSST, and Euclid?

The above-mentioned questions fall under the regime of transfer learning. In this approach, a model trained for one task is adapted to a different task, typically by fine-tuning the model with a smaller training set \citep{Yosinski}. In computer vision, DL models such as CNNs or transformers typically aim to detect edges in the first layers and learn how to integrate these edges to understand task-specific features in the latter layers. Hence, the features learned in the first few layers of a DL model are general, and the weights of the initial layers can be effectively transferred from one task to another, facilitating faster and more efficient learning.

Recognizing the need for advanced machine learning (ML) techniques in the era of big data, the astronomy community has also explored the potential of transfer learning. For example, \citet{Ackermann} utilized a CNN model initially trained on everyday object images from the ImageNet dataset \citep{Image_net_dataset} and retrained it to detect galaxy mergers. Similarly, \citet{Wei} and \citet{Hannon} applied DL models pre-trained on ImageNet for star cluster classification. Previous studies have explored transfer learning across surveys for tasks such as galaxy morphological classification \citep{Sanchez}, LSBG classification \citep{Tanoglidis2} and galaxy merger identification \citep{bickley2024effectimagequalitygalaxy} using CNNs. However, CNN performance drops significantly when applied to a different survey, requiring fine-tuning to achieve satisfactory results.

In this paper, we successfully apply transfer learning to identify LSBGs in the Abell 194 cluster using the deep data from dedicated Hyper Suprime-Cam (HSC) observations. We train two different ensemble transformer models using data from DES data release 1 (DES DR 1) and apply them to the HSC data, which is 2 magnitudes deeper than DES DR 1, to identify LSBGs.  We chose Abell 194 for our analysis due to its coverage by DES and the availability of known LSBG and UDG samples from other studies \citep{Tanoglidis1, Zaritsky_2023, haressh_lsbs}. 

The primary goal of this paper is twofold: to efficiently identify LSBGs and to analyze their properties. The first component focuses on the methodology centred around transfer learning for cross-survey applications. While primarily aimed at LSBG identification, this methodology serves as a versatile tool for the astrophysical community, with potential applications in surveys like Euclid \citep{laureijs2011euclid} and LSST \citep{Ivezi__2019}. The second and more significant component explores the scientific implications of our findings. Using the identified LSBG and UDG samples, we investigate how the cluster environment influences their morphological and physical properties.

Throughout this paper, we adopt the following notations: the apparent magnitude is $m$ (mag), half-light radius is $r_{\mathrm{eff}}$ (arcsecond), central surface brightness is \muo{} (\magperarcsec{}), mean surface brightness within the $r_{\mathrm{\textit{eff}}}$ is $\overline{\mu}_{\mathrm{eff}}$ (\magperarcsec{}), S\'ersic index is $n$, axis ratio is $q$, and position angle is PA (degrees). All $r_{\mathrm{\textit{eff}}}$ values reported are non-circularized, whereas both $\mu_0$ and $\overline{\mu}_{\mathrm{eff}}$ have been corrected for inclination.  The PA is defined with north at  $0^\circ$ and east at $+90^\circ$.
When specifying the measurement in a particular band, we will use a subscript, such as $r_{\mathrm{eff},g}$ or $r_{\mathrm{eff},r}$ for the \textit{g} and \textit{r}-bands, respectively. 
For comparison purposes, we follow the LSBG definition from \citet{Tanoglidis1}, which is based on the extinction-corrected \textit{g}-band mean surface brightness ($\overline{\mu}_{\mathrm{eff},g} > 24.2$ \magperarcsec{}) and half-light radius ($r_{\mathrm{eff},g} > 2.5\arcsec$). Similarly, for defining UDGs, we adopt $\mu_{0,g} > 24.0$\footnote{\citet{vanDokkum} assumed a constant $n = 1$ for all measurements, whereas we allow $n$ to vary.} \magperarcsec{} and $r_{\mathrm{eff},g} > 1.5$ kpc, following \citet{Dokkum} and \citet{Roman_2017}.

We adopt the cosmological parameters of ($h_0, \Omega_M,\Omega_{\lambda}$) = (0.697, 0.282,0.718) following \citet{Hinshaw}. This corresponds to a luminosity distance of 77.7 Mpc, an angular diameter distance of 75.0 Mpc, and a scale of 1 arcsecond corresponding to 0.364 kpc for Abell 194 at a redshift $z = 0.0178$ \citep{Girardi_abell_a94, Rines_abell_194_z}. \citet{Rines_abell_194_z} estimated the Abell 194 cluster to have a virial radius ($R_{200}$) of 0.9824 Mpc and a virial mass $(M_{200})$ of 7.6 $\times$ $10^{13} M_{\odot}$ where $R_{200}$ is the radius at which the average density of a galaxy cluster is 200 times the critical density of the universe at that redshift.

The paper is organized as follows: Sect. \ref{data} discusses the data and Sect. \ref{method} provides a brief overview of the methodology, including the model architecture, training, and visual inspection. The results are presented in Sect. \ref{results}, followed by a discussion of the results and the properties of the newly identified LSBGs in Sect. \ref{discs}. Finally, Sect. \ref{conclusion} summarizes the conclusions of our analysis.

\section{Data}\label{data}
\subsection{Training data from DES}\label{traing_des_data}
 The Dark Energy Survey (DES; \citealt{DR1, DESDR2}) covers $\sim5000 \text{ deg}^2$ of the southern Galactic cap in the optical and near-infrared wavelength using the Dark Energy Camera (DECam) on the 4-m Blanco Telescope at the Cerro Tololo Inter-American Observatory (CTIO). The DES used \textit{g,r, i,z, Y} photometric bands with approximately 10 overlapping dithered exposures in each filter (90 sec in \textit{griz}-bands and 45 sec in \textit{Y}-band). The median surface brightness depth at $3\sigma$ for a $10"\times10"$ region for \textit{g}-band is $28.26\substack{+0.09 \\ -0.13}$ \magperarcsec{} and for \textit{r}-band is $27.86\substack{+0.10 \\ -0.15}$ \magperarcsec{} where the upper and lower
bounds represent the 16th and 84th percentiles of the distribution over DES tiles \citep{Tanoglidis1}. 
 
For training, validating, and testing of the models, we used the labelled dataset of LSBGs and contaminants identified from DES by \citet{Tanoglidis1} and extended by \cite{haressh_lsbs}. The catalogue for the contaminants was created based on the publicly available dataset, which consists of 20\,000 contaminants\footnote{\url{https://github.com/dtanoglidis/DeepShadows/blob/main/Datasets}}.  However, \citet{haressh_lsbs} have shown that some of the contaminants listed in this catalogue are not contaminants but are, in fact, LSBGs. After removing these sources from the catalogue, we had 18\,468 contaminants remaining in the list, which were labelled 0. Since an unbalanced training dataset creates a biased ML model, we randomly selected 18\,532 LSBGs(comparable to the number of contaminants) from the extended sample of 27\,873 LSBGs\footnote{23\,790 from \citet{Tanoglidis1} and 4\,083 from \citet{haressh_lsbs}}, which were assigned a label 1. 

\begin{figure}[h]
  \centering
  \begin{subfigure}{\linewidth}
    \centering
    \includegraphics[width=\linewidth, keepaspectratio]{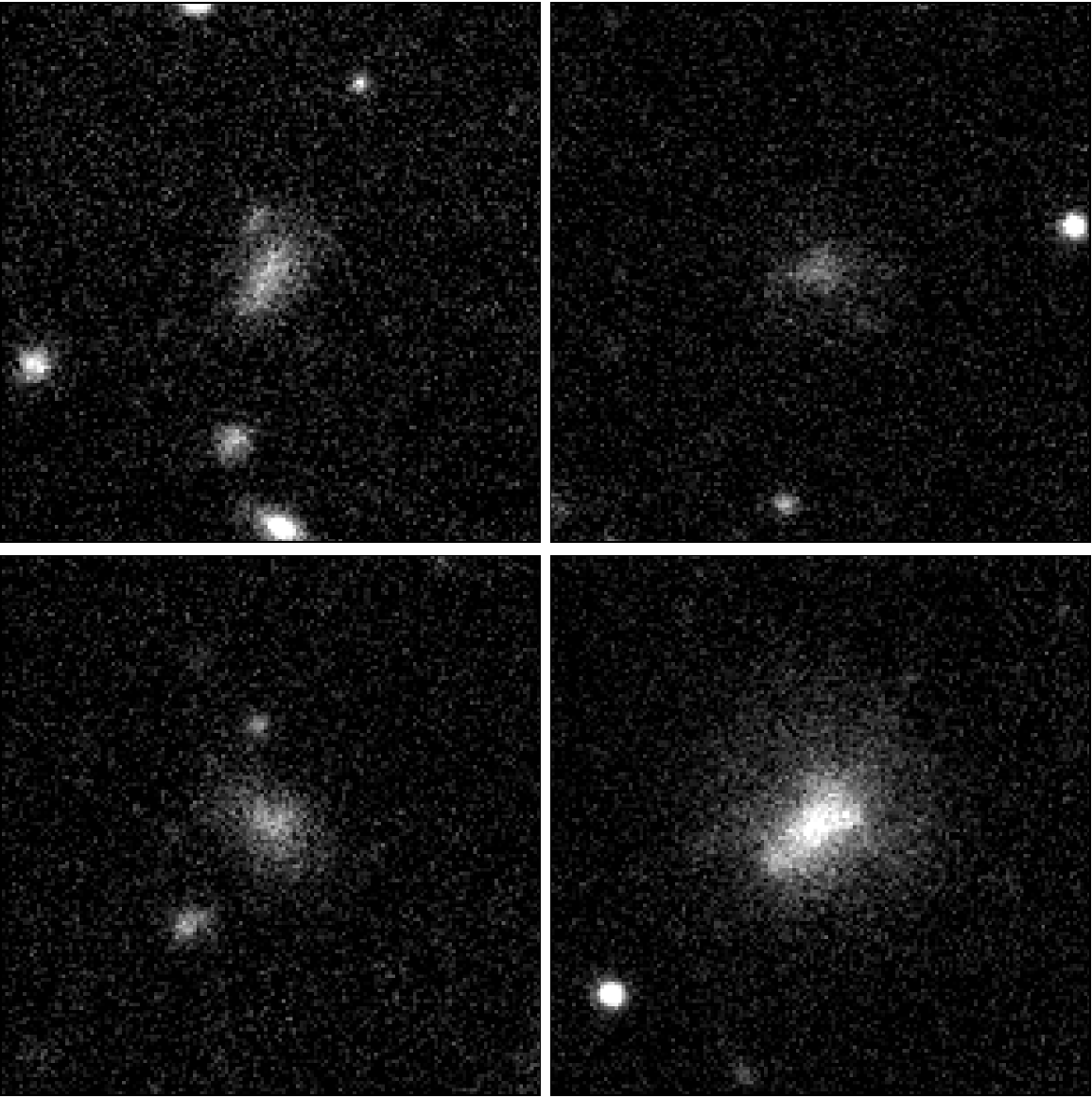}
    \caption{Examples of LSBGs in DES DR1.}
    \label{lsb_train*}
  \end{subfigure}

  \begin{subfigure}{\linewidth}
    \centering
    \includegraphics[width=\linewidth, keepaspectratio]{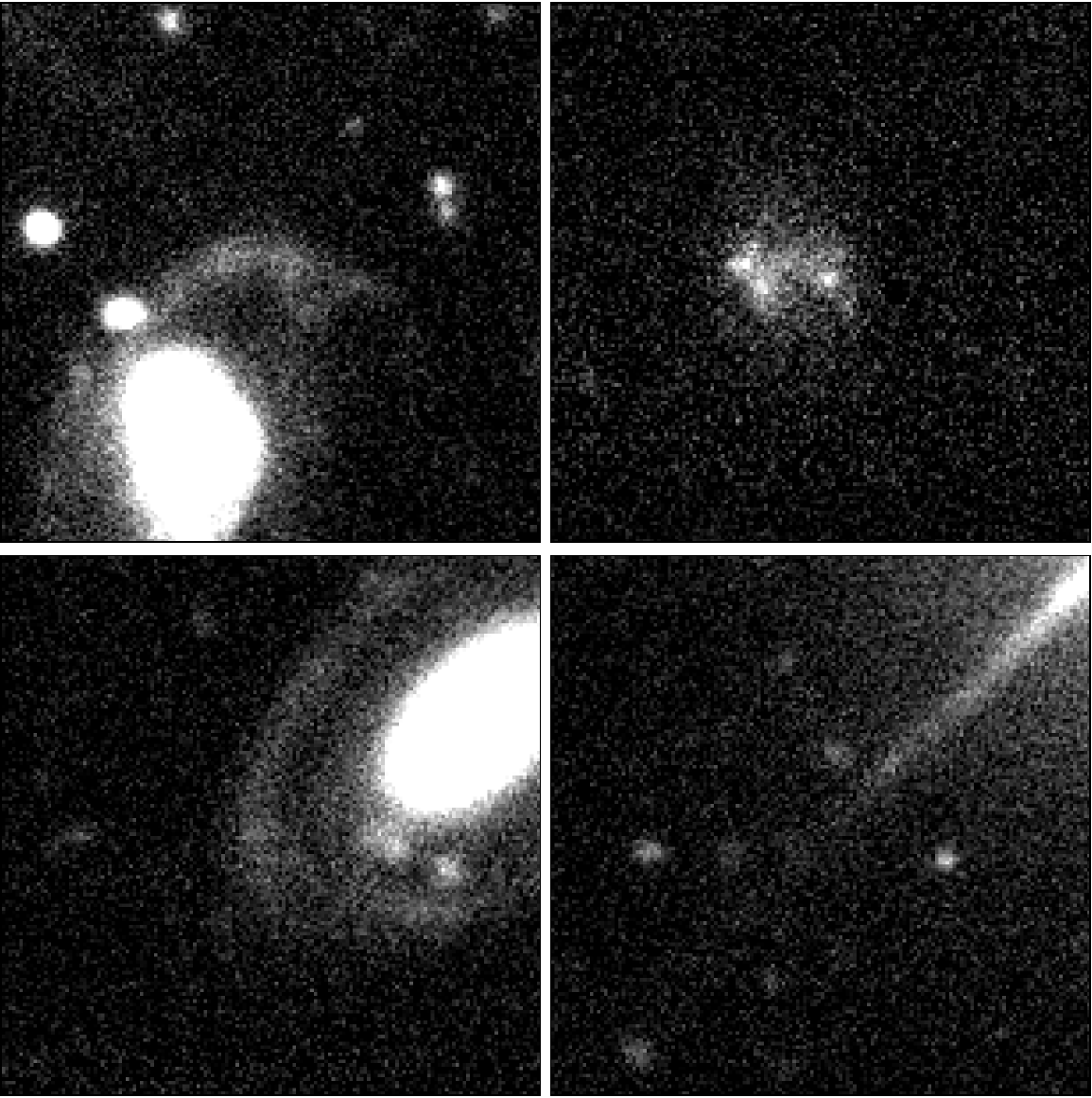}
    \caption{Examples of contaminants in DES DR1.}
    \label{art_train}
  \end{subfigure}

  \caption{\textit{g}-band cutouts of four examples of LSBGs (\ref{lsb_train*}) and contaminants (\ref{art_train}) used in the training data. Each cutout corresponds to a $40\arcsec \times 40\arcsec$ ($152 \times 152$ pixels) region of the sky centred on the LSBG or artefact.}
  \label{fig:training}
\end{figure}

We generated multi-band cutouts for each object using DES DR 1 image in the \textit{g}-band and \textit{r}-band. Each cutout covers a $40\arcsec \times 40\arcsec$ region of the sky (equivalent to $152 \times 152$ pixels) and is centred on the coordinates of the object (either an LSBG or an artefact). The cutouts were resized to $64 \times 64$ pixels using {\tt skimage.transform}\footnote{\url{https://scikit-image.org/docs/stable/api/skimage.transform.html}} Python package to reduce the computational cost. This was done while maintaining the same sky coverage by adjusting the pixel scale to 
0.625 arcseconds per pixel. The cutouts of $g \text{ and } r$-bands were stacked together to create the dataset for training the models. Our training catalogue contains 38,500 objects, comprising 18,532 LSBGs and 18,468 contaminants. Before training, we randomly split the full sample into a training set, a validation set, and a test set consisting of 32,000, 2,000, and 4,500 objects, respectively. Each of these subsets maintained similar proportions of LSBGs and contaminants to ensure balanced representation. Examples of LSBGs and contaminants in the training set are shown in Fig. \ref{fig:training}.

\subsection{HSC data of Abell 194 cluster}\label{HSC_data}
The HSC is an imaging camera covering an area of 1.77 degree$^2$ situated at the prime focus of the Subaru telescope with a pixel scale of $0.168\arcsec$ \citep{HSC_software, Hsc_cam1, Hsc_cam2, HSC_cam3, HSC_cam4}.
The Abell 194 cluster was observed with HSC using the HSC-g and HSC-r2 filters in a single pointing with dithered exposures. The r2 filter is an improved version of the r filter \citep{Hsc_cam2} and is referred to simply as the \textit{r}-band throughout the text from this point onward.
The \textit{r}-band observations were conducted on October 3, 2016, and the \textit{g}-band observations took place on December 24, 2016.
The dithering pattern was optimized to ensure that gaps between CCDs do not overlap in many exposures. Each dithered integration (referred to as a “visit”) began with the rotator angle set to 0 degrees to maintain consistent flat patterns near the edge of the field of view. The typical seeing size, represented by the full-width at half maximum (FWHM) of the point spread function (PSF), was approximately $1\arcsec$ during the observing runs.
 For easier data handling, the observed field was divided into square patches, each $12\arcmin$ on a side ($4200 \times 4200$ pixels), with a $17\arcsec$ (100 pixels) overlap between adjacent patches. Hence, resulting in a total of 94 patches. 


The data were reduced using the HSC pipeline, hscPipe\citep{Bosch_HSC_cam} version 8.5.3, in a standard way. hscPipe is based on software developed for the LSST project \citep{Ivezic_LSST_design, Juric_LSST_design}. Astrometric and photometric calibrations were performed using the Pan-STARRS1 catalogue \citep{Magnier_PanSTARS, Chambers_PANSTARRs}, which is the standard with the HSC pipeline.

To estimate the median surface brightness depth at $3\sigma$ for a $10\arcsec\times10\arcsec$ region for the HSC data, we followed the procedure described by \citet{Roman_surface_brightness_depth}. We randomly selected 1,000 sky apertures, each covering a $10\arcsec \times 10\arcsec$ region, from each of the 94 tiles (totalling 94,000 apertures) and fitted the counts in each aperture with a Gaussian profile. The $\sigma$ values from each $10\arcsec$ aperture were used to estimate the surface brightness depth at the $3\sigma_{10\times10}$ level for a $10\arcsec \times 10\arcsec$ region using the relation.
 \citep{Roman_surface_brightness_depth}:
\begin{equation}
    \mu_{lim}(3\sigma_{10\times10}) = -2.5\times log\left(\frac{3\sigma}{pix\times10}\right)+ZP
\end{equation}
Here, \textit{pix} is the pixel scale (0.168 arcsec/pix) of the HSC data, and ZP is the zero point of the HSC data, which is calibrated to 27.0\footnote{This value may be inaccurate by a few per cent due to aperture correction adjustments.} mag. We estimated the median surface brightness depth at $3\sigma$ for a $10\arcsec\times10\arcsec$ region to be $30.5\pm0.2$ \magperarcsec{} in $g$-band and $29.7 \pm 0.2$ \magperarcsec{} for $r$-band. Here, the error corresponds to the median absolute deviation. This is $\sim 2$ magnitude deeper than DES data which has a median surface brightness depth of $28.26\substack{+0.09 \\ -0.13}$ \magperarcsec{} in the \textit{g}-band and $27.86\substack{+0.10 \\ -0.15}$ \magperarcsec{} in the \textit{r}-band \citep{Tanoglidis1}.

\subsection{GALEX}
The Galaxy Evolution Explorer (GALEX; \citealt{Martin_galaex}) is a NASA small explorer mission that imaged the sky in far-ultraviolet (FUV; 1344–1786 $\AA$) and near-ultraviolet (NUV; 1771–2831 $\AA$) bands. The telescope has a 1.25-degree field of view and a resolution with an FWHM of $4.2\arcsec$ and $5.3\arcsec$ for the FUV and NUV bands, respectively. Each pixel in the intensity map from the GALEX corresponds to $1.5\arcsec$ in the angular scale \citep{Morrissey}.

For our analysis, we use the intensity maps of all the available GALEX data within a 2-degree search radius of the Abell 194 cluster from the MAST database\footnote{\url{https://archive.stsci.edu/}}. The intensity maps that partially overlap were coadded using their corresponding exposure maps with the {\tt reproject} Python package. The final co-added image has a variable exposure along the field of the Abell 194 cluster with a median exposure time of $4274\pm1232$ seconds in the NUV and $3943\pm1393$ seconds in the FUV band, where the error denotes the median absolute deviation.  

\section{Methodology}\label{method}

\subsection{Transformer Models}
Transformers are DL models that use the attention mechanism to analyze correlations within input features for predictions or decisions \citep{vaswani2017attention}. In this work, we employ two types of transformer models: LSBG Detection Transformers (LSBG DETR) and LSBG Vision Transformers (LSBG ViT). These architectures were previously used for identifying LSBGs in DES DR1 by \citet{haressh_lsbs}. Here, we apply the same models to HSC data of the Abell 194 cluster, enabling a direct comparison with the results from \citet{haressh_lsbs}. Details about the training process can be found in Appendix \ref{training_on_des}, and the performance of the individual models on the DES data is described in Appendix \ref{model_perfomance_on_DES}. A comprehensive discussion of the transformer models is available in \citet{haressh_lsbs}.

\subsection{Transfer learning}
Generally, transfer learning refers to the practice of re-using a pre-trained model for a new task. In this work, we test transfer learning using the transformer models presented in \citet{haressh_lsbs}. These transformer models are trained to classify LSBGs and contaminants from the DES data (Sect. \ref{traing_des_data}). We apply these models to a deeper dataset of the Abell 194 cluster observed with HSC. This dataset is two magnitudes deeper than the training data and has a different resolution compared to DES. Our goal is to evaluate how well the transformer models can distinguish LSBGs from contaminants in this deeper dataset.

When using transfer learning between surveys, it is important to adjust for differences in the instruments, as these cause variations in photometric zero points, pixel scales, and FWHM of PSF. For example, observing the same sky area with two instruments with different zero points and pixel scales will produce images with different pixel values in count units. For a DL model, the input parameter space (in this case, pixel values) is crucial for ensuring good model performance. Hence, a model trained on one survey could not be directly used on another survey and needs standardisation. 

To standardize the data, we convert the pixel values of the images from counts per second to surface brightness units. This ensures that the average pixel values over a region remain consistent, enabling direct comparison and standardization of image data at the same wavelength across different surveys. Hence, before training and testing our transformer models, we convert each pixel value to its surface brightness ($\mu Jy \text{ arcsec}^{-2}$) for DES DR 1 and HSC data, respectively.
However, it is important to note that this standardization of data does not address the difference in PSF values between different surveys. It should be noted that we apply this conversion only for the application of the ML models, and all the measurements and subsequent analysis are done on the original data.

\subsection{Object detection and preselection of LSBG candidates in Abell 194}\label{object_detect}
First, we generate a master catalogue of LSBG candidates by running {\tt SExtractor} \citep{sxtractor} on the image patches of the HSC data. Since the $r$-band is expected to better trace the mass distribution of LSBGs compared to the $g$-band, we use sky-subtracted \textit{r}-band images for detection. However, for comparing with prior LSBG and UDG catalogues in Abell 194 \citep{Tanoglidis1, Zaritsky_2023, haressh_lsbs}, which used a $g$-band definition, we adopt their definitions for consistency. For source detection, we set the {\tt DETECT\_THRESH} to $1.5\sigma$ ($\sim 30.5$ \magperarcsec{} for \textit{r}-band) and {\tt DETECT\_MINAREA} to 49 pixels ($\sim $1.38 arcsec$^2$), with these conservative values chosen to ensure all LSBGs are included.

We further apply selection cuts on the output catalogue based on the \re{}, \mue{}, and the $q$ as measured by the {\tt SExtractor} to reduce the LSBG candidate sample. Since we are looking for extended LSBGs, we removed the bright and point objects from the full catalogue. The selection cuts applied to create the initial LSBG candidate sample are as follows:
\begin{itemize}
   \item We set \mue{} to be in the range of 24.0 to 31.0 \magperarcsec{}. The brighter limit is 0.2 magnitudes brighter than our LSBG definition to ensure no faint LSBGs are excluded. The faint end limit is fainter than the $3\sigma$ surface brightness detection threshold of the HSC data to ensure the preliminary catalogue includes all potential faint LSBGs.  
\item \re{} is restricted to $2\arcsec<$\re{}$<20\arcsec$, with the lower limit relaxed to maximize the detection of extended LSBGs. At the distance of Abell 194, this corresponds to 0.73 kpc < \re{} < 7.3 kpc.
\item The axis ratio $q$ ({\tt B\_IMAGE}/{\tt A\_IMAGE}) is set in the range $0.3 < q \leq 1.0$, following \citet{Greco} and \citet{Tanoglidis1}. This removes contaminants such as highly elongated- diffraction spikes.
\item We remove the known foreground and background sources based on the redshift from SDSS Data Release 16 \citep[SDSS DR16;][]{SDSS16}.
\end{itemize}


For each remaining source, we generated cutouts in \textit{g} and \textit{r}-bands. Each cutout is a $ 40 {\arcsec} \times  40 {\arcsec}$ ($238 \times 238 $ pixels) region of the sky and is centred on the source. For sources near the edges of patches, we co-added the overlapping regions from nearby patches to obtain a $40{\arcsec} \times 40{\arcsec}$ cutout. 
To classify the central source in each cutout as an LSBG or artefact, we resized the HSC data cutouts from $238 \times 238$ pixels to $64 \times 64$ pixels to match the input size of the transformer models. Similar to the DES data, the cutouts in each band were resized using the Python package {\tt skimage.transform}. The resized $g$- and $r$-band cutouts were stacked for testing with the models. However, the resized cutouts were used only for model testing, while the original cutouts were utilized for S\'ersic fitting, visual inspection, and aperture photometry.

\begin{figure*}[t]
\includegraphics[width=525 pt,keepaspectratio]{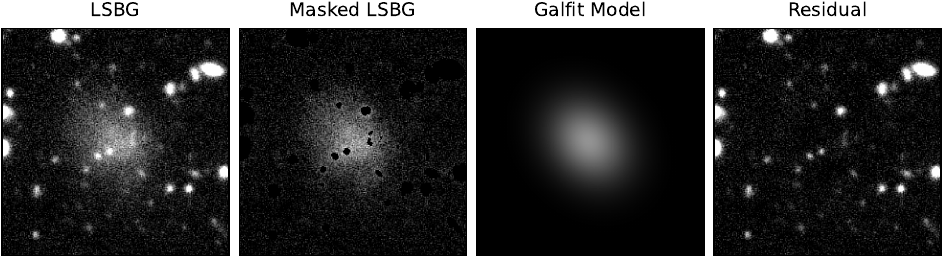}
\caption{Image of an LSBG (RA = $21.82879^\circ $, DEC=-1.81381$^\circ$) as observed in HSC in the \textit{g}-band, the image with all the objects other than LSBG masked, corresponding S\'ersic model fitted by {\tt GALFIT} and the residual are shown respectively from left to right.}
\label{fig:lsb_masked}
\end{figure*}
\subsection{Masks}\label{masks_creation}
Compact objects on top of LSBGs can influence the accuracy of S\'ersic profile fitting (as discussed in Sect. \ref{sersic_fitting}). To obtain precise S\'ersic parameters, we masked these objects following the procedures outlined by \citet{Bautista}. We run the {\tt SExtractor} on each cutout three times by adjusting the control parameters to detect different types of contaminants, such as the bright sources, faint sources lying outside the LSBG, and faint sources lying on top of the LSBG. The {\tt CHECKIMAGE} image that contains only the detected objects is used to create the masks for each kind of object, and all three masks are combined to create the final mask. 

The first step is to remove the bright objects from the cutouts. The bright objects can be removed by setting a higher detection threshold without any limits on the size. We set {\tt DETECT\_THRESH} as 23.0 \magperarcsec{} and masked all the bright pixels. The second step is to remove faint, small objects that are far from the central LSBG. For this, we set a maximum area threshold ({\tt DETECT\_MAXAREA}) of 200 pixels at a detection threshold of 27.5 \magperarcsec{}.

To identify smaller objects on top of LSBGs, we use the unsharp masking technique described in \citet{Bautista}. This involves smoothing the original \textit{g}-band image by convolving it with a Gaussian kernel, then subtracting the smoothed image from the original. After subtraction, features with sizes similar to or smaller than the smoothing kernel remain in the subtracted images. {\tt SExtractor} is then used to detect and remove these small objects with an optimal configuration of FWHM of the Gaussian kernel and {\tt DETECT\_MAXAREA}. The combination of these parameters varies depending on the sizes of the LSBGs and the overlapping small objects. While most masks ($\sim 40\%$) were created using FWHM = $2.0\arcsec$ and {\tt DETECT\_MAXAREA} = 200, the same configurations can lead to poor masks for certain LSBGs. Through trial and error, we determined that using FWHM values of [$2.0\arcsec, 1.25\arcsec, 2.50\arcsec$] and {\tt DETECT\_MAXAREA} values of [200, 100] can produce accurate masks. Notably, the masks used in this study are more accurate than those from \citet{Tanoglidis1} and \citet{haressh_lsbs}, as they did not follow the same approach.

\subsection{S\'ersic fitting}\label{sersic_fitting}
We used {\tt GALFIT} \citep{GALFIT} to fit a single-component S\'ersic profile to the sources identified as LSBGs by the transformers in order to re-evaluate their \mue{} and \re{}. We opted for a single-component S\'ersic fitting to align with the methodology of \citet{Tanoglidis1} and \citet{haressh_lsbs}. However, it should be noted that S\'ersic fitting may not always capture the complete light profile of a galaxy.

Since an error in the sky determination will significantly impact the S\'ersic parameters, we subtract the local sky from each cut-out prior to the S\'ersic fitting \citep{Bautista}. The local sky is estimated as $2.5 \times$ median - $1.5\times$ mean of the flux from the masked cutout. However, if the median is greater than the mean, then the mean is treated as the local sky background \citep{Da_Costa}. Similar to \citet{Bautista}, when calculating the total flux from each cutout, a circular region of $10\arcsec$ centred on the LSBG candidate, in addition to the masked region, is excluded to avoid including the light from the LSBG. The PSF image is generated by running PSFEX version 3.22.1 \citep{Bertin_psfex} on the patch from which the cutout is taken.

The \textit{g}-band images were fitted first. We used the {\tt MAG\_AUTO} and {\tt FLUX\_RADIUS} (with PHOT\_FLUXFRAC=0.5) values obtained from the {\tt SExtractor} as an initial guess for $m$ and $r_{\mathrm{eff}}$. We also used {\tt A\_IMAGE} and {\tt B\_IMAGE} to calculate the initial axis ratio $q$ ({\tt B\_IMAGE}/{\tt A\_IMAGE}), which was allowed to vary in the range of $0.3 < q \leq 1.0$. The S\'ersic index ($n$) was initialized at 1 and was allowed to vary only within the range of $0.2 < n < 5.0$. The position angle was initialized at $90\deg$ and was allowed to vary without any restrictions to find the optimal angle. The initial centre was set at {\tt X} = 119 pixels and {\tt Y} = 119 pixels, which was allowed to vary $\pm 5$ pixels (0.84 \arcsec) in both directions. These conditions for the S\'ersic fit were based on \citet{Greco} and \citet{Tanoglidis1}. An example of an LSBG fitted with {\tt GALFIT} is shown in Fig. \ref{fig:lsb_masked}. During the \textit{r}-band fitting, we fixed the parameters $X$, $Y$, $n$, $q$, and PA to the values obtained from the \textit{g}-band fitting, allowing only the magnitude and radius to vary.

After the fitting, we excluded all the sources with poor/failed fits with a reduced $\chi^{2}>3$. We excluded cases where the estimated $n$, $q$, {\tt X}, and {\tt Y} values did not converge or were at the edge of the specified range, as most of these objects ($\sim 90\%$) were later found to be contaminants. For the remaining galaxies, we re-applied our \textit{g}-band sample selection criteria of $\overline{\mu}_{\mathrm{eff}}>24.2$ mag arcsec$^{-2}$ and $r_{\mathrm{eff}}>2.5 {\arcsec}$. 
The $\overline{\mu}_{\mathrm{eff}}$ and \muo{} values were calculated using the relations \citep{Graham_sersic}:
\begin{equation}
    \centering
 \overline{\mu}_{\mathrm{eff}} = m + 2.5\log_{10}(2\pi r_{\mathrm{eff}}^2 q),
\end{equation}
and 
\begin{equation}
    \centering
    \mu_{0} = \overline{\mu}_{\mathrm{eff}} + 2.5\log_{10}\bigg(\frac{n}{b^{2n}}\Gamma(2n)\bigg),
\end{equation}
where $b$ is defined as $2\gamma(2n,b) = \Gamma(2n)$. Here $\Gamma$ and $\gamma$ represent the complete and incomplete gamma functions, respectively, and the parameters were obtained from  {\tt GALFIT}. For all our measurements, we also applied a foreground galactic extinction correction using the \citet{Schlegel1998} maps normalised by \citet{Schlafly2011} and a \citet{Fitzpatrick1999} dust extinction law. 
\subsection{Visual inspection}
We used visual inspection as the final step to improve the purity of the sample. However, for upcoming large-scale surveys like LSST, visual inspection may not be feasible due to the volume of data. In such cases, alternative approaches, such as accepting a sample with low contamination (~5\%) or using crowd science, may be necessary.
Here we considered candidates identified by either the LSBG DETR or LSBT ViT models and those that met the LSBG selection criteria with the updated {\tt GALFIT} parameters for visual inspection. This refined sample was independently visually inspected by two authors. 

To aid in visual inspection, we used the color images of the LSBG candidate downloaded from the DESI Legacy Imaging Surveys Sky Viewer \citep{DESI} as well as the \textit{g}-band images from the HSC. Furthermore, the \textit{g}-band S\'ersic models from {\tt GALFIT} were also inspected visually to ensure the quality of the fits. Each candidate was then categorised into three classes based on the {\tt GALFIT} model fit and the image: LSBG, non-LSBG (contaminants), or misfitted LSBGs. 
If both visual inspectors agreed that the created masks and the resulting S\'ersic model were accurate, the source was classified as an LSBG. However, if one visual inspector deemed the masks inaccurate or the S\'ersic model's radius did not match the source, it was classified as a misfitted LSBG and refitted with different initial conditions or improved masks. Finally, if the candidate does not look like an LSBG, we classify it as a contaminant or non-LSBG. 

\subsection{Aperture photometry}\label{aperture_photometry}
As mentioned above, S\'ersic fitting cannot always capture the complete light profile of a galaxy. Hence, we perform aperture photometry using {\tt photutils}\footnote{\url{https://photutils.readthedocs.io/en/stable/}} to estimate the apparent magnitude within a circular aperture of radius $8\arcsec$ ($\sim2.9$ kpc) in the \textit{g}, \textit{r}, $NUV$ and $FUV$ bands.
The aperture size was selected based on the maximum of the half-light radius distribution from {\tt GALFIT}. This ensures that the size of the LSBGs is always smaller than the aperture size.

The masks used for the S\'ersic fit were also applied in the aperture photometry. However, since the resolution of the NUV and FUV data ($5\arcsec$) is larger than the size of most LSBGs, we unmasked point objects overlying the LSBGs to avoid unnecessarily masking the data. These masks, originally created from the \textit{g}-band images, were resized to match the pixel scale of the GALEX data. For NUV and FUV photometry, the error from the sky background is estimated by choosing an annulus with an inner radius of $10\arcsec$ and an outer radius of $20\arcsec$ and estimating the sky background from this region. Only sources with $S/N>3\sigma$ were considered as a confident detection, and for the cases with $S/N<3\sigma$, the $3\sigma$ value was used to report the upper limit for the NUV and FUV magnitude.
The aperture magnitudes were corrected for foreground Galactic extinction similarly as discussed in Sect. \ref{sersic_fitting}. 

\section{Results}\label{results}

\subsection{Search for LSBGs with HSC in Abell 194 cluster} 
By running the {\tt SExtractor} patch by patch with the parameters specified in Sect. \ref{object_detect}, we identified 170\,328 sources. To reduce the sample size, we apply the selection criteria mentioned in Sect. \ref{object_detect}. 
After applying the selection cuts to \re{}, \mue{}, and $q$, we have 991 sources detected from the Abell 194, which could be a potential LSBG.  However, a large fraction of these sources are likely to be instrumental or physical contaminants rather than true LSBGs.

After applying the preselection as described in section \ref{object_detect}, we have a crude candidate catalogue of LSBGs. Further, we cross-matched the crude catalogue with SDSS DR 16 catalogue \citep{SDSS16} to remove the foreground and background galaxies compared to the Abell 194 cluster. We identified 14 galaxies with known redshifts that do not match the redshift of Abell 194 ($z=0.0178$), and these were removed from the sample. This reduced the size of the crude catalogue to 977 objects. The remaining objects in this catalogue could be an LSBG, an artefact, or a non-LSBG (a faint galaxy but not faint enough to be classified as an LSBG as per the definition that we use in this work).  

 To separate faint galaxies from contaminants, we apply our LSBG DETR and LSBG ViT models separately to all the objects that passed the selection criteria. LSBG DETR models identified 258 LSBG candidates, and LSBG ViT models identified 261 LSBG candidates independently. Among these, only 247 sources were classified as LSBG candidates by both models. To maximise the number of LSBGs, an input classified as an LSBG by either one of the ensembles is treated as an LSBG candidate and passed on for the subsequent analysis. Combining the outputs from both samples resulted in 272 LSBG candidates, which all underwent single-component S\'ersic profile fitting using {\tt GALFIT}. After the fitting, we reapplied the selection criteria to screen the LSBG candidates. We applied selection cuts to the \textit{n, q,} and centre positions X and Y of the fitter parameters as described in Sect. \ref{sersic_fitting}.
These criteria were used to eliminate any poor fits and contaminants, if present. Two of the authors independently visually inspected the masks and fitted galaxy profiles to ensure the quality of the fit. Finally, we had 159 LSBGs, with 11 detected by only the LSBG ViT ensemble model.

\begin{figure}[h]
    \centering
    \includegraphics[width=\linewidth]{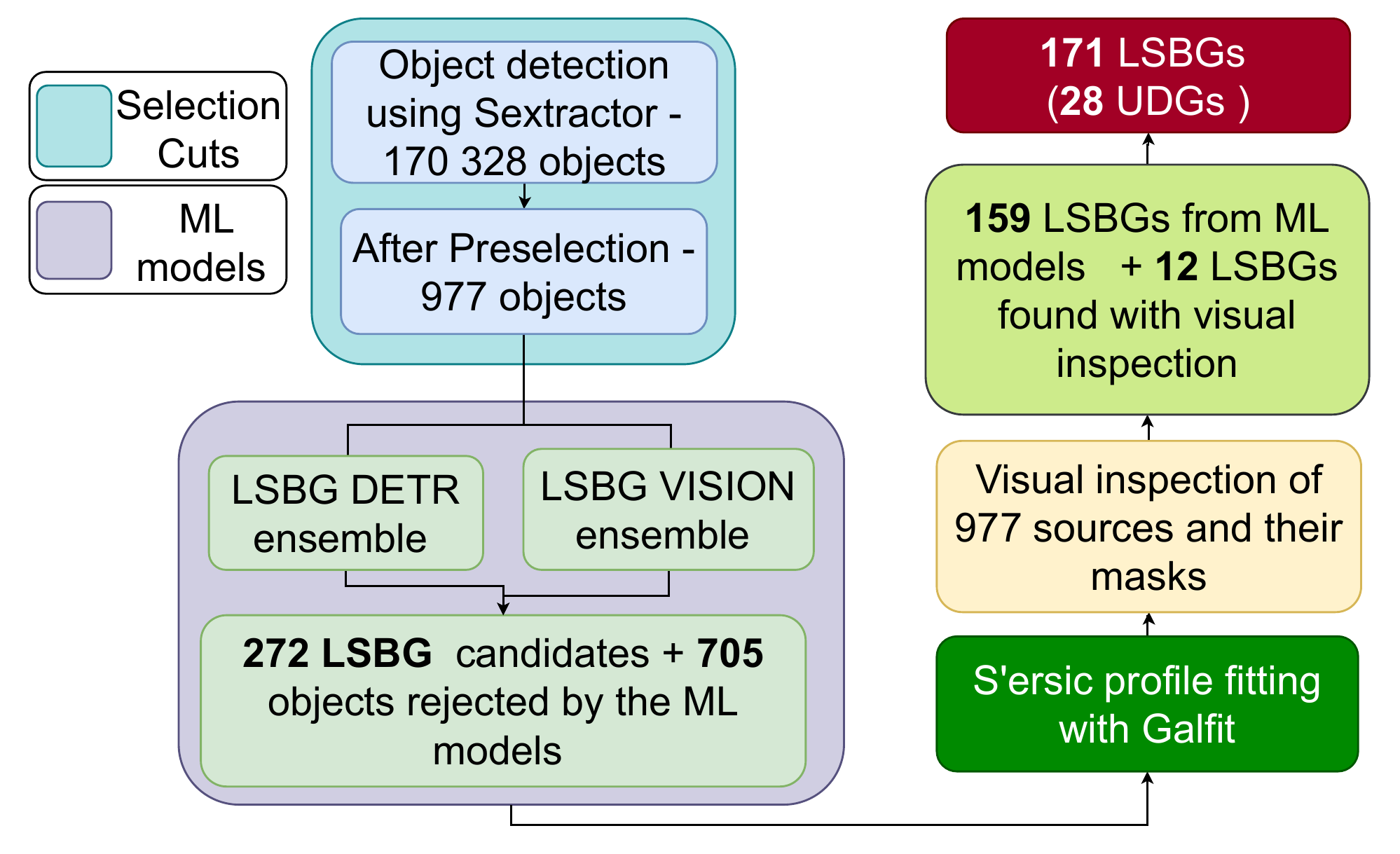}
    \caption{Schematic diagram showing the sequential selection steps used to find the LSBGs and UDGs in Abell 194.}
    \label{fig:candidate_slection_a194}
\end{figure}

\begin{figure*}
\begin{subfigure}{\linewidth}
  \includegraphics[width=\linewidth]{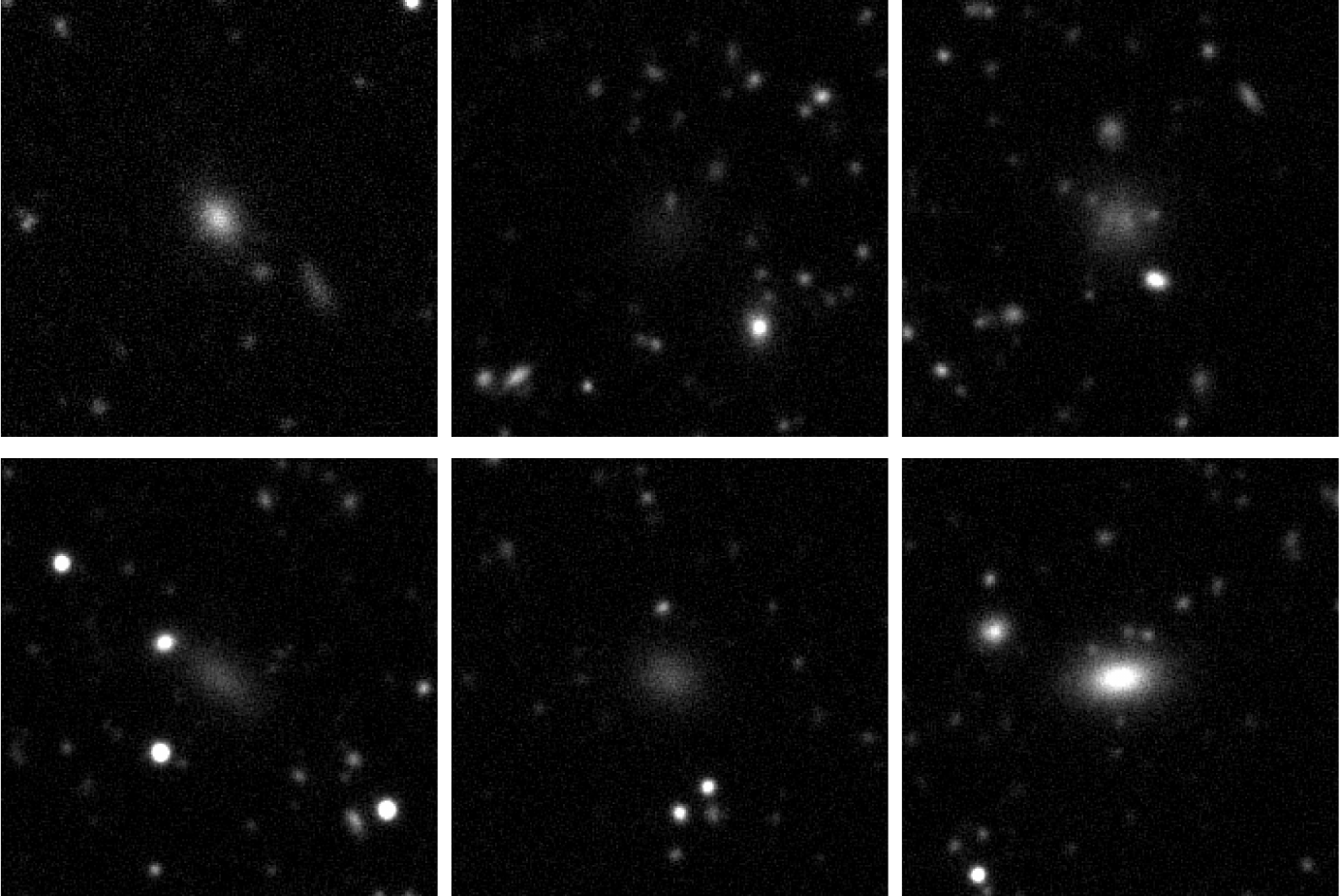}
\caption{Six examples of LSBGs identified from Abell 194.}
  \label{fig:lsbs_hsc}
\end{subfigure}
\begin{subfigure}{\linewidth}
  \includegraphics[width=\linewidth]{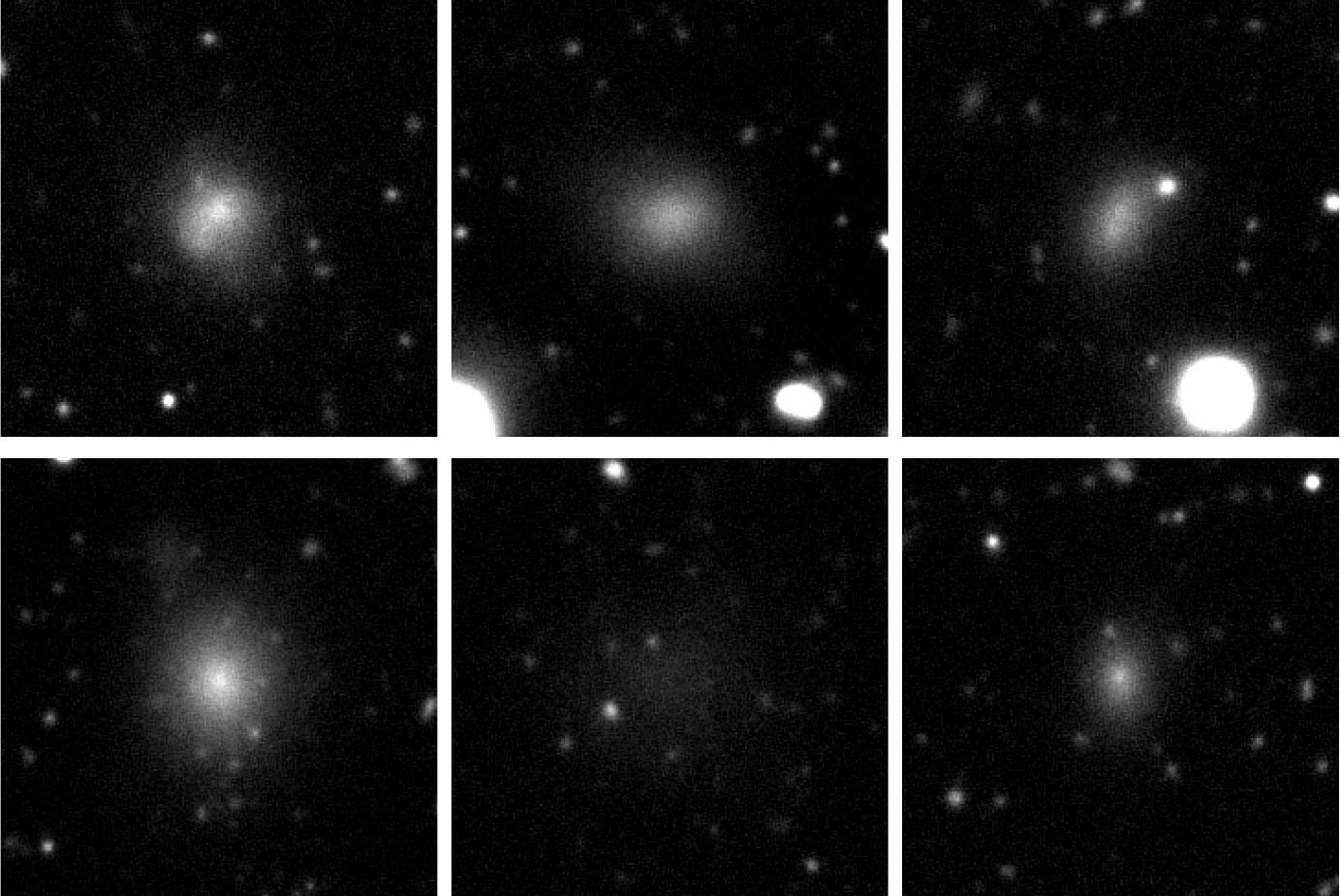}
\caption{Six examples of UDGs identified from Abell 194.}
  \label{fig:udgs_hsc}
\end{subfigure}
\caption{The top panel (\ref{fig:lsbs_hsc}) shows six examples of LSBGs identified from our study. The bottom panel (\ref{fig:udgs_hsc}) shows six examples of UDGs identified from our study. Each cutout of the LSBG and UDG corresponds to a $ 40 {\arcsec} \times  40 {\arcsec}$ $(152 \times 152$ pixels) region of the sky centred around the LSBG or the UDG in the \textit{g}-band. }\label{lsbs_udgs}
 \end{figure*}

To estimate the number of the LSBGs missed by our model, we also repeated the {\tt GALFIT} and visual inspection of all the 705 objects rejected by the ensemble models. This was done to estimate the number of FNs predicted by the model and to minimize their occurrence in the future by understanding why they were missed.  Our model missed 12 LSBGs, making the total number of LSBGs around the Abell 194 to 171. Thus, our model achieved a true positive rate (TPR) of $\sim 93\%$ on the HSC dataset without any fine-tuning.
The schematic diagram showing the sequential selection steps used to create LSBGs and UDGs in Abell 194 is shown in Fig. \ref{fig:candidate_slection_a194}. A sample catalogue comprising the properties of the identified LSBGs in this work is shown in Table \ref{table:HSC_catalog}. Six examples of LSBGs and UDGs identified from our study are shown in Fig. \ref{lsbs_udgs}.  


\citet{Girardi_abell_a94} and \citet{Rines_abell_194_z} have estimated the comoving radius of Abell 194 as 0.69 h$^{-1}$ Mpc. Plugging in the cosmological parameters, we calculate the virial radius, $r_{200}$ of Abell 194 to be 0.9824 Mpc.  12 among the 171 LSBGs are located slightly beyond the virial radius of the cluster, with the most distant LSBG being 1.15 Mpc away from the cluster centre. However, since LSBGs and UDGs have also been observed outside the virial radius in other clusters \citep{Venhola_dwarfs, Venhola_udgs, Junais2022}, for simplicity, we treat all 171 LSBGs as part of the Abell 194 cluster. Based on the redshift of Abell 194, 28 of the 171 identified LSBGs, with \re{} > 1.5 kpc and $\mu_{0,g} > 24.0$ \magperarcsec{}, are classified as UDGs. Only 2 of these 28 UDGs were missed by the ML models and required visual re-identification, consistent with a TPR of 93\%.

\begin{figure*}
\begin{subfigure}{0.49\linewidth}
  \includegraphics[width=\linewidth]{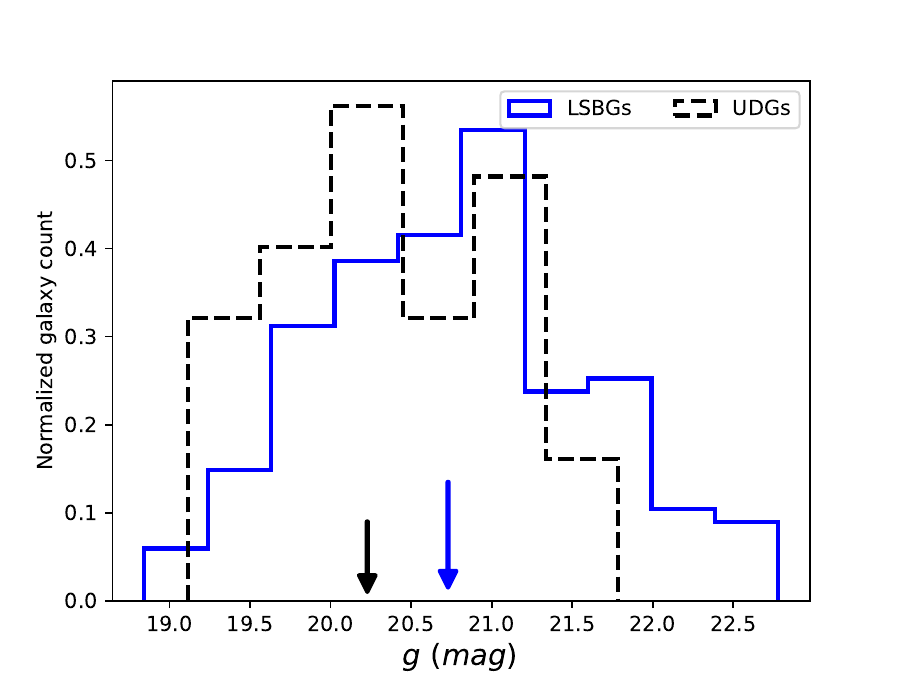}
\caption{}
  \label{fig:lsb_g_mag}
\end{subfigure}
\begin{subfigure}{0.49\linewidth}
  \includegraphics[width=\linewidth]{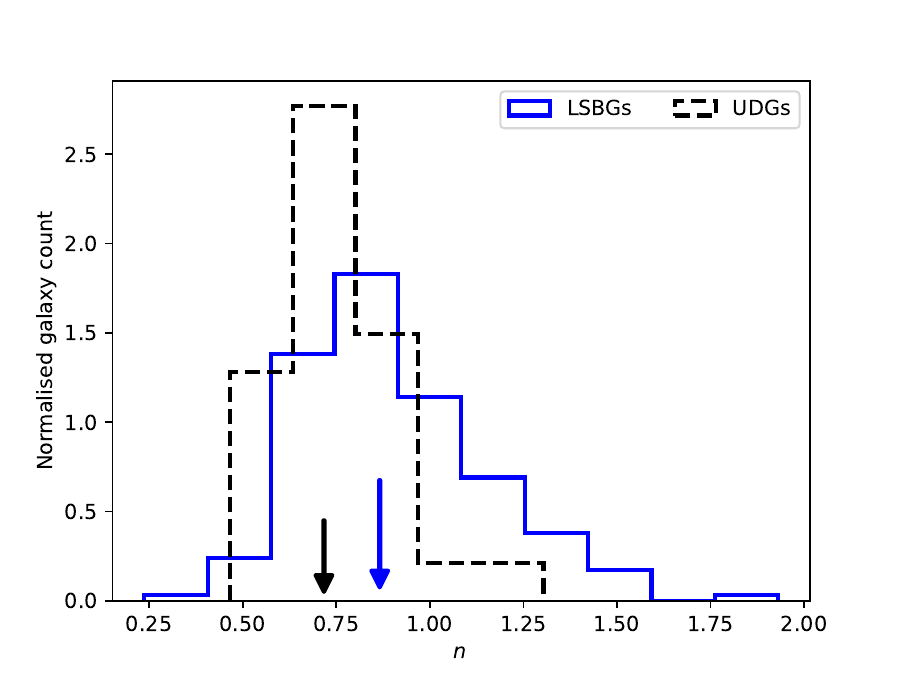}
\caption{}
  \label{fig:lsbn}
\end{subfigure}
\caption{The normalized distributions of the \textit{g}-band magnitude and S\'ersic index for the LSBGs and UDGs identified from HSC data in this study are shown in the left (\ref{fig:lsb_g_mag}) and right (\ref{fig:lsbn}) panels, respectively. The long black arrow shows the median of the UDG distribution, and the shorter blue arrow shows the median of the LSBG distribution in both plots.}
 \end{figure*}

The distribution of the \textit{g}-band magnitudes of LSBGs and UDGs is shown in Fig. \ref{fig:lsb_g_mag}. The UDGs have a median \textit{g}-band magnitude of $20.2\pm0.5$ mag, which is brighter than the median value of $20.7\pm0.6$ mag for the LSBGs. Here, the median absolute deviation represents the error associated with the median. This suggests that our sample includes very faint LSBGs that are not large enough to be classified as UDGs.
The distribution of the Sérsic index (\textit{n}) for LSBGs and UDGs is shown in Fig. \ref{fig:lsbn}. The Sérsic index for LSBGs predominantly lies between 0.5 and 1.5, with a median value of $0.87\pm0.14$. In contrast, the UDGs have a Sérsic index distribution with a median of $0.72\pm0.09$, which is slightly lower than that of the LSBGs. A detailed discussion of the properties of the LSBGs and UDGs identified in this work is presented in Sect. \ref{discs}. 

As discussed in Sect. \ref{aperture_photometry}, aperture photometry provides a more reliable estimate of apparent magnitude. We estimated the apparent magnitudes in the \textit{g}, \textit{r}, $NUV$, and $FUV$ bands, applying corrections for galactic extinction as outlined in Sect. \ref{aperture_photometry}. 20 LSBGs in our sample had confident detections in $NUV$ $(S/N>3\sigma)$, whereas only 15 LSBGs had confident detection in the $FUV$. For the rest of the sample, we estimated the upper limits of the $NUV$ and $FUV$ magnitude based on their $3\sigma$ values within the aperture. The aperture magnitudes were used to estimate the colors $g-r$, $NUV-r$, and $FUV-NUV$. The aperture magnitude of the \textit{r}-band was used to estimate the absolute magnitude ($M_r$) in the \textit{r}-band. Using the stellar mass-to-light ratio vs color relation for LSBGs from \citet{Du_2020}, we estimated the stellar masses ($M_\star$) of our sample based on their aperture g-r color. Similarly, the stellar mass surface density ($\Sigma_{\star}$) was calculated using the stellar mass-to-light ratio and the mean surface brightness ($\overline{\mu}_r$) in the \textit{r}-band, following Eq. 1 of \citet{Chamba2022}. A detailed discussion of the physical and cluster-centric properties of the LSBGs and UDGs identified in this study is provided in Sect. \ref{discs}.

\subsection{Transformers for cross-survey LSBG detection}

Here we explore the effectiveness of transfer learning for identifying LSBGs in the HSC dataset using transformer models from \citet{haressh_lsbs} trained on DES DR1. 
On the HSC dataset, they correctly identified 159 of 171 LSBGs, with a 93\% TPR, without fine-tuning. This demonstrates the potential of transfer learning with transformer models.

\begin{figure}[h]
\centering
\includegraphics[width=\linewidth,keepaspectratio]{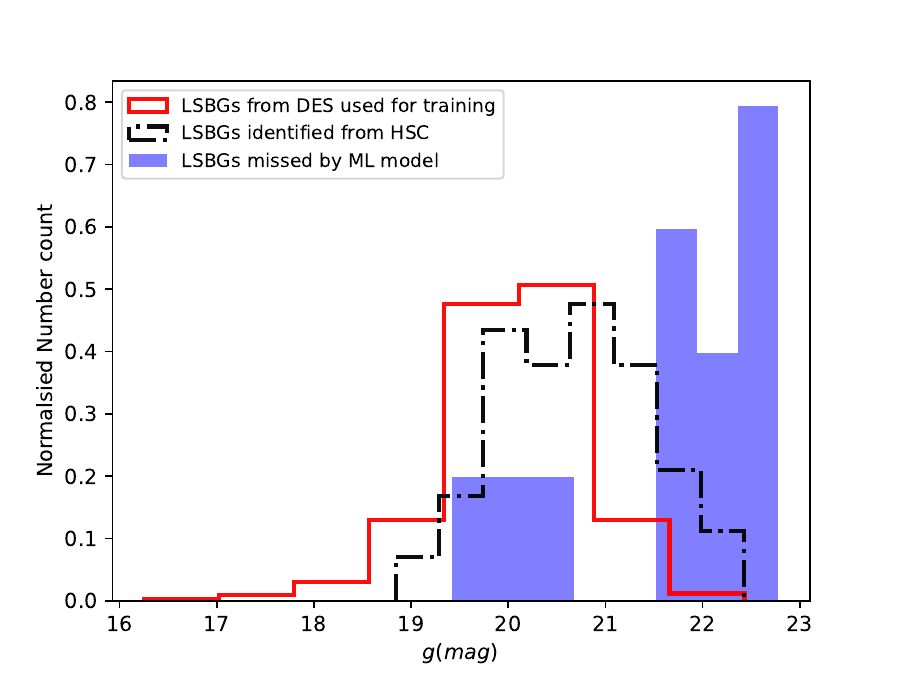}
\caption{Normalized distribution of \textit{g}-band magnitudes for the LSBGs used to train the ensemble models, LSBGs identified by the ensemble models in the HSC data of the Abell 194 cluster, and LSBGs missed by the ensemble models in the HSC data of the Abell 194 cluster.}
\label{LSBS_old_new_missed_mag_g}
\end{figure}

Although the model missed 12 LSBGs, which is a small fraction of the total $(\sim7\%)$, it is important to closely examine these missed cases to enhance model performance in the future. The magnitude distribution of the LSBGs in the training data, LSBGs identified and missed by the models, are presented in Fig. \ref{LSBS_old_new_missed_mag_g}. Here, the magnitude distribution of the training sample and LSBGs identified by the ensemble models share the same lower limit ($g \sim 22.5$ mag). However, only $\sim 1.7\%$ of LSBGs in the training data have $21.5 < g < 22.5$. Consequently, 9 out of the 12 missed LSBGs had $g<21.5$ mag and the performance of the models in identifying LSBGs with $g \geq 21.5$ mag is suboptimal. This suggests that the inability of the model to identify them was due to the underrepresentation of such faint LSBGs in the training data. Nevertheless, potentially, one could refine the ensemble models by retraining them with very faint LSBGs ($g>21.5$ mag) from simulations or small-scale deep surveys, thereby extending the models' performance to a fainter regime.

\begin{figure}[h]
\centering
\begin{subfigure}{0.49\linewidth}\label{fig:missed_LSBS_fig1}
\centering
\includegraphics[width=\linewidth,keepaspectratio]{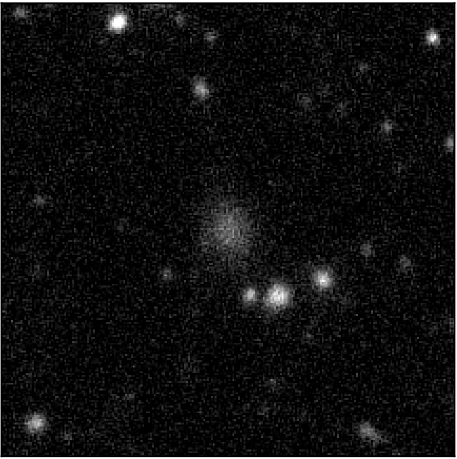}
\caption{}
\end{subfigure}
  \begin{subfigure}{0.49\linewidth}
\centering
\includegraphics[width=\linewidth,keepaspectratio]{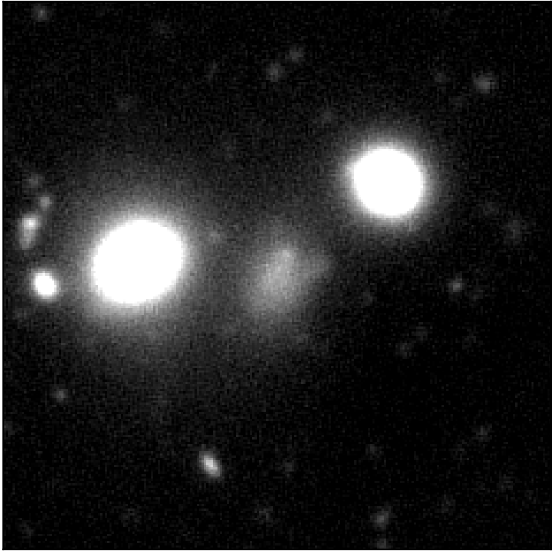}\label{fig:missed_LSBS_fig2}
\caption{}
\end{subfigure}
\caption{Example \textit{g}-band images of LSBGs missed by the ensemble models. The left panel shows an LSBG, which is fainter than the training sample, and the right panel shows an LSBG near bright galaxies.}
\label{fig:missed_LSBS_fig}
\end{figure}

Similarly, analysing the other 3 LSBGs missed by the ensemble model, which had $g<21$ mag, it was found these galaxies were very close to a bright galaxy. This caused the model to confuse them as part of a bright galaxy and classify them as a non-LSBG. The presence of very bright objects introduces a bias in the prediction probabilities of the ensemble models, and it was also found that the effect is greater if the bright object is close to the centre of the cutout. Examples of the LSBGs missed by the model are shown in Fig. \ref{fig:missed_LSBS_fig}. 

A total of 101 objects initially classified as LSBGs were reclassified as non-LSBGs after further analysis using {\tt GALFIT} and visual inspection. The rejected sample predominantly comprised three types of objects: faint galaxies that did not meet the LSBG criteria, with \mue{} in the range 24.0–24.2 \magperarcsec{} and \re{} between 2\arcsec and 2.5\arcsec (approximately 60\%), very faint blended sources (about 10\%), and instrumental contaminants from HSC (around 30\%). Simple selection cuts on the morphological parameters can remove the first class. Similarly, the third class of objects could be removed by having a representative of instrumental contaminants in the training set. Alternatively, applying selection cuts based on the convergence of the S\'ersic fit reduces these contaminants, as most do not converge. The second class of contaminants are generally very faint and are hard to remove. While their numbers are small compared to LSBGs, they still require visual inspection. Notably, even human inspectors will have trouble distinguishing these faint blended sources from LSBGs. For instance, 14 of the 17 sources where the two authors disagreed during independent visual inspection had \mue{} > 26.0 \magperarcsec{}. 


Large-scale surveys like LSST and Euclid are expected to detect  $10^5+$ LSBGs, necessitating efficient methods to analyze large volumes of data. Successful implementation of transfer learning will be pivotal in this effort, allowing the application of knowledge from existing datasets to process data from upcoming surveys effectively. Models with high TPR will ensure better completeness by capturing a larger fraction of genuine LSBGs, while low false positive rates (FPR) will enhance sample purity by minimizing contamination. Generally, TPR and FPR should be balanced based on the classification problem. In unbalanced datasets, reducing FPR is key, while maximizing TPR helps when contaminants can be pre-filtered. Furthermore, advanced ML methodologies will be essential for addressing the challenges posed by the scale and complexity of future datasets.
\section{Discussion}\label{discs}
\subsection{Comparison with previous catalogues of LSBGs and UDGs in Abell 194}

In total, \citet{Tanoglidis1} and \citet{haressh_lsbs} identified 96 LSBGs in Abell 194, which we call the \textit{LSBGs-DES} sample. Among these, 19 LSBGs could be classified as UDGs based on their \re{} and $\mu_{0,g}$ measured from DES. Similarly, \citet{Zaritsky_2023} identified 14 UDGs in Abell 194, with 9 overlapping both the \textit{LSBGs-DES} sample and the subset of 19 UDG candidates from \textit{LSBGs-DES}.
In our final sample, the ensemble model identified 13 UDGs from \citet{Zaritsky_2023}. One missing UDG was found later when reviewing the rejected sources. Similarly, 95 out of 96 from the \textit{LSBGs-DES} sample were re-identified by the ensemble model. One LSBG, missed by the model, was subsequently found during visual inspection. 

However, the final catalogue contains only 84 LSBGs from the \textit{LSBGs-DES} sample. The remaining 12 were reclassified as non-LSBGs: 11 galaxies were slightly brighter (24.0 <\mue{}$<24.2$) than our threshold, and one LSBG was found to have a higher redshift (z=0.15) when cross-matched with the SDSS DR 16 sample. 
Additionally, 7 UDG candidates based on DES measurements were reclassified as non-UDGs due to having a physical radius smaller than 1.5 kpc, and one due to a $ \mu_{0,g} $ 
brighter than 24.0 \magperarcsec{}. Similarly, 2 UDGs from \citet{Zaritsky_2023} were reclassified as non-UDGs which was also a subset of the \textit{LSBGs-DES} sample: one due to a $ \mu_{0,g} $ brighter than 24.0 \magperarcsec{}, and the other due to a physical radius smaller than 1.5 kpc.

One of the primary reasons for rejecting some LSBGs and UDGs from previous catalogues is the improved masking strategy implemented in this study. These improved masks effectively removed small contaminants overlaid on the galaxies, as shown in Fig. \ref{fig:lsb_masked}. In shallow data, these contaminants might not be clearly visible, but with deeper data, these contaminants are clearly visible and significantly affect the estimation of the morphological parameters such as S\'ersic index and half-light radius during a S\'ersic fit. 
The measurements from DES were performed without masking the contaminants that may reside on top of the LSBGs. As these contaminants are very compact objects, the presence of these contaminants tends to increase the S\'ersic index of the LSBG during fitting.
Similarly, because of the additional light from these contaminants, the \re{} of these galaxies also tends to be over-estimated. However, the presence of these contaminants does not significantly affect the magnitude of the galaxy. These trends were also observed by \citet{Bautista} for the estimated S\'ersic parameters of UDGs from the Coma cluster.



An additional factor contributing to the deviation in the S\'ersic parameters between this study and those reported by DES is the local sky subtraction. Previous searches for LSBGs in DES by \citet{Tanoglidis1} and \citet{haressh_lsbs} did not consider the local sky subtraction. The sky background in publicly available DES data is estimated with regions of size $\sim 1'$ \citep{Morganson, Bernstein_DES_sky_bg}. This is considerably larger than the size of the LSBGs we investigate in this work. Sky subtraction over a large area tends to over-subtract the sky due to the influence of bright objects. This can result in LSBGs appearing fainter than their actual surface brightness. Hence, failing to correct for local sky subtraction can lead to inaccurate S\'ersic fits, as also highlighted by \citet{Bautista}.

Despite excluding some LSBGs from the previous DES sample, the number of LSBGs in the Abell 194 cluster has increased from 84 to 171. This rise is partly due to the deeper sensitivity of the HSC data, which is two magnitudes deeper than DES, but not solely because of it. Preselection criteria applied with {\tt SExtractor} also played a significant role. For instance, the ratio of \re{} estimated by {\tt SExtractor} to {\tt GALFIT} has a median value of 0.8, indicating that {\tt SExtractor} underestimates the radius. A similar ratio was found in the LSBG sample by \citet{Tanoglidis1}. As a result, we relaxed our preselection to {\tt FLUX\_RADIUS\_G} $>2.0\arcsec$, compared to the $>2.5\arcsec$ threshold used by \citet{Tanoglidis1} and \citet{haressh_lsbs}. Without this adjustment, only 25 new LSBGs would have been identified, which is a 30\% increase. For UDGs, preselection had less impact, as only 3 of 28 UDGs had {\tt FLUX\_RADIUS\_G} between $2-2.5\arcsec$. However, if the cluster were slightly farther away ($z \gtrsim 0.025$), the preselection effects would become significant, as the angular size of a UDG would approach $2.5\arcsec$.

\begin{figure}[h]
\centering
\includegraphics[width=\linewidth,keepaspectratio]{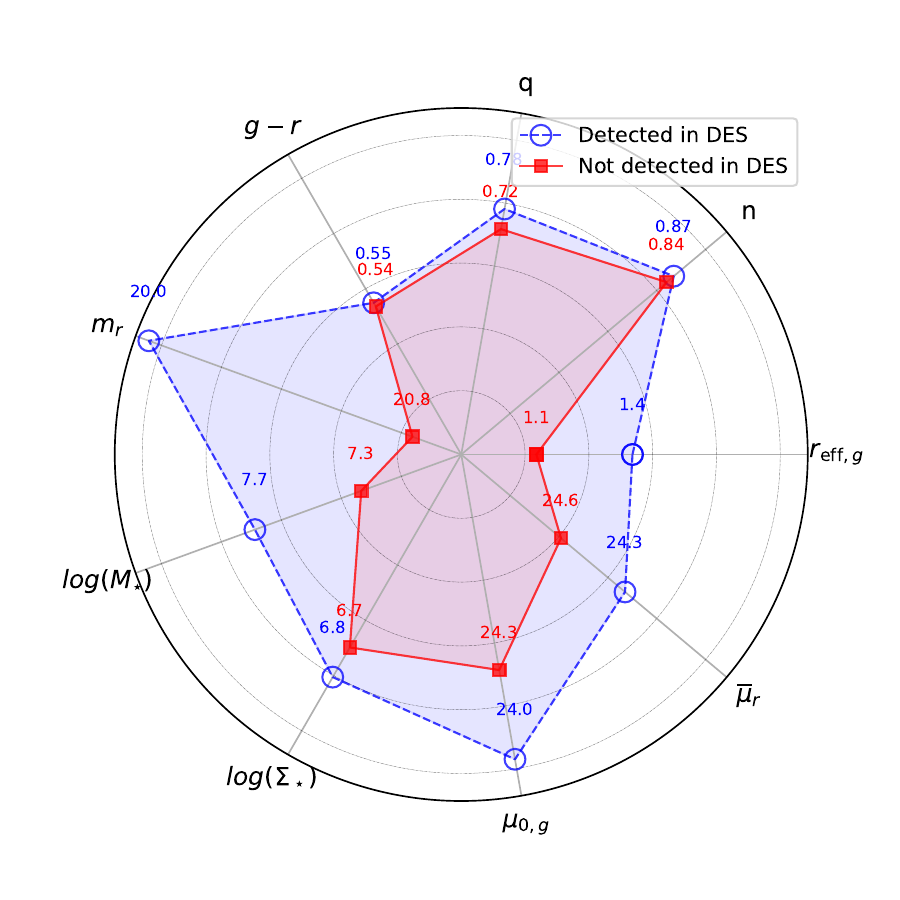}
\caption{Comparison of morphological and physical properties of LSBGs identified from DES and the LSBGs identified only in HSC. The median of $g-r$ (mag), $q$, $n$, $r_{\mathrm{eff},g}$ (kpc), $\overline{\mu}_r$ (mag arcsec$^{-2}$), $\mu_{0,g}$ (mag arcsec$^{-2}$), log $\Sigma_{\star}$ ($M_{\odot}$ kpc$^{-2}$), log $M_{\star}$ ($M_{\odot}$) and $m_{r}$ (mag) are shown in the clockwise order.}
\label{new_vs_old_lsbs}
\end{figure}

The radar plot comparing the properties of the \textit{LSBGs-DES} sample and the new LSBGs identified with HSC in Abell 194 is shown in Fig. \ref{new_vs_old_lsbs}. In terms of color $(g-r$), $q$ and $n$, both samples have the same median indicating that both samples are similar in these properties. As mentioned earlier, the new sample of LSBGs has many small LSBGs in size, which is evident from the median values of the LSBGs detected from DES ($1.4\pm 0.2$ kpc) and not detected in DES ($1.1\pm0.1$ kpc). Similarly, as expected, the new sample of LSBGs is fainter than the \textit{LSBGs-DES} sample. This is evident from the median values of $\overline{\mu}_r$, $\mu_{0,g}$, log$\Sigma_{\star}$, log$M_{\star}$, and $r_{aperture}$ when comparing LSBGs detected in DES to those not detected in DES. 

\subsection{Properties of LSBG and UDG population of Abell 194 cluster}
As mentioned in Sect. \ref{introduction}, UDGs are a subclass within LSBGs, characterized by their extended half-light radii ($r_e \geq 1.5$ kpc) and high central surface brightness ($\mu_0 > 24$ mag/arcsec$^2$) in the \textit{g}-band \citep{Dokkum}. Assuming all the LSBGs in this work share the same redshift as Abell 194, we estimate that our sample has 28 UDGs and 143 non-UDGs. However, some may be foreground or background galaxies, and individual redshifts are required to confirm their UDG status.

\subsubsection{Abundance of UDGs}
\begin{figure}[h]
\centering
\includegraphics[width=\linewidth,keepaspectratio]{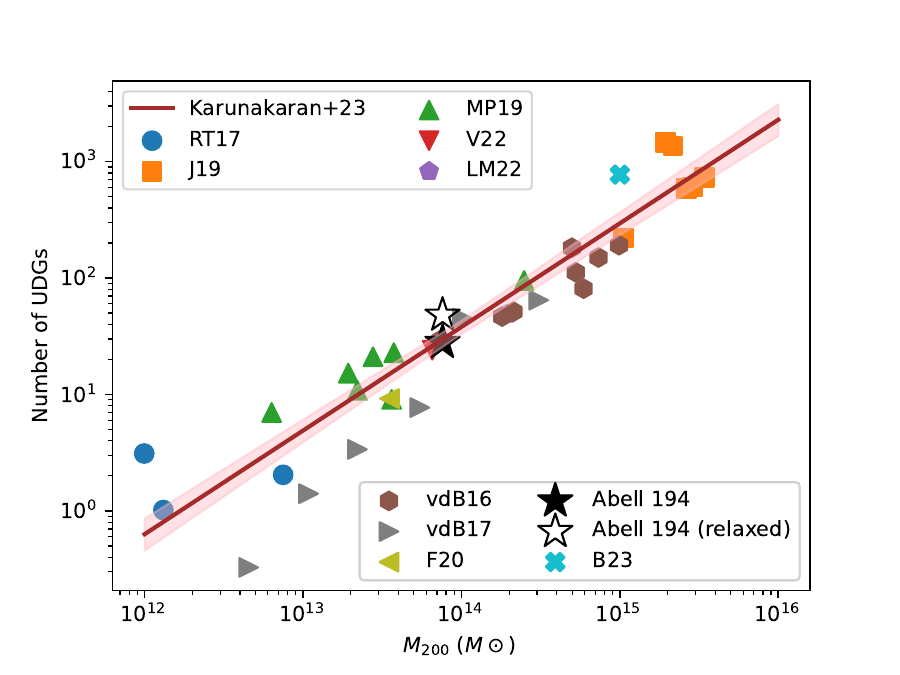}
\caption{The number of UDGs as a function of halo mass includes data from various studies. Cluster UDGs are represented by brown hexagons (\citealt{Burg}), orange squares (\citealt{Janssens}), green upward triangles (\citealt{Mancera_udg}), red downward triangles (\citealt{Venhola_udgs}), purple pentagons (\citealt{2022AMarca}), and cyan crosses (\citealt{Bautista}). Group UDGs are represented by grey right-pointing triangles (\citealt{vanderBurg2017}), blue circles (\citealt{Roman_2017}), and olive left-pointing triangles (\citealt{Forbes}). The UDGs reported in this work are shown as a black star, while the UDG sample with a relaxed definition, similar to \citet{Karunakaran}, is represented by a hollow black star.}
\label{N_udgs_halo_mass}
\end{figure}

Since \citet{Dokkum} found a substantial population of UDGs within the Coma cluster, the subsequent studies have identified numerous UDGs in various galaxy clusters \citep{Koda, Yagi, Mihos, Lim, Marca, Bautista}. Further investigations have demonstrated that not only clusters, groups are also abundant with UDGs \citep{Merritt,Roman_2017, muller, Somalwar,Forbes} and their abundance has a near-linear correlation in the log scale with the mass of the dark matter halo of the cluster \citep{vanderBurg2017, Mancera, Karunakaran}. \citet{Karunakaran} estimated that the number of UDGs scale as a function of the halo mass following the relation $N = (38\pm5)(\frac{M_{200}}{10^{14}})^{0.89\pm 0.04}$. For the Abell 194 cluster, which has a halo mass of 7.6 $\times$ $10^{13} M{\odot}$, this relation predicts $30 \pm 4$ UDGs.  The number of UDGs as a function of host halo mass, including the Abell 194, is shown in Fig. \ref{N_udgs_halo_mass}. 

However, since the UDG definition in the literature is not unique, the sample used by \citet{Karunakaran} is brighter ($\overline{\mu}_g>24.0 \text{ mag arcsec}^{-2}$) than our UDG sample which has ${\mu_{0,g}}>24.0 \text{ mag arcsec}^{-2}$. Using the same definition as \citet{Karunakaran}, we identified 45 LSBGs that meet the criteria to be considered UDGs, referred to as the 'relaxed' UDG sample. This number exceeds the value predicted by \citet{Karunakaran}, as shown in Fig. \ref{N_udgs_halo_mass}. As our primary objective does not focus on constraining the relationship between halo mass and the number of UDGs, we do not re-calibrate this relation and have reserved it for potential future work.

\subsubsection{UDG surface number density}
\begin{figure}[h]
\centering
\includegraphics[width=\linewidth,keepaspectratio]{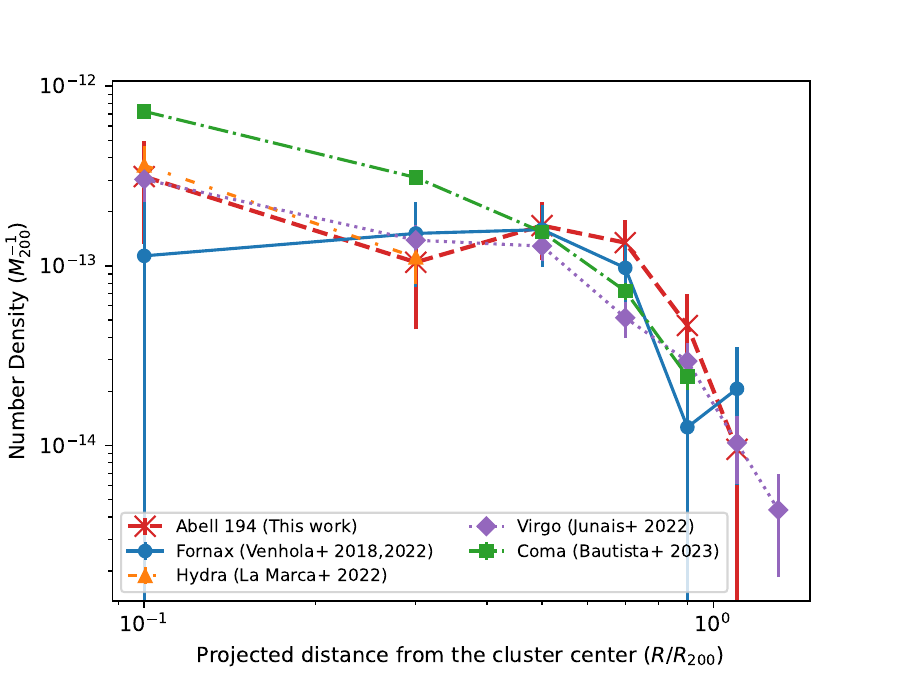}
\caption{The radial surface density profile of cluster UDGs, normalized by cluster mass ($M_{200}$), as reported in this work and the literature. UDGs are from \citealt{Venhola_dwarfs, Venhola_udgs} (Fornax; blue circles), \citealt{2022AMarca} (Hydra; orange triangles), \citealt{Junais2022} (Virgo; purple rhombus), and \citealt{Bautista} (Coma; green squares).}
\label{udg_surface_number_density}
\end{figure}

The spatial distribution of the UDGs in the clusters can be used to probe how the environment affects the formation of the UDG population. The radial surface density profile of UDGs from Abell 194, as well as the UDGs from the literature, is shown in Fig. \ref{udg_surface_number_density}. Given that more massive clusters tend to host a greater number of UDGs, as indicated in Fig. \ref{N_udgs_halo_mass}, we normalized the radial surface density of each cluster by its mass to standardize the comparison of UDG distributions across different clusters. In addition, since the UDG definition in the literature is not unique, we selected the UDG samples in each cluster based on the definition used by \citet{Dokkum}. 

The normalized surface number density distribution of UDGs in the cluster follows the same trend as shown in Fig. \ref{N_udgs_halo_mass}. Since the number of UDGs scales with cluster mass, the surface number density normalized by cluster mass remains consistent across different clusters, as evident in Fig. \ref{udg_surface_number_density}. Furthermore, the spatial variation of the normalized UDG surface number density is uniform beyond $0.3R_{200}$ for various clusters. However, near the cluster centre, the values show variation, indicating that UDG formation in the inner region is influenced by factors other than halo mass. A more detailed study is needed to identify these additional factors.

We also observe that the Abell 194, Hydra \citep{2022AMarca}, and Virgo \citep{Junais2022}\footnote{We use the sample of 114 UDGs reported in Appendix B of \citet{Junais2022}}. clusters exhibit similar normalized UDG surface number density distributions near the cluster centre ($<0.3R_{200}$), suggesting that UDG formation in these clusters may follow a similar pattern. However, further investigation and follow-up studies are needed to confirm this. In contrast, the Fornax cluster \citep{Venhola_dwarfs, Venhola_udgs} shows a decline in the normalized number density near the cluster centre, which could indicate that UDGs either do not survive in the inner region of the Fornax cluster or they have not been detected yet.

\subsubsection{LSBGs vs UDGs}

\begin{figure}[h]
\centering
\includegraphics[width=\linewidth,keepaspectratio]{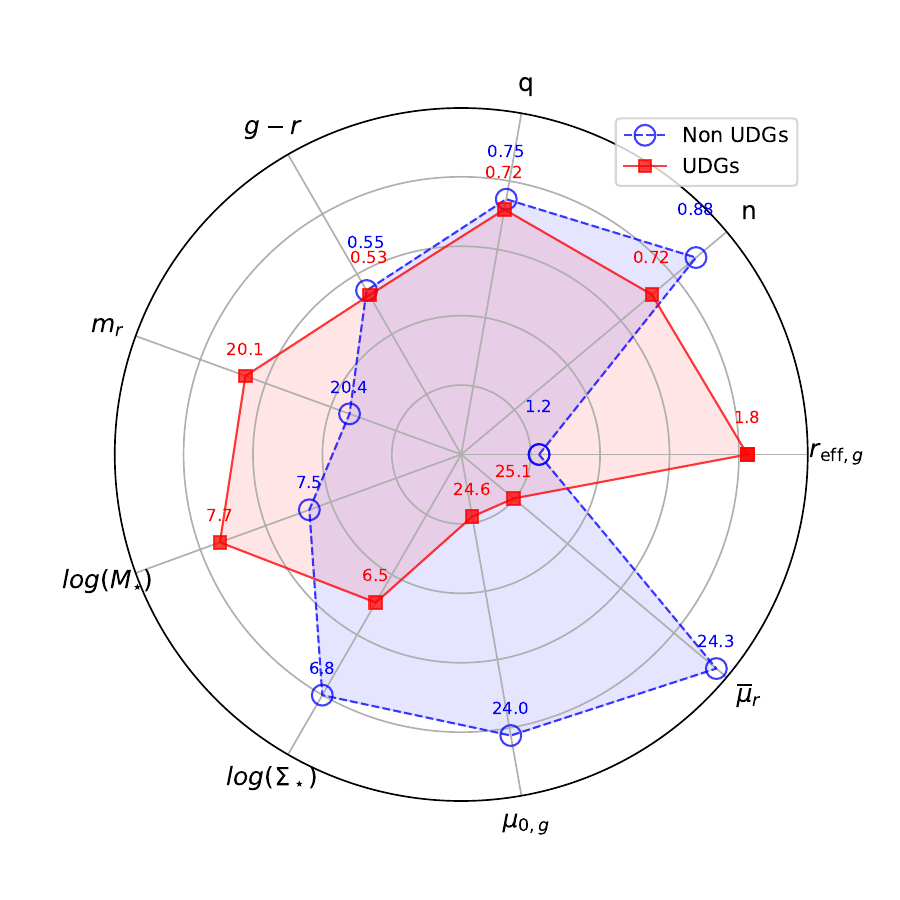}
\caption{Comparison of morphological and physical properties of UDGs and non-UDGs. The median of $g-r$ (mag), $q$, $n$, $r_{\mathrm{eff},g}$ (kpc), $\overline{\mu}_r$ (mag arcsec$^{-2}$), $\mu_{0,g}$ (mag arcsec$^{-2}$), log $\Sigma_{\star}$ ($M_{\odot}$ kpc$^{-2}$), log $M_{\star}$ ($M_{\odot}$) and $m_{r}$ (mag) are shown in the clockwise order.}
\label{udg_non_udg}
\end{figure}

As the UDGs are considered a subclass of LSBGs, one might be curious about the difference in the properties of UDGs and non-UDGs (LSBGs that did not satisfy the condition to be a UDG). The radar plot comparing the median distribution of all the properties estimated in this work for the UDGs and non-UDGs is shown in Fig. \ref{udg_non_udg}. Comparing the median value of $g-r$ color, we can see that both UDGs ($0.53\pm0.04$) and non-UDGs $0.55\pm0.03$ have similar color distribution. The median color of the UDGs reported here is consistent with the median color of the UDG sample presented in \citealt{Mancera_udg} (0.59) and \citealt{Bautista} (0.55). Similarly, in terms of axis ratio also, both the UDGs ($0.72\pm0.09$) and non-UDGs ($0.75\pm0.14$) have a similar distribution, indicating that they have a slightly elongated shape.
Coming to the S\'ersic index, the non-UDGs ($0.88\pm0.14$) and UDGs ($0.72\pm0.09$) have comparable median values. These values are comparable with the values reported for the UDGs in \citealt{Mancera_udg} (0.96) and \citealt{Bautista} (1.0).

The median \re{} for UDGs ($1.8\pm0.2$ kpc) is larger than that of non-UDGs ($1.2\pm0.2$ kpc), which is a consequence of the definition of UDGs. UDGs also have fainter median $\mu_{0,g}$ ($24.6\pm0.5$ \magperarcsec{} vs. $24.0\pm 0.5$ \magperarcsec{}) and $\overline{\mu}_r$ ($25.1 \pm 0.5$\magperarcsec{} vs. $24.3 \pm 0.5$ \magperarcsec{}) compared to non-UDGs. The stellar mass surface density of UDGs ($10^{6.5\pm0.2}$ M$_{\odot}$ kpc$^{-2}$) is slightly lower than that of non-UDGs ($10^{6.8\pm0.2}$ M$_{\odot}$ kpc$^{-2}$). These trends in  $\mu_{0,g}$,  $\overline{\mu}_r$ and M$_{\odot}$ are consistent with the UDGs being fainter extended sources.  However, UDGs have higher median total stellar mass and a brighter median apparent magnitude in the \textit{r}-band, indicating they contain more stellar mass and emit more light, but their extended structure results in fainter surface brightness.

\subsubsection{LSBGs, UDGs and dwarf galaxies}
The morphological parameters, such as the S\'ersic index of LSBGs and UDGs, show a distribution similar to that of dwarf ellipticals reported by \citet{Poulain}. Simulations further indicate that the dark matter halos of UDGs and dwarf galaxies have comparable masses \citep{Amorisco, Cintio, Sales, Jiang_2019, Benavides_2023}. These results suggest that the UDGs in our sample may simply be extended dwarf galaxies rather than a distinct population of galaxies.

\begin{figure}[h]
\centering
\includegraphics[width=\linewidth,keepaspectratio]{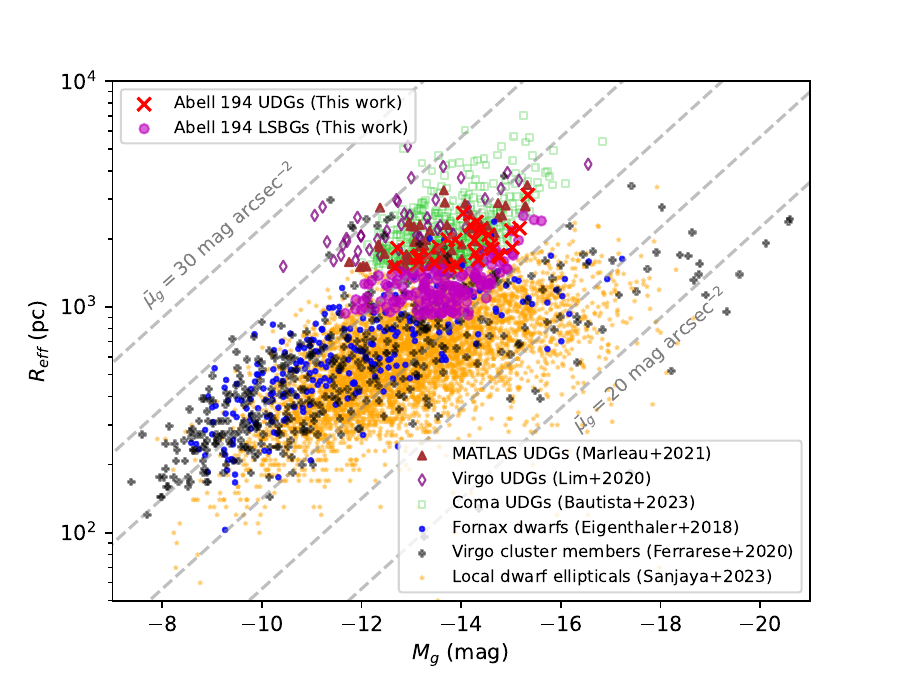}
\caption{The size-luminosity relation for local dwarf ellipticals, LSBGs, and UDGs from various studies is shown, with \re{} (in parsecs) on the y-axis and $M_g$ (mag) in the $g$-band on the x-axis. Red crosses and magenta circles represent the LSBGs and UDGs from this work. Brown triangles, purple diamonds, and lime green squares indicate UDGs from the MATLAS survey \citep{Marleau}, the Virgo cluster \citep{Lim}, and the Coma cluster \citep{Bautista}. The blue circle marks the Fornax dwarf from \citep{Eigenthaler_2018}, orange stars represent local dwarf ellipticals from \citep{Paudel_sanjaya}, and black pentagons show low-mass Virgo cluster members from \citep{Ferrarese_2020}. The grey lines represent constant surface brightness values, ranging from $\overline{\mu}_g = 20$ to $\overline{\mu}_g = 30$ mag arcsec$^{-2}$, with each subsequent line increasing by 2 magnitudes.}
\label{size_luminosity}
\end{figure}

Different galaxy populations can be distinguished by their structural parameters and scaling relations, which vary among galaxy families \citep{Kormendy_1985, Kormendy_2009}. A comparison of the size-luminosity relations for local dwarf ellipticals \citep{Eigenthaler_2018, Ferrarese_2020, Paudel_sanjaya} and UDGs \citep{Lim, Marleau, Bautista} from the literature, along with the LSBG and UDG populations identified in this work, is presented in Fig. \ref{size_luminosity}.

The overlap between dwarf galaxies and UDGs in the $R_{\mathrm{eff}}-M_g$ plane suggests that UDGs are an extended subset of dwarf galaxies, occupying the diffuse, low surface brightness end of the size-luminosity relation. These findings also align with previous studies \citep{Conselice, Marleau, Wang_2023, Bautista, zoller_coma}. UDGs from different clusters also occupy similar regions in this plane, indicating that their formation channels are likely consistent across clusters. Nonetheless, environmental factors may also play a role, as Coma, the most massive cluster, hosts the largest UDGs, while the Virgo cluster contains the faintest UDGs. Since these UDG samples are taken from different surveys, this conclusion may be biased by the surface brightness limits of these surveys as well.

The majority of LSBGs in our sample exhibit structural parameters similar to those of local dwarf galaxies, suggesting a close connection between these populations. This is further supported by the findings of \citet{Lazar_dwarf_morphology}, who observed that the red-to-blue fraction of dwarf galaxies closely mirrors that of LSBGs, reinforcing the idea that most LSBGs are indeed dwarf galaxies. Furthermore, the lack of a clear boundary in the size-luminosity plane between dwarf galaxies, LSBGs, and UDGs implies that these populations may form a continuous spectrum rather than distinct classes.

\subsection{Cluster-centric properties of LSBGs and UDGs}
\begin{figure}[h]
\centering
\includegraphics[width=\linewidth,keepaspectratio]{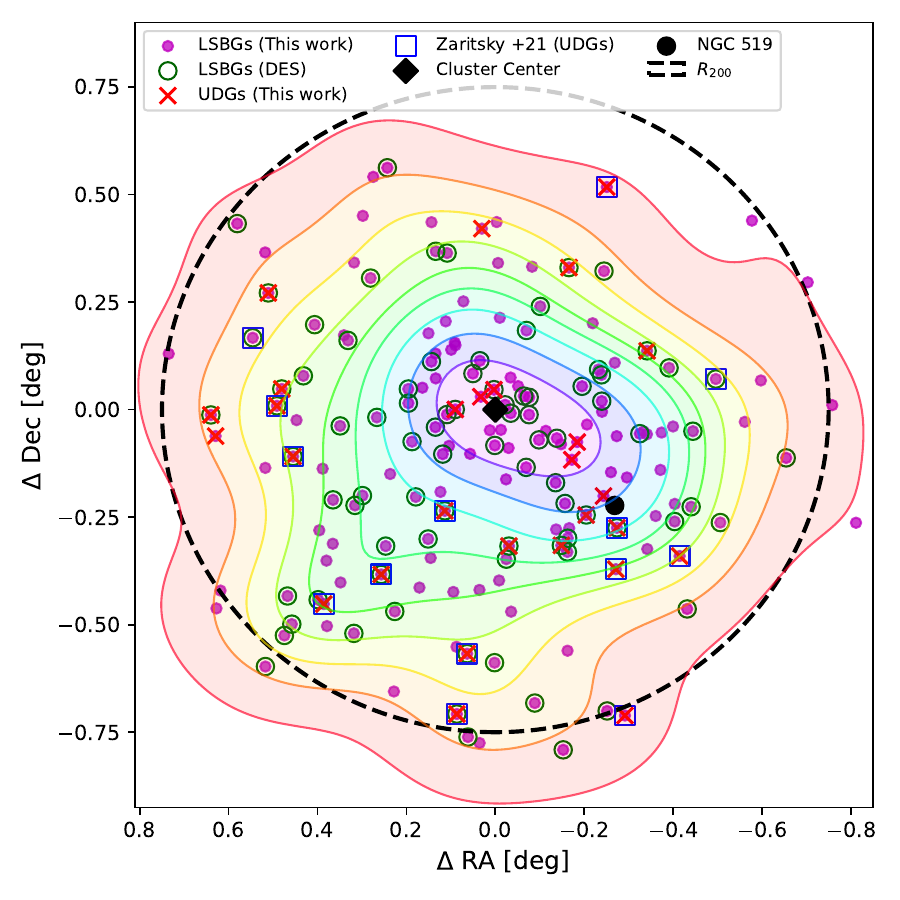}
\caption{Spatial distribution of LSBGs and UDGs in the Abell 194 cluster. LSBGs identified in this work are shown as magenta circles, and UDGs from this work are shown as red crosses. LSBGs identified in DES by \citet{Tanoglidis1} and \citet{haressh_lsbs} are marked with dark green circles, and UDGs reported in \citet{Zaritsky_2023} are represented by blue squares. The KDE highlights regions with varying concentrations of LSBGs, ranging from blue to red, with blue indicating areas of higher density. The black dashed circle, the black diamond, and the black dot represent the virial radius, the cluster centre, and NGC 519, respectively.}
\label{LSB_UDG_scatter_plot}
\end{figure}

The spatial distribution of LSBGs and UDGs in the Abell 194, along with the spatial kernel density estimate (KDE) of LSBGs, is presented in Fig. \ref{LSB_UDG_scatter_plot}. The KDE provides a smooth estimate of the density of these galaxies across the cluster, highlighting areas with higher concentrations of LSBGs. The Abell 194 has a tail of bright galaxies from the cluster centre to the direction of NGC 519, which are found to be falling towards the cluster centre  \citep{Tempel_cluster_memebership}. The spatial density of LSBGs at the cluster centre aligns with the bright galaxies in the direction of NGC 519, suggesting that LSBGs may follow the bright galaxies toward the cluster centre and could be their satellite galaxies.

\subsubsection{Trends in morphological and physical properties with projected cluster-centric distance}
\begin{figure}[h]
\centering
\includegraphics[width=\linewidth,keepaspectratio]{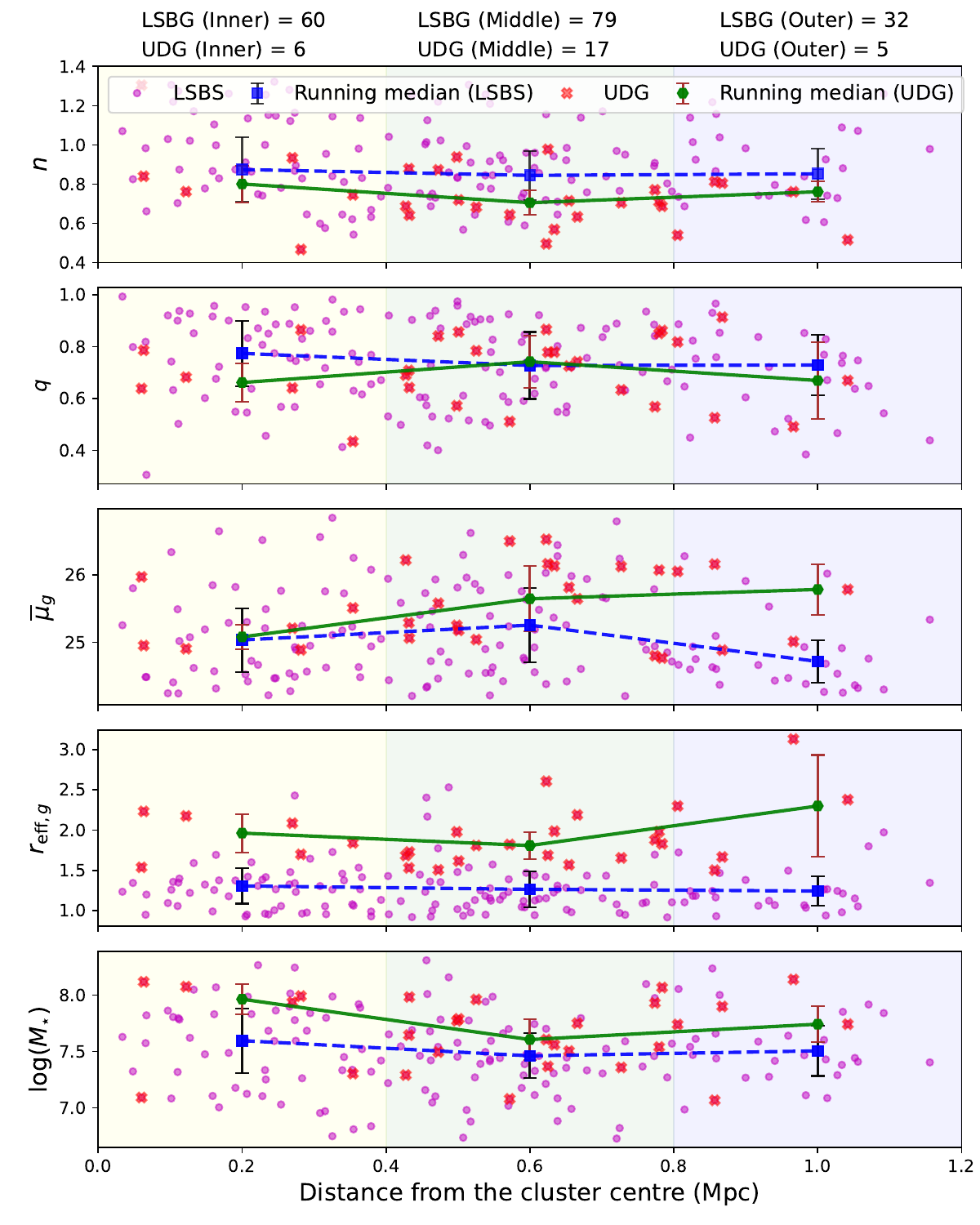}
\caption{Morphological and physical properties of LSBS and UDGs as a function of cluster centric distance. The x-axis shows the distance from the cluster centre in Mpc.  The LSBGs and UDGs identified in this work are shown in magenta dots and red crosses, respectively. The running median of LSBGs and UDGs are calculated with a bin size of 0.4 Mpc and are represented as blue dotted and green lines, respectively. The inner, middle and outer regions of the cluster are shaded with yellow, green and blue, respectively.}
\label{morphological_properties}
\end{figure}

The projected distance can serve as a proxy for density within clusters and can be used to analyse how the environment affects the structural properties of an LSBG or UDG. The structural and physical properties of the LSBGs and UDGs as a function of the cluster-centric distance are shown in Fig. \ref{morphological_properties}. 
To facilitate comparison, we assign the area within a radius of less than 0.4 Mpc around the cluster centre as the inner region of the cluster, while the region beyond a radius of 0.8 Mpc is considered the outer skirts of the cluster. The region in between is considered the middle region of the cluster. 

The median value of the S\'ersic index shows no significant trend for both LSBGs and UDGs as a function of the cluster-centric distance. Similarly, the median axis ratio of LSBGs tends to be relatively flat throughout the regions. In contrast, the median axis ratio of UDGs tends to be $\sim 0.6\pm0.1$ near the inner region of the cluster, increases to $\sim 0.7\pm0.1$ in the middle, and decreases again towards the outer region of the cluster. 
Our observations on the S\'ersic index and axis ratio align well with the observations by \citet{Roman_2017, Mancera_udg} and are in good agreement with expectations from models of dwarf galaxies that have undergone harassment and tidal interaction processes \citep{Moore1996, Aguerri}.

The surface brightness distribution shown in Fig. \ref{morphological_properties} indicates that LSBGs and UDGs in the inner regions of the cluster are brighter than those in the middle region. The absence of faint LSBGs and UDGs in the inner regions could be due to the destruction of low-mass galaxies by the more massive galaxies at the cluster centre. Another possible reason for this could be the non-detection of the faint sources due to the contamination from the bright sources and the intra-cluster light. However, it should also be noted that the outer region of the cluster shows a lack of very faint LSBGs. A detailed follow-up study with improved methodology is necessary to confirm if this is a statistical bias or reflects a physical phenomenon.

The size distribution of UDGs in the cluster shows that UDGs in the inner and middle regions have similar sizes. In contrast, UDGs in the outer skirts of the cluster tend to be larger, potentially because they are already extended UDGs falling into the cluster. In terms of stellar mass distribution, UDGs in the inner region of the cluster had relatively higher stellar masses compared to those in the middle and outer regions. In contrast, the running median for LSBGs remained flat across all the cluster regions.

\subsubsection{Trends in color with projected cluster-centric distance}

\begin{figure}[h]
\centering
\includegraphics[width=\linewidth,keepaspectratio]{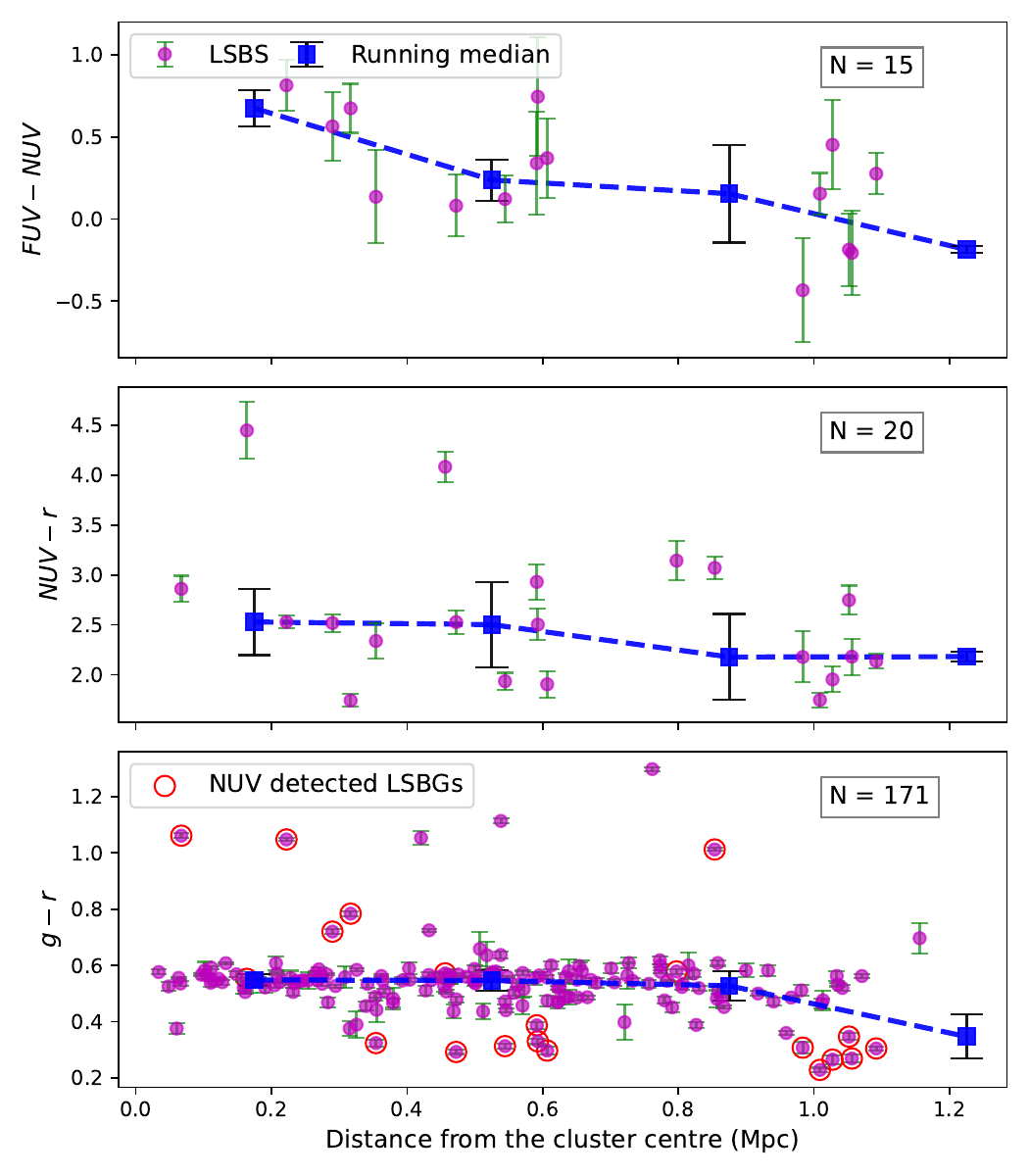}
\caption{$FUV-NUV$ (top panel), $NUV-r$ (middle panel) and $g-r$ (bottom panel) colors of the LSBGs presented in this works as a function of cluster centric distance. The number of LSBGs used is also shown in each subplot since not all the LSBGs had $3\sigma$ detection in $NUV$ and $FUV$ bands}
\label{color_comparisons}
\end{figure}

Recently, \citet{Venhola_optical_fornax} found that the early type dwarf galaxy population in the Fornax cluster becomes redder as we go towards the cluster centre in the $u-X$ (where $X\in {g,r, i})$ color. UDGs also have shown a similar trend in becoming bluer as going towards the outskirts of the cluster centre  \citep{Roman_2017, Mancera_udg, Junais2022}. The $FUV-NUV$, $NUV-r$ and $g-r$ colors of our sample as a function of the cluster centric distance are shown in Fig. \ref{color_comparisons}. 

We observe that LSBGs become progressively redder in $FUV-NUV$ color as they approach the cluster centre. The median $FUV-NUV$ color in the inner region of the cluster is approximately 0.5 mag redder than those in the outer region, suggesting that the LSBGs near the cluster centre in Abell 194 are quenched. Since $FUV-NUV$ color is sensitive to recent star formation history, this indicates that LSBGs on the outskirts may have undergone more recent star formation than those near the cluster centre. However, these conclusions are limited by the small sample size and should be confirmed with future studies.

Additionally, the median $NUV - r$ color shows a weak trend of becoming redder towards the cluster centre, though this trend is less pronounced than in \textit{FUV - NUV}. Nonetheless, the reddest LSBGs in \textit{NUV - r} are found in the inner cluster region, further supporting the idea that LSBGs near the centre of Abell 194 are quenched. Similar trends in UV-based colors have been reported by \citet{Venhola_optical_fornax} and \citet{Junais2022}.

In contrast, our sample of LSBGs shows no clear trend in $g-r$ color (which traces the old stellar population) with respect to cluster-centric distance up to 0.8 Mpc. However, LSBGs in the outer regions of the cluster appear bluer by 0.1 mag compared to those in the inner and middle regions. We also note that LSBGs with NUV detection in the inner region are redder in $g-r$ color than those in the middle and outer regions of the cluster.

Recently, \citet{Singh_UDG_nuv} studied the UV properties of UDGs in the Coma cluster using GALEX data and concluded that most UDGs in the cluster are quiescent, showing no signs of recent star formation. However, a subsequent analysis by \citet{Lee_UV_udgs} suggested that some UDGs near the cluster centre may exhibit signs of recent star formation. In our sample, we also observed a few LSBGs near the cluster centre that had detectable UV emissions. 

\begin{figure}[h]
\centering
\includegraphics[width=\linewidth,keepaspectratio]{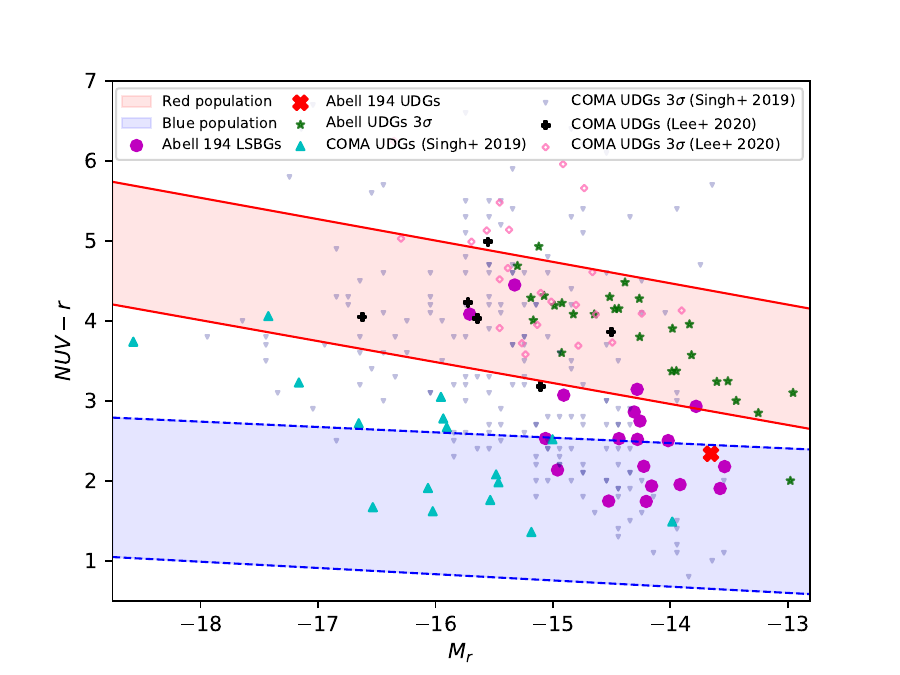}
\caption{The $NUV-r$ color is plotted against the absolute magnitude in the \textit{r}-band. The red region denotes the space occupied by the red-quiescent galaxy population, while the blue region represents the blue galaxy population \citep{Singh_UDG_nuv}. Magenta dots represent LSBGs, and red crosses represent UDGs with NUV detection in our sample. Green stars indicate the upper limits of $NUV-r$ for UDGs in Abell 194 without NUV detection. Cyan upward triangles and black plus symbols represent UDGs from the Coma cluster reported by \cite{Singh_UDG_nuv} and \citet{Lee_UV_udgs}, respectively. Downward navy blue triangles and pink rhombuses denote the upper limits of $NUV-r$ for UDGs in the Coma cluster from \cite{Singh_UDG_nuv} and \citet{Lee_UV_udgs}, respectively.}
\label{NUV_r}
\end{figure}

In Fig, \ref{NUV_r}, we compare the distribution of the LSBGs and UDGs from our sample and the UDGs from the Coma cluster from \citet{Singh_UDG_nuv} and \citet{Lee_UV_udgs} in the color-magnitude space ($NUV-r$ vs $M_r$). In our sample, we have only one UDG with NUV detection, which belongs to the blue population. Similar to the population of Coma UDGs from \citet{Lee_UV_udgs}, we have four LSBGs belonging to the red population with NUV detection. It should also be noted that the population of LSBGs and UDGs in our sample are much fainter than the population of UDGs presented in \citet{Singh_UDG_nuv} and \citet{Lee_UV_udgs}.

\section{Conclusion}\label{conclusion}
In this paper, we explore the potential of transfer learning with the transformer models presented in \citet{haressh_lsbs}. We trained the transformer ensemble models on the LSBGs and contaminants from DES DR 1, which is presented in \citet{Tanoglidis1} and \citet{Tanoglidis2} and updated by \citet{haressh_lsbs}. Subsequently, the trained transformer models are applied to identify LSBGs from the data of the Abell 194 cluster, observed with HSC, which is two magnitudes deeper than DES DR1. Since the instrument used to image in DES DR 1 is different from the HSC, we standardize the data from both instruments by converting their units into surface brightness units ($\mu \text{Jy arcsec}^{-2}$). 

After fitting S\'ersic profiles and conducting visual inspections on the LSBG candidate sample identified by the transformer models, we identified 159 LSBGs. However, the transformer models failed to identify 12 LSBGs, which were found through visual inspection of the rejected sample. Among the 12 LSBGs missed by the ensemble models:
\begin{itemize}
    \item 9 LSBGs are fainter than the training sample ($g<21.5$ mag) and lack representation in the training set.
    \item 3 LSBGs have bright objects near their centres.
\end{itemize}
Thus, our final sample has 171 LSBGs, including 28 UDGs. The transformer models achieved a TPR of 93\%, comparable to the 96\% TPR obtained on the DES dataset (see Appendix \ref{model_perfomance_on_DES}). The reported metrics, even without fine-tuning, confirm the success of transfer learning, suggesting these models could be effectively applied to deeper datasets from LSST and Euclid.

\citet{Tanoglidis1} and \citet{haressh_lsbs} have identified 96 LSBGs in Abell 194, referred to as \textit{LSBGs-DES} sample. Similarly, \citet{Zaritsky_2023} identified 14 UDGs in Abell 194, among which 9 are part of the \textit{LSBGs-DES} sample. From the deep data of Abell 194, our models identified 95 out of 96 known LSBGs and 13 out of 14 known UDGs. In this work, we found that improved masks and local sky subtraction significantly influence the parameters such as the S\'ersic index, similar to the finding of \citet{Bautista}.  In the new sample, 12 LSBGs identified from the DES are not included as they are slightly brighter (0.1 mag) than our selection cut.
In the final sample, we have:
\begin{itemize}
    \item 84 LSBGs also detected in DES by \citet{Tanoglidis1} and \citet{haressh_lsbs}.
    \item new 87 LSBGs identified with the deep HSC observations.
\end{itemize}

The newly estimated number of UDGs in Abell 194, based on data from the HSC, supports the argument in the literature that the number of UDGs in a cluster is linearly correlated with cluster mass on a log scale \citep{vanderBurg2017, Mancera, Karunakaran}. However, we found that the number of UDGs in Abell 194 exceeds the predicted value from \citet{Karunakaran}. Since we lacked sufficient new data points to statistically refine the relation, further analysis was left for future work. Additionally, this relation may be influenced by the limiting magnitude of the surveys used to identify UDGs. In the future, we plan to account for the limiting magnitudes when constraining these relations. The surface number density of UDGs, normalized by cluster mass, is consistent across different clusters, further indicating that UDG abundance is correlated with cluster mass.

The S\'ersic index distribution of the UDGs in this work is consistent with values reported in the literature \citep{Koda, Yagi, Roman_2017, vanderBurg2017, Venhola_udgs, Bautista, haressh_lsbs}. Additionally, the S\'ersic index distribution of UDGs is similar to that of dwarf elliptical galaxies, with UDGs occupying the diffuse, low surface brightness end of dwarf galaxies in the $R_{\mathrm{eff}}-M_g$ plane. This suggests that UDGs may be an extended subset of the dwarf galaxy population.

No significant trends in $n$ are observed for LSBGs and UDGs as a function of the cluster-centric distance. LSBGs and UDGs near the cluster centre tend to be brighter, and there is an under-representation of faint LSBGs in the outer parts of the cluster, which needs further investigation. 

Applying aperture photometry to the available GALEX data of Abell 194, 15 LSBGs in our sample show confident detections ($S/N \geq 3\sigma$) in both NUV and FUV, while 5 LSBGs have detections only in NUV. These NUV and FUV detections indicate recent star formation in these galaxies.
The $FUV-NUV$ color becomes redder towards the centre of the cluster. This trend is consistent with the observations of \citet{Venhola_optical_fornax} and \citet{Junais2022}. Examining the relation between $NUV-r$ color and the absolute magnitude in the \textit{r} band, we found that most NUV-emitting LSBGs are in the blue cloud, with only four LSBGs being red in $NUV-r$.

LSST and Euclid are expected to detect $10^5+$ LSBGs, requiring efficient data analysis. Transfer learning enables knowledge transfer possible for these upcoming surveys. We aim to build pipelines with high TPR and low FPR that ensure better completeness and purity. Advanced ML methods are the key to building these pipelines. 
\begin{acknowledgement}

We sincerely appreciate the valuable feedback from the anonymous referee, which greatly contributed to enhancing the quality of this paper. This research is based on data collected at the Subaru Telescope, which is operated by the National Astronomical Observatory of Japan. We are honored and grateful for the opportunity of observing the Universe from Maunakea, which has the cultural, historical, and natural significance in Hawaii.
This research was supported by the Polish National Science centre grant 2023/50/A/ST9/00579. 
We thank Samuel Boissier for helpful discussions and suggestions. This article/publication is based upon work from COST Action CA21136 – “Addressing observational tensions in cosmology with systematics and fundamental physics (CosmoVerse)”, supported by COST (European Cooperation in Science and Technology). 
J. is funded by the European Union (MSCA EDUCADO, GA 101119830 and WIDERA ExGal-Twin, GA 101158446). 
J.K. acknowledges support from NSF through grants AST-1812847, AST-2006600, and AST-2406608.
M.R. acknowledges support from the Narodowe Centrum Nauki (UMO-2020/38/E/ST9/00077).
D.D. acknowledge support from the National Science Center (NCN) grant SONATA (UMO-2020/39/D/ST9/00720).

This project used public archival data from the Dark Energy Survey (DES). Funding for the DES Projects has been provided by the U.S. Department of Energy, the U.S. National Science Foundation, the Ministry of Science and Education of Spain, the Science and Technology FacilitiesCouncil of the United Kingdom, the Higher Education Funding Council for England, the National Center for Supercomputing Applications at the University of Illinois at Urbana-Champaign, the Kavli Institute of Cosmological Physics at the University of Chicago, the Center for Cosmology and Astro-Particle Physics at the Ohio State University, the Mitchell Institute for Fundamental Physics and Astronomy at Texas A\&M University, Financiadora de Estudos e Projetos, Funda{\c c}{\~a}o Carlos Chagas Filho de Amparo {\`a} Pesquisa do Estado do Rio de Janeiro, Conselho Nacional de Desenvolvimento Cient{\'i}fico e Tecnol{\'o}gico and the Minist{\'e}rio da Ci{\^e}ncia, Tecnologia e Inova{\c c}{\~a}o, the Deutsche Forschungsgemeinschaft, and the Collaborating Institutions in the Dark Energy Survey.
The Collaborating Institutions are Argonne National Laboratory, the University of California at Santa Cruz, the University of Cambridge, Centro de Investigaciones Energ{\'e}ticas, Medioambientales y Tecnol{\'o}gicas-Madrid, the University of Chicago, University College London, the DES-Brazil Consortium, the University of Edinburgh, the Eidgen{\"o}ssische Technische Hochschule (ETH) Z{\"u}rich,  Fermi National Accelerator Laboratory, the University of Illinois at Urbana-Champaign, the Institut de Ci{\`e}ncies de l'Espai (IEEC/CSIC), the Institut de F{\'i}sica d'Altes Energies, Lawrence Berkeley National Laboratory, the Ludwig-Maximilians Universit{\"a}t M{\"u}nchen and the associated Excellence Cluster Universe, the University of Michigan, the National Optical Astronomy Observatory, the University of Nottingham, The Ohio State University, the OzDES Membership Consortium, the University of Pennsylvania, the University of Portsmouth, SLAC National Accelerator Laboratory, Stanford University, the University of Sussex, and Texas A\&M University.
Based in part on observations at Cerro Tololo Inter-American Observatory, National Optical Astronomy Observatory, which is operated by the Association of Universities for Research in Astronomy (AURA) under a cooperative agreement with the National Science Foundation.

\end{acknowledgement}
\bibliographystyle{aa}
\bibliography{aa}

\begin{thebibliography}{132}
\expandafter\ifx\csname natexlab\endcsname\relax\def\natexlab#1{#1}\fi

\bibitem[{{Abbott} {et~al.}(2018){Abbott}, {Abdalla}, {Allam}, {Amara}, {Annis}, {Asorey}, {Avila}, {Ballester}, {Banerji}, {Barkhouse}, {Baruah}, {Baumer}, {Bechtol}, {Becker}, {Benoit-L{\'e}vy}, {Bernstein}, {Bertin}, {Blazek}, {Bocquet}, {Brooks}, {Brout}, {Buckley-Geer}, {Burke}, {Busti}, {Campisano}, {Cardiel-Sas}, {Carnero Rosell}, {Carrasco Kind}, {Carretero}, {Castander}, {Cawthon}, {Chang}, {Chen}, {Conselice}, {Costa}, {Crocce}, {Cunha}, {D'Andrea}, {da Costa}, {Das}, {Daues}, {Davis}, {Davis}, {De Vicente}, {DePoy}, {DeRose}, {Desai}, {Diehl}, {Dietrich}, {Dodelson}, {Doel}, {Drlica-Wagner}, {Eifler}, {Elliott}, {Evrard}, {Farahi}, {Fausti Neto}, {Fernandez}, {Finley}, {Flaugher}, {Foley}, {Fosalba}, {Friedel}, {Frieman}, {Garc{\'\i}a-Bellido}, {Gaztanaga}, {Gerdes}, {Giannantonio}, {Gill}, {Glazebrook}, {Goldstein}, {Gower}, {Gruen}, {Gruendl}, {Gschwend}, {Gupta}, {Gutierrez}, {Hamilton}, {Hartley}, {Hinton}, {Hislop}, {Hollowood}, {Honscheid}, {Hoyle}, {Huterer}, {Jain}, {James}, {Jeltema},
  {Johnson}, {Johnson}, {Kacprzak}, {Kent}, {Khullar}, {Klein}, {Kovacs}, {Koziol}, {Krause}, {Kremin}, {Kron}, {Kuehn}, {Kuhlmann}, {Kuropatkin}, {Lahav}, {Lasker}, {Li}, {Li}, {Liddle}, {Lima}, {Lin}, {L{\'o}pez-Reyes}, {MacCrann}, {Maia}, {Maloney}, {Manera}, {March}, {Marriner}, {Marshall}, {Martini}, {McClintock}, {McKay}, {McMahon}, {Melchior}, {Menanteau}, {Miller}, {Miquel}, {Mohr}, {Morganson}, {Mould}, {Neilsen}, {Nichol}, {Nogueira}, {Nord}, {Nugent}, {Nunes}, {Ogando}, {Old}, {Pace}, {Palmese}, {Paz-Chinch{\'o}n}, {Peiris}, {Percival}, {Petravick}, {Plazas}, {Poh}, {Pond}, {Porredon}, {Pujol}, {Refregier}, {Reil}, {Ricker}, {Rollins}, {Romer}, {Roodman}, {Rooney}, {Ross}, {Rykoff}, {Sako}, {Sanchez}, {Sanchez}, {Santiago}, {Saro}, {Scarpine}, {Scolnic}, {Serrano}, {Sevilla-Noarbe}, {Sheldon}, {Shipp}, {Silveira}, {Smith}, {Smith}, {Smith}, {Soares-Santos}, {Sobreira}, {Song}, {Stebbins}, {Suchyta}, {Sullivan}, {Swanson}, {Tarle}, {Thaler}, {Thomas}, {Thomas}, {Troxel}, {Tucker}, {Vikram}, {Vivas},
  {Walker}, {Wechsler}, {Weller}, {Wester}, {Wolf}, {Wu}, {Yanny}, {Zenteno}, {Zhang}, {Zuntz}, {DES Collaboration}, {Juneau}, {Fitzpatrick}, {Nikutta}, {Nidever}, {Olsen}, {Scott}, \& {NOAO Data Lab}}]{DR1}
{Abbott}, T.~M.~C., {Abdalla}, F.~B., {Allam}, S., {et~al.} 2018, \apjs, 239, 18

\bibitem[{{Abbott} {et~al.}(2021){Abbott}, {Adam{\'o}w}, {Aguena}, {Allam}, {Amon}, {Annis}, {Avila}, {Bacon}, {Banerji}, {Bechtol}, {Becker}, {Bernstein}, {Bertin}, {Bhargava}, {Bridle}, {Brooks}, {Burke}, {Carnero Rosell}, {Carrasco Kind}, {Carretero}, {Castander}, {Cawthon}, {Chang}, {Choi}, {Conselice}, {Costanzi}, {Crocce}, {da Costa}, {Davis}, {De Vicente}, {DeRose}, {Desai}, {Diehl}, {Dietrich}, {Drlica-Wagner}, {Eckert}, {Elvin-Poole}, {Everett}, {Evrard}, {Ferrero}, {Fert{\'e}}, {Flaugher}, {Fosalba}, {Friedel}, {Frieman}, {Garc{\'\i}a-Bellido}, {Gaztanaga}, {Gelman}, {Gerdes}, {Giannantonio}, {Gill}, {Gruen}, {Gruendl}, {Gschwend}, {Gutierrez}, {Hartley}, {Hinton}, {Hollowood}, {Honscheid}, {Huterer}, {James}, {Jeltema}, {Johnson}, {Kent}, {Kron}, {Kuehn}, {Kuropatkin}, {Lahav}, {Li}, {Lidman}, {Lin}, {MacCrann}, {Maia}, {Manning}, {Maloney}, {March}, {Marshall}, {Martini}, {Melchior}, {Menanteau}, {Miquel}, {Morgan}, {Myles}, {Neilsen}, {Ogando}, {Palmese}, {Paz-Chinch{\'o}n}, {Petravick},
  {Pieres}, {Plazas}, {Pond}, {Rodriguez-Monroy}, {Romer}, {Roodman}, {Rykoff}, {Sako}, {Sanchez}, {Santiago}, {Scarpine}, {Serrano}, {Sevilla-Noarbe}, {Smith}, {Smith}, {Soares-Santos}, {Suchyta}, {Swanson}, {Tarle}, {Thomas}, {To}, {Tremblay}, {Troxel}, {Tucker}, {Turner}, {Varga}, {Walker}, {Wechsler}, {Weller}, {Wester}, {Wilkinson}, {Yanny}, {Zhang}, {Nikutta}, {Fitzpatrick}, {Jacques}, {Scott}, {Olsen}, {Huang}, {Herrera}, {Juneau}, {Nidever}, {Weaver}, {Adean}, {Correia}, {de Freitas}, {Freitas}, {Singulani}, {Vila-Verde}, \& {Linea Science Server}}]{DESDR2}
{Abbott}, T.~M.~C., {Adam{\'o}w}, M., {Aguena}, M., {et~al.} 2021, \apjs, 255, 20

\bibitem[{{Ackermann} {et~al.}(2018){Ackermann}, {Schawinski}, {Zhang}, {Weigel}, \& {Turp}}]{Ackermann}
{Ackermann}, S., {Schawinski}, K., {Zhang}, C., {Weigel}, A.~K., \& {Turp}, M.~D. 2018, \mnras, 479, 415

\bibitem[{{Aguerri} \& {Gonz{\'a}lez-Garc{\'\i}a}(2009)}]{Aguerri}
{Aguerri}, J.~A.~L. \& {Gonz{\'a}lez-Garc{\'\i}a}, A.~C. 2009, \aap, 494, 891

\bibitem[{{Ahumada} {et~al.}(2020){Ahumada}, {Allende Prieto}, {Almeida}, {Anders}, {Anderson}, {Andrews}, {Anguiano}, {Arcodia}, {Armengaud}, {Aubert}, {Avila}, {Avila-Reese}, {Badenes}, {Balland}, {Barger}, {Barrera-Ballesteros}, {Basu}, {Bautista}, {Beaton}, {Beers}, {Benavides}, {Bender}, {Bernardi}, {Bershady}, {Beutler}, {Bidin}, {Bird}, {Bizyaev}, {Blanc}, {Blanton}, {Boquien}, {Borissova}, {Bovy}, {Brandt}, {Brinkmann}, {Brownstein}, {Bundy}, {Bureau}, {Burgasser}, {Burtin}, {Cano-D{\'\i}az}, {Capasso}, {Cappellari}, {Carrera}, {Chabanier}, {Chaplin}, {Chapman}, {Cherinka}, {Chiappini}, {Doohyun Choi}, {Chojnowski}, {Chung}, {Clerc}, {Coffey}, {Comerford}, {Comparat}, {da Costa}, {Cousinou}, {Covey}, {Crane}, {Cunha}, {Ilha}, {Dai}, {Damsted}, {Darling}, {Davidson}, {Davies}, {Dawson}, {De}, {de la Macorra}, {De Lee}, {Queiroz}, {Deconto Machado}, {de la Torre}, {Dell'Agli}, {du Mas des Bourboux}, {Diamond-Stanic}, {Dillon}, {Donor}, {Drory}, {Duckworth}, {Dwelly}, {Ebelke}, {Eftekharzadeh}, {Davis
  Eigenbrot}, {Elsworth}, {Eracleous}, {Erfanianfar}, {Escoffier}, {Fan}, {Farr}, {Fern{\'a}ndez-Trincado}, {Feuillet}, {Finoguenov}, {Fofie}, {Fraser-McKelvie}, {Frinchaboy}, {Fromenteau}, {Fu}, {Galbany}, {Garcia}, {Garc{\'\i}a-Hern{\'a}ndez}, {Garma Oehmichen}, {Ge}, {Geimba Maia}, {Geisler}, {Gelfand}, {Goddy}, {Gonzalez-Perez}, {Grabowski}, {Green}, {Grier}, {Guo}, {Guy}, {Harding}, {Hasselquist}, {Hawken}, {Hayes}, {Hearty}, {Hekker}, {Hogg}, {Holtzman}, {Horta}, {Hou}, {Hsieh}, {Huber}, {Hunt}, {Ider Chitham}, {Imig}, {Jaber}, {Jimenez Angel}, {Johnson}, {Jones}, {J{\"o}nsson}, {Jullo}, {Kim}, {Kinemuchi}, {Kirkpatrick}, {Kite}, {Klaene}, {Kneib}, {Kollmeier}, {Kong}, {Kounkel}, {Krishnarao}, {Lacerna}, {Lan}, {Lane}, {Law}, {Le Goff}, {Leung}, {Lewis}, {Li}, {Lian}, {Lin}, {Long}, {Longa-Pe{\~n}a}, {Lundgren}, {Lyke}, {Mackereth}, {MacLeod}, {Majewski}, {Manchado}, {Maraston}, {Martini}, {Masseron}, {Masters}, {Mathur}, {McDermid}, {Merloni}, {Merrifield}, {M{\'e}sz{\'a}ros}, {Miglio}, {Minniti},
  {Minsley}, {Miyaji}, {Mohammad}, {Mosser}, {Mueller}, {Muna}, {Mu{\~n}oz-Guti{\'e}rrez}, {Myers}, {Nadathur}, {Nair}, {Nandra}, {Correa do Nascimento}, {Nevin}, {Newman}, {Nidever}, {Nitschelm}, {Noterdaeme}, {O'Connell}, {Olmstead}, {Oravetz}, {Oravetz}, {Osorio}, {Pace}, {Padilla}, {Palanque-Delabrouille}, {Palicio}, {Pan}, {Pan}, {Parker}, {Paviot}, {Peirani}, {Ram{\'r}ez}, {Penny}, {Percival}, {Perez-Fournon}, {P{\'e}rez-R{\`a}fols}, {Petitjean}, {Pieri}, {Pinsonneault}, {Poovelil}, {Povick}, {Prakash}, {Price-Whelan}, {Raddick}, {Raichoor}, {Ray}, {Rembold}, {Rezaie}, {Riffel}, {Riffel}, {Rix}, {Robin}, {Roman-Lopes}, {Rom{\'a}n-Z{\'u}{\~n}iga}, {Rose}, {Ross}, {Rossi}, {Rowlands}, {Rubin}, {Salvato}, {S{\'a}nchez}, {S{\'a}nchez-Menguiano}, {S{\'a}nchez-Gallego}, {Sayres}, {Schaefer}, {Schiavon}, {Schimoia}, {Schlafly}, {Schlegel}, {Schneider}, {Schultheis}, {Schwope}, {Seo}, {Serenelli}, {Shafieloo}, {Shamsi}, {Shao}, {Shen}, {Shetrone}, {Shirley}, {Silva Aguirre}, {Simon}, {Skrutskie}, {Slosar},
  {Smethurst}, {Sobeck}, {Sodi}, {Souto}, {Stark}, {Stassun}, {Steinmetz}, {Stello}, {Stermer}, {Storchi-Bergmann}, {Streblyanska}, {Stringfellow}, {Stutz}, {Su{\'a}rez}, {Sun}, {Taghizadeh-Popp}, {Talbot}, {Tayar}, {Thakar}, {Theriault}, {Thomas}, {Thomas}, {Tinker}, {Tojeiro}, {Toledo}, {Tremonti}, {Troup}, {Tuttle}, {Unda-Sanzana}, {Valentini}, {Vargas-Gonz{\'a}lez}, {Vargas-Maga{\~n}a}, {V{\'a}zquez-Mata}, {Vivek}, {Wake}, {Wang}, {Weaver}, {Weijmans}, {Wild}, {Wilson}, {Wilson}, {Wolthuis}, {Wood-Vasey}, {Yan}, {Yang}, {Y{\`e}che}, {Zamora}, {Zarrouk}, {Zasowski}, {Zhang}, {Zhao}, {Zhao}, {Zheng}, {Zheng}, {Zhu}, \& {Zou}}]{SDSS16}
{Ahumada}, R., {Allende Prieto}, C., {Almeida}, A., {et~al.} 2020, \apjs, 249, 3

\bibitem[{{Aihara} {et~al.}(2018){Aihara}, {Armstrong}, {Bickerton}, {Bosch}, {Coupon}, {Furusawa}, {Hayashi}, {Ikeda}, {Kamata}, {Karoji}, {Kawanomoto}, {Koike}, {Komiyama}, {Lang}, {Lupton}, {Mineo}, {Miyatake}, {Miyazaki}, {Morokuma}, {Obuchi}, {Oishi}, {Okura}, {Price}, {Takata}, {Tanaka}, {Tanaka}, {Tanaka}, {Uchida}, {Uraguchi}, {Utsumi}, {Wang}, {Yamada}, {Yamanoi}, {Yasuda}, {Arimoto}, {Chiba}, {Finet}, {Fujimori}, {Fujimoto}, {Furusawa}, {Goto}, {Goulding}, {Gunn}, {Harikane}, {Hattori}, {Hayashi}, {He{\l}miniak}, {Higuchi}, {Hikage}, {Ho}, {Hsieh}, {Huang}, {Huang}, {Imanishi}, {Iwata}, {Jaelani}, {Jian}, {Kashikawa}, {Katayama}, {Kojima}, {Konno}, {Koshida}, {Kusakabe}, {Leauthaud}, {Lee}, {Lin}, {Lin}, {Mandelbaum}, {Matsuoka}, {Medezinski}, {Miyama}, {Momose}, {More}, {More}, {Mukae}, {Murata}, {Murayama}, {Nagao}, {Nakata}, {Niida}, {Niikura}, {Nishizawa}, {Oguri}, {Okabe}, {Ono}, {Onodera}, {Onoue}, {Ouchi}, {Pyo}, {Shibuya}, {Shimasaku}, {Simet}, {Speagle}, {Spergel}, {Strauss}, {Sugahara},
  {Sugiyama}, {Suto}, {Suzuki}, {Tait}, {Takada}, {Terai}, {Toba}, {Turner}, {Uchiyama}, {Umetsu}, {Urata}, {Usuda}, {Yeh}, \& {Yuma}}]{Aihara}
{Aihara}, H., {Armstrong}, R., {Bickerton}, S., {et~al.} 2018, \pasj, 70, S8

\bibitem[{{Amorisco} \& {Loeb}(2016)}]{Amorisco}
{Amorisco}, N.~C. \& {Loeb}, A. 2016, \mnras, 459, L51

\bibitem[{{Bautista} {et~al.}(2023){Bautista}, {Koda}, {Yagi}, {Komiyama}, \& {Yamanoi}}]{Bautista}
{Bautista}, J. M.~G., {Koda}, J., {Yagi}, M., {Komiyama}, Y., \& {Yamanoi}, H. 2023, \apjs, 267, 10

\bibitem[{{Benavides} {et~al.}(2023{\natexlab{a}}){Benavides}, {Sales}, {Abadi}, {Marinacci}, {Vogelsberger}, \& {Hernquist}}]{Benavides}
{Benavides}, J.~A., {Sales}, L.~V., {Abadi}, M.~G., {et~al.} 2023{\natexlab{a}}, \mnras, 522, 1033

\bibitem[{{Benavides} {et~al.}(2023{\natexlab{b}}){Benavides}, {Sales}, {Abadi}, {Marinacci}, {Vogelsberger}, \& {Hernquist}}]{Benavides_2023}
{Benavides}, J.~A., {Sales}, L.~V., {Abadi}, M.~G., {et~al.} 2023{\natexlab{b}}, \mnras, 522, 1033

\bibitem[{{Bernstein} {et~al.}(2018){Bernstein}, {Abbott}, {Armstrong}, {Burke}, {Diehl}, {Gruendl}, {Johnson}, {Li}, {Rykoff}, {Walker}, {Wester}, \& {Yanny}}]{Bernstein_DES_sky_bg}
{Bernstein}, G.~M., {Abbott}, T.~M.~C., {Armstrong}, R., {et~al.} 2018, \pasp, 130, 054501

\bibitem[{{Bertin}(2013)}]{Bertin_psfex}
{Bertin}, E. 2013, {PSFEx: Point Spread Function Extractor}, Astrophysics Source Code Library, record ascl:1301.001

\bibitem[{{Bertin} \& {Arnouts}(1996)}]{sxtractor}
{Bertin}, E. \& {Arnouts}, S. 1996, \aaps, 117, 393

\bibitem[{Bickley {et~al.}(2024)Bickley, Wilkinson, Ferreira, Ellison, Bottrell, \& Jyoti}]{bickley2024effectimagequalitygalaxy}
Bickley, R.~W., Wilkinson, S., Ferreira, L., {et~al.} 2024, The effect of image quality on galaxy merger identification with deep learning

\bibitem[{{Bosch} {et~al.}(2018{\natexlab{a}}){Bosch}, {Armstrong}, {Bickerton}, {Furusawa}, {Ikeda}, {Koike}, {Lupton}, {Mineo}, {Price}, {Takata}, {Tanaka}, {Yasuda}, {AlSayyad}, {Becker}, {Coulton}, {Coupon}, {Garmilla}, {Huang}, {Krughoff}, {Lang}, {Leauthaud}, {Lim}, {Lust}, {MacArthur}, {Mandelbaum}, {Miyatake}, {Miyazaki}, {Murata}, {More}, {Okura}, {Owen}, {Swinbank}, {Strauss}, {Yamada}, \& {Yamanoi}}]{HSC_software}
{Bosch}, J., {Armstrong}, R., {Bickerton}, S., {et~al.} 2018{\natexlab{a}}, \pasj, 70, S5

\bibitem[{{Bosch} {et~al.}(2018{\natexlab{b}}){Bosch}, {Armstrong}, {Bickerton}, {Furusawa}, {Ikeda}, {Koike}, {Lupton}, {Mineo}, {Price}, {Takata}, {Tanaka}, {Yasuda}, {AlSayyad}, {Becker}, {Coulton}, {Coupon}, {Garmilla}, {Huang}, {Krughoff}, {Lang}, {Leauthaud}, {Lim}, {Lust}, {MacArthur}, {Mandelbaum}, {Miyatake}, {Miyazaki}, {Murata}, {More}, {Okura}, {Owen}, {Swinbank}, {Strauss}, {Yamada}, \& {Yamanoi}}]{Bosch_HSC_cam}
{Bosch}, J., {Armstrong}, R., {Bickerton}, S., {et~al.} 2018{\natexlab{b}}, \pasj, 70, S5

\bibitem[{{Bothun} {et~al.}(1997){Bothun}, {Impey}, \& {McGaugh}}]{Bothun}
{Bothun}, G., {Impey}, C., \& {McGaugh}, S. 1997, \pasp, 109, 745

\bibitem[{{Bullock} \& {Boylan-Kolchin}(2017)}]{Bullock_Boylan}
{Bullock}, J.~S. \& {Boylan-Kolchin}, M. 2017, \araa, 55, 343

\bibitem[{{Bullock} {et~al.}(2001){Bullock}, {Kolatt}, {Sigad}, {Somerville}, {Kravtsov}, {Klypin}, {Primack}, \& {Dekel}}]{Bullock}
{Bullock}, J.~S., {Kolatt}, T.~S., {Sigad}, Y., {et~al.} 2001, \mnras, 321, 559

\bibitem[{{Burkholder} {et~al.}(2001){Burkholder}, {Impey}, \& {Sprayberry}}]{Burkholder}
{Burkholder}, V., {Impey}, C., \& {Sprayberry}, D. 2001, \aj, 122, 2318

\bibitem[{{Cabrera-Vives} {et~al.}(2017){Cabrera-Vives}, {Reyes}, {F{\"o}rster}, {Est{\'e}vez}, \& {Maureira}}]{2017ApJ...836...97C}
{Cabrera-Vives}, G., {Reyes}, I., {F{\"o}rster}, F., {Est{\'e}vez}, P.~A., \& {Maureira}, J.-C. 2017, \apj, 836, 97

\bibitem[{{Chamba} {et~al.}(2022){Chamba}, {Trujillo}, \& {Knapen}}]{Chamba2022}
{Chamba}, N., {Trujillo}, I., \& {Knapen}, J.~H. 2022, \aap, 667, A87

\bibitem[{{Chambers} {et~al.}(2016){Chambers}, {Magnier}, {Metcalfe}, {Flewelling}, {Huber}, {Waters}, {Denneau}, {Draper}, {Farrow}, {Finkbeiner}, {Holmberg}, {Koppenhoefer}, {Price}, {Rest}, {Saglia}, {Schlafly}, {Smartt}, {Sweeney}, {Wainscoat}, {Burgett}, {Chastel}, {Grav}, {Heasley}, {Hodapp}, {Jedicke}, {Kaiser}, {Kudritzki}, {Luppino}, {Lupton}, {Monet}, {Morgan}, {Onaka}, {Shiao}, {Stubbs}, {Tonry}, {White}, {Ba{\~n}ados}, {Bell}, {Bender}, {Bernard}, {Boegner}, {Boffi}, {Botticella}, {Calamida}, {Casertano}, {Chen}, {Chen}, {Cole}, {Deacon}, {Frenk}, {Fitzsimmons}, {Gezari}, {Gibbs}, {Goessl}, {Goggia}, {Gourgue}, {Goldman}, {Grant}, {Grebel}, {Hambly}, {Hasinger}, {Heavens}, {Heckman}, {Henderson}, {Henning}, {Holman}, {Hopp}, {Ip}, {Isani}, {Jackson}, {Keyes}, {Koekemoer}, {Kotak}, {Le}, {Liska}, {Long}, {Lucey}, {Liu}, {Martin}, {Masci}, {McLean}, {Mindel}, {Misra}, {Morganson}, {Murphy}, {Obaika}, {Narayan}, {Nieto-Santisteban}, {Norberg}, {Peacock}, {Pier}, {Postman}, {Primak}, {Rae}, {Rai},
  {Riess}, {Riffeser}, {Rix}, {R{\"o}ser}, {Russel}, {Rutz}, {Schilbach}, {Schultz}, {Scolnic}, {Strolger}, {Szalay}, {Seitz}, {Small}, {Smith}, {Soderblom}, {Taylor}, {Thomson}, {Taylor}, {Thakar}, {Thiel}, {Thilker}, {Unger}, {Urata}, {Valenti}, {Wagner}, {Walder}, {Walter}, {Watters}, {Werner}, {Wood-Vasey}, \& {Wyse}}]{Chambers_PANSTARRs}
{Chambers}, K.~C., {Magnier}, E.~A., {Metcalfe}, N., {et~al.} 2016, arXiv e-prints, arXiv:1612.05560

\bibitem[{Clevert {et~al.}(2016)Clevert, Unterthiner, \& Hochreiter}]{Clevert}
Clevert, D., Unterthiner, T., \& Hochreiter, S. 2016, in 4th International Conference on Learning Representations, {ICLR} 2016, San Juan, Puerto Rico, May 2-4, 2016, Conference Track Proceedings, ed. Y.~Bengio \& Y.~LeCun

\bibitem[{{Conselice}(2018)}]{Conselice}
{Conselice}, C.~J. 2018, Research Notes of the American Astronomical Society, 2, 43

\bibitem[{{Conselice} {et~al.}(2003){Conselice}, {Gallagher}, \& {Wyse}}]{Conselice_udgs}
{Conselice}, C.~J., {Gallagher}, John~S., I., \& {Wyse}, R. F.~G. 2003, \aj, 125, 66

\bibitem[{{Da Costa}(1992)}]{Da_Costa}
{Da Costa}, G.~S. 1992, in Astronomical Society of the Pacific Conference Series, Vol.~23, Astronomical CCD Observing and Reduction Techniques, ed. S.~B. {Howell}, 90

\bibitem[{{Dalcanton} {et~al.}(1997){Dalcanton}, {Spergel}, {Gunn}, {Schmidt}, \& {Schneider}}]{Dalcanton}
{Dalcanton}, J.~J., {Spergel}, D.~N., {Gunn}, J.~E., {Schmidt}, M., \& {Schneider}, D.~P. 1997, \aj, 114, 635

\bibitem[{{de Blok} {et~al.}(2001){de Blok}, {McGaugh}, {Bosma}, \& {Rubin}}]{de_Blok}
{de Blok}, W.~J.~G., {McGaugh}, S.~S., {Bosma}, A., \& {Rubin}, V.~C. 2001, \apjl, 552, L23

\bibitem[{Deng {et~al.}(2009)Deng, Dong, Socher, Li, Li, \& Fei-Fei}]{Image_net_dataset}
Deng, J., Dong, W., Socher, R., {et~al.} 2009, in 2009 IEEE Conference on Computer Vision and Pattern Recognition, 248--255

\bibitem[{{Dey} {et~al.}(2019){Dey}, {Schlegel}, {Lang}, {Blum}, {Burleigh}, {Fan}, {Findlay}, {Finkbeiner}, {Herrera}, {Juneau}, {Landriau}, {Levi}, {McGreer}, {Meisner}, {Myers}, {Moustakas}, {Nugent}, {Patej}, {Schlafly}, {Walker}, {Valdes}, {Weaver}, {Y{\`e}che}, {Zou}, {Zhou}, {Abareshi}, {Abbott}, {Abolfathi}, {Aguilera}, {Alam}, {Allen}, {Alvarez}, {Annis}, {Ansarinejad}, {Aubert}, {Beechert}, {Bell}, {BenZvi}, {Beutler}, {Bielby}, {Bolton}, {Brice{\~n}o}, {Buckley-Geer}, {Butler}, {Calamida}, {Carlberg}, {Carter}, {Casas}, {Castander}, {Choi}, {Comparat}, {Cukanovaite}, {Delubac}, {DeVries}, {Dey}, {Dhungana}, {Dickinson}, {Ding}, {Donaldson}, {Duan}, {Duckworth}, {Eftekharzadeh}, {Eisenstein}, {Etourneau}, {Fagrelius}, {Farihi}, {Fitzpatrick}, {Font-Ribera}, {Fulmer}, {G{\"a}nsicke}, {Gaztanaga}, {George}, {Gerdes}, {Gontcho}, {Gorgoni}, {Green}, {Guy}, {Harmer}, {Hernandez}, {Honscheid}, {Huang}, {James}, {Jannuzi}, {Jiang}, {Joyce}, {Karcher}, {Karkar}, {Kehoe}, {Kneib}, {Kueter-Young}, {Lan},
  {Lauer}, {Le Guillou}, {Le Van Suu}, {Lee}, {Lesser}, {Perreault Levasseur}, {Li}, {Mann}, {Marshall}, {Mart{\'\i}nez-V{\'a}zquez}, {Martini}, {du Mas des Bourboux}, {McManus}, {Meier}, {M{\'e}nard}, {Metcalfe}, {Mu{\~n}oz-Guti{\'e}rrez}, {Najita}, {Napier}, {Narayan}, {Newman}, {Nie}, {Nord}, {Norman}, {Olsen}, {Paat}, {Palanque-Delabrouille}, {Peng}, {Poppett}, {Poremba}, {Prakash}, {Rabinowitz}, {Raichoor}, {Rezaie}, {Robertson}, {Roe}, {Ross}, {Ross}, {Rudnick}, {Safonova}, {Saha}, {S{\'a}nchez}, {Savary}, {Schweiker}, {Scott}, {Seo}, {Shan}, {Silva}, {Slepian}, {Soto}, {Sprayberry}, {Staten}, {Stillman}, {Stupak}, {Summers}, {Sien Tie}, {Tirado}, {Vargas-Maga{\~n}a}, {Vivas}, {Wechsler}, {Williams}, {Yang}, {Yang}, {Yapici}, {Zaritsky}, {Zenteno}, {Zhang}, {Zhang}, {Zhou}, \& {Zhou}}]{DESI}
{Dey}, A., {Schlegel}, D.~J., {Lang}, D., {et~al.} 2019, \aj, 157, 168

\bibitem[{{Di Cintio} {et~al.}(2017){Di Cintio}, {Brook}, {Dutton}, {Macci{\`o}}, {Obreja}, \& {Dekel}}]{Cintio}
{Di Cintio}, A., {Brook}, C.~B., {Dutton}, A.~A., {et~al.} 2017, \mnras, 466, L1

\bibitem[{{Dom{\'\i}nguez S{\'a}nchez} {et~al.}(2019){Dom{\'\i}nguez S{\'a}nchez}, {Huertas-Company}, {Bernardi}, {Kaviraj}, {Fischer}, {Abbott}, {Abdalla}, {Annis}, {Avila}, {Brooks}, {Buckley-Geer}, {Carnero Rosell}, {Carrasco Kind}, {Carretero}, {Cunha}, {D'Andrea}, {da Costa}, {Davis}, {De Vicente}, {Doel}, {Evrard}, {Fosalba}, {Frieman}, {Garc{\'\i}a-Bellido}, {Gaztanaga}, {Gerdes}, {Gruen}, {Gruendl}, {Gschwend}, {Gutierrez}, {Hartley}, {Hollowood}, {Honscheid}, {Hoyle}, {James}, {Kuehn}, {Kuropatkin}, {Lahav}, {Maia}, {March}, {Melchior}, {Menanteau}, {Miquel}, {Nord}, {Plazas}, {Sanchez}, {Scarpine}, {Schindler}, {Schubnell}, {Smith}, {Smith}, {Soares-Santos}, {Sobreira}, {Suchyta}, {Swanson}, {Tarle}, {Thomas}, {Walker}, \& {Zuntz}}]{Sanchez}
{Dom{\'\i}nguez S{\'a}nchez}, H., {Huertas-Company}, M., {Bernardi}, M., {et~al.} 2019, \mnras, 484, 93

\bibitem[{{Driver}(1999)}]{Driver}
{Driver}, S.~P. 1999, \apjl, 526, L69

\bibitem[{Du {et~al.}(2020)Du, Cheng, Zheng, \& Wu}]{Du_2020}
Du, W., Cheng, C., Zheng, Z., \& Wu, H. 2020, The Astronomical Journal, 159, 138

\bibitem[{{Eigenthaler} {et~al.}(2018){Eigenthaler}, {Puzia}, {Taylor}, {Ordenes-Brice{\~n}o}, {Mu{\~n}oz}, {Ribbeck}, {Alamo-Mart{\'\i}nez}, {Zhang}, {{\'A}ngel}, {Capaccioli}, {C{\^o}t{\'e}}, {Ferrarese}, {Galaz}, {Grebel}, {Hempel}, {Hilker}, {Lan{\c{c}}on}, {Mieske}, {Miller}, {Paolillo}, {Powalka}, {Richtler}, {Roediger}, {Rong}, {S{\'a}nchez-Janssen}, \& {Spengler}}]{Eigenthaler_2018}
{Eigenthaler}, P., {Puzia}, T.~H., {Taylor}, M.~A., {et~al.} 2018, \apj, 855, 142

\bibitem[{{Ferrarese} {et~al.}(2020){Ferrarese}, {C{\^o}t{\'e}}, {MacArthur}, {Durrell}, {Gwyn}, {Duc}, {S{\'a}nchez-Janssen}, {Santos}, {Blakeslee}, {Boselli}, {Boyer}, {Cantiello}, {Courteau}, {Cuillandre}, {Emsellem}, {Erben}, {Gavazzi}, {Guhathakurta}, {Huertas-Company}, {Jord{\'a}n}, {Lan{\c{c}}on}, {Liu}, {Mei}, {Mihos}, {Peng}, {Puzia}, {Roediger}, {Schade}, {Taylor}, {Toloba}, \& {Zhang}}]{Ferrarese_2020}
{Ferrarese}, L., {C{\^o}t{\'e}}, P., {MacArthur}, L.~A., {et~al.} 2020, \apj, 890, 128

\bibitem[{{Fitzpatrick}(1999)}]{Fitzpatrick1999}
{Fitzpatrick}, E.~L. 1999, \pasp, 111, 63

\bibitem[{{Forbes} {et~al.}(2020){Forbes}, {Dullo}, {Gannon}, {Couch}, {Iodice}, {Spavone}, {Cantiello}, \& {Schipani}}]{Forbes}
{Forbes}, D.~A., {Dullo}, B.~T., {Gannon}, J., {et~al.} 2020, \mnras, 494, 5293

\bibitem[{{Furusawa} {et~al.}(2018){Furusawa}, {Koike}, {Takata}, {Okura}, {Miyatake}, {Lupton}, {Bickerton}, {Price}, {Bosch}, {Yasuda}, {Mineo}, {Yamada}, {Miyazaki}, {Nakata}, {Koshida}, {Komiyama}, {Utsumi}, {Kawanomoto}, {Jeschke}, {Noumaru}, {Schubert}, {Iwata}, {Finet}, {Fujiyoshi}, {Tajitsu}, {Terai}, \& {Lee}}]{Hsc_cam1}
{Furusawa}, H., {Koike}, M., {Takata}, T., {et~al.} 2018, \pasj, 70, S3

\bibitem[{{Girardi} {et~al.}(1998){Girardi}, {Giuricin}, {Mardirossian}, {Mezzetti}, \& {Boschin}}]{Girardi_abell_a94}
{Girardi}, M., {Giuricin}, G., {Mardirossian}, F., {Mezzetti}, M., \& {Boschin}, W. 1998, \apj, 505, 74

\bibitem[{Glorot \& Bengio(2010)}]{Glorot}
Glorot, X. \& Bengio, Y. 2010, in {JMLR} Proceedings, Vol.~9, Proceedings of the Thirteenth International Conference on Artificial Intelligence and Statistics, {AISTATS} 2010, Chia Laguna Resort, Sardinia, Italy, May 13-15, 2010, ed. Y.~W. Teh \& D.~M. Titterington (JMLR.org), 249--256

\bibitem[{{Graham} \& {Driver}(2005)}]{Graham_sersic}
{Graham}, A.~W. \& {Driver}, S.~P. 2005, \pasa, 22, 118

\bibitem[{{Greco} {et~al.}(2018){Greco}, {Greene}, {Strauss}, {Macarthur}, {Flowers}, {Goulding}, {Huang}, {Kim}, {Komiyama}, {Leauthaud}, {Leisman}, {Lupton}, {Sif{\'o}n}, \& {Wang}}]{Greco}
{Greco}, J.~P., {Greene}, J.~E., {Strauss}, M.~A., {et~al.} 2018, \apj, 857, 104

\bibitem[{{Grespan} {et~al.}(2024){Grespan}, {Thuruthipilly}, {Pollo}, {Lochner}, {Biesiada}, \& {Etsebeth}}]{Grespan}
{Grespan}, M., {Thuruthipilly}, H., {Pollo}, A., {et~al.} 2024, \aap, 688, A34

\bibitem[{{Haberzettl} {et~al.}(2007){Haberzettl}, {Bomans}, \& {Dettmar}}]{Haberzettl}
{Haberzettl}, L., {Bomans}, D.~J., \& {Dettmar}, R.~J. 2007, \aap, 471, 787

\bibitem[{{Hannon} {et~al.}(2023){Hannon}, {Whitmore}, {Lee}, {Thilker}, {Deger}, {Huerta}, {Wei}, {Mobasher}, {Klessen}, {Boquien}, {Dale}, {Chevance}, {Grasha}, {Sanchez-Blazquez}, {Williams}, {Scheuermann}, {Groves}, {Kim}, {Kruijssen}, \& {The Phangs-HST Team}}]{Hannon}
{Hannon}, S., {Whitmore}, B.~C., {Lee}, J.~C., {et~al.} 2023, \mnras, 526, 2991

\bibitem[{{Hinshaw} {et~al.}(2013){Hinshaw}, {Larson}, {Komatsu}, {Spergel}, {Bennett}, {Dunkley}, {Nolta}, {Halpern}, {Hill}, {Odegard}, {Page}, {Smith}, {Weiland}, {Gold}, {Jarosik}, {Kogut}, {Limon}, {Meyer}, {Tucker}, {Wollack}, \& {Wright}}]{Hinshaw}
{Hinshaw}, G., {Larson}, D., {Komatsu}, E., {et~al.} 2013, \apjs, 208, 19

\bibitem[{Huang {et~al.}(2023)Huang, Chen, Chang, Lin, Hsu, Thengane, \& Lin}]{Huang_tranformer}
Huang, K.-W., Chen, G. C.-F., Chang, P.-W., {et~al.} 2023, in Computer Vision -- ECCV 2022 Workshops, ed. L.~Karlinsky, T.~Michaeli, \& K.~Nishino (Cham: Springer Nature Switzerland), 143--153

\bibitem[{{Hwang} {et~al.}(2023){Hwang}, {Sabiu}, {Park}, \& {Hong}}]{Hwang_VIT}
{Hwang}, S.~Y., {Sabiu}, C.~G., {Park}, I., \& {Hong}, S.~E. 2023, \jcap, 2023, 075

\bibitem[{{Ivezic} {et~al.}(2008){Ivezic}, {Axelrod}, {Brandt}, {Burke}, {Claver}, {Connolly}, {Cook}, {Gee}, {Gilmore}, {Jacoby}, {Jones}, {Kahn}, {Kantor}, {Krabbendam}, {Lupton}, {Monet}, {Pinto}, {Saha}, {Schalk}, {Schneider}, {Strauss}, {Stubbs}, {Sweeney}, {Szalay}, {Thaler}, {Tyson}, \& {LSST Collaboration}}]{Ivezic_LSST_design}
{Ivezic}, Z., {Axelrod}, T., {Brandt}, W.~N., {et~al.} 2008, Serbian Astronomical Journal, 176, 1

\bibitem[{Ivezi{\' c} {et~al.}(2019)Ivezi{\' c}, Kahn, Tyson, Abel, Acosta, Allsman, Alonso, AlSayyad, Anderson, Andrew, \& et~al.}]{Ivezi__2019}
Ivezi{\' c}, {\v{Z}}., Kahn, S.~M., Tyson, J.~A., {et~al.} 2019, \apj, 873, 111

\bibitem[{{Janssens} {et~al.}(2019){Janssens}, {Abraham}, {Brodie}, {Forbes}, \& {Romanowsky}}]{Janssens}
{Janssens}, S.~R., {Abraham}, R., {Brodie}, J., {Forbes}, D.~A., \& {Romanowsky}, A.~J. 2019, \apj, 887, 92

\bibitem[{{Jia} {et~al.}(2023){Jia}, {Sun}, {Li}, {Song}, {Ning}, {Wei}, \& {Luo}}]{Jia}
{Jia}, P., {Sun}, R., {Li}, N., {et~al.} 2023, \aj, 165, 26

\bibitem[{{Jiang} {et~al.}(2019){Jiang}, {Dekel}, {Freundlich}, {Romanowsky}, {Dutton}, {Macci{\`o}}, \& {Di Cintio}}]{Jiang_2019}
{Jiang}, F., {Dekel}, A., {Freundlich}, J., {et~al.} 2019, \mnras, 487, 5272

\bibitem[{{Junais} {et~al.}(2022){Junais}, {Boissier}, {Boselli}, {Ferrarese}, {C{\^o}t{\'e}}, {Gwyn}, {Roediger}, {Lim}, {Peng}, {Cuillandre}, {Longobardi}, {Fossati}, {Hensler}, {Koda}, {Bautista}, {Boquien}, {Ma{\l}ek}, {Amram}, \& {Roehlly}}]{Junais2022}
{Junais}, {Boissier}, S., {Boselli}, A., {et~al.} 2022, \aap, 667, A76

\bibitem[{{Junais} {et~al.}(2024){Junais}, {Weilbacher}, {Epinat}, {Boissier}, {Galaz}, {Johnston}, {Puzia}, {Amram}, \& {Ma{\l}ek}}]{Junais_malin2}
{Junais}, {Weilbacher}, P.~M., {Epinat}, B., {et~al.} 2024, \aap, 681, A100

\bibitem[{{Juri{\'c}} {et~al.}(2017){Juri{\'c}}, {Kantor}, {Lim}, {Lupton}, {Dubois-Felsmann}, {Jenness}, {Axelrod}, {Aleksi{\'c}}, {Allsman}, {AlSayyad}, {Alt}, {Armstrong}, {Basney}, {Becker}, {Becla}, {Biswas}, {Bosch}, {Boutigny}, {Kind}, {Ciardi}, {Connolly}, {Daniel}, {Daues}, {Economou}, {Chiang}, {Fausti}, {Fisher-Levine}, {Freemon}, {Gris}, {Hernandez}, {Hoblitt}, {Ivezi{\'c}}, {Jammes}, {Jevremovi{\'c}}, {Jones}, {Kalmbach}, {Kasliwal}, {Krughoff}, {Lurie}, {Lust}, {MacArthur}, {Melchior}, {Moeyens}, {Nidever}, {Owen}, {Parejko}, {Peterson}, {Petravick}, {Pietrowicz}, {Price}, {Reiss}, {Shaw}, {Sick}, {Slater}, {Strauss}, {Sullivan}, {Swinbank}, {Van Dyk}, {Vuj{\v{c}}i{\'c}}, {Withers}, \& {Yoachim}}]{Juric_LSST_design}
{Juri{\'c}}, M., {Kantor}, J., {Lim}, K.~T., {et~al.} 2017, in Astronomical Society of the Pacific Conference Series, Vol. 512, Astronomical Data Analysis Software and Systems XXV, ed. N.~P.~F. {Lorente}, K.~{Shortridge}, \& R.~{Wayth}, 279

\bibitem[{{Karunakaran} \& {Zaritsky}(2023)}]{Karunakaran}
{Karunakaran}, A. \& {Zaritsky}, D. 2023, \mnras, 519, 884

\bibitem[{{Kawanomoto} {et~al.}(2018){Kawanomoto}, {Uraguchi}, {Komiyama}, {Miyazaki}, {Furusawa}, {Finet}, {Hattori}, {Wang}, {Yasuda}, \& {Suzuki}}]{Hsc_cam2}
{Kawanomoto}, S., {Uraguchi}, F., {Komiyama}, Y., {et~al.} 2018, \pasj, 70, 66

\bibitem[{Kingma \& Ba(2015)}]{kingma2017adam}
Kingma, D.~P. \& Ba, J. 2015, in 3rd International Conference on Learning Representations, {ICLR} 2015, San Diego, CA, USA, May 7-9, 2015, Conference Track Proceedings, ed. Y.~Bengio \& Y.~LeCun

\bibitem[{{Koda} {et~al.}(2015){Koda}, {Yagi}, {Yamanoi}, \& {Komiyama}}]{Koda}
{Koda}, J., {Yagi}, M., {Yamanoi}, H., \& {Komiyama}, Y. 2015, \apjl, 807, L2

\bibitem[{{Komiyama} {et~al.}(2018){Komiyama}, {Obuchi}, {Nakaya}, {Kamata}, {Kawanomoto}, {Utsumi}, {Miyazaki}, {Uraguchi}, {Furusawa}, {Morokuma}, {Uchida}, {Miyatake}, {Mineo}, {Fujimori}, {Aihara}, {Karoji}, {Gunn}, \& {Wang}}]{HSC_cam4}
{Komiyama}, Y., {Obuchi}, Y., {Nakaya}, H., {et~al.} 2018, \pasj, 70, S2

\bibitem[{{Kormendy}(1985)}]{Kormendy_1985}
{Kormendy}, J. 1985, \apj, 295, 73

\bibitem[{{Kormendy} {et~al.}(2009){Kormendy}, {Fisher}, {Cornell}, \& {Bender}}]{Kormendy_2009}
{Kormendy}, J., {Fisher}, D.~B., {Cornell}, M.~E., \& {Bender}, R. 2009, \apjs, 182, 216

\bibitem[{{La Marca} {et~al.}(2022{\natexlab{a}}){La Marca}, {Iodice}, {Cantiello}, {Forbes}, {Rejkuba}, {Hilker}, {Arnaboldi}, {Greggio}, {Spiniello}, {Mieske}, {Venhola}, {Spavone}, {D'Ago}, {Raj}, {Ragusa}, {Mirabile}, {Rampazzo}, {Peletier}, {Paolillo}, {Challapa}, \& {Schipani}}]{Marca}
{La Marca}, A., {Iodice}, E., {Cantiello}, M., {et~al.} 2022{\natexlab{a}}, \aap, 665, A105

\bibitem[{{La Marca} {et~al.}(2022{\natexlab{b}}){La Marca}, {Iodice}, {Cantiello}, {Forbes}, {Rejkuba}, {Hilker}, {Arnaboldi}, {Greggio}, {Spiniello}, {Mieske}, {Venhola}, {Spavone}, {D'Ago}, {Raj}, {Ragusa}, {Mirabile}, {Rampazzo}, {Peletier}, {Paolillo}, {Challapa}, \& {Schipani}}]{2022AMarca}
{La Marca}, A., {Iodice}, E., {Cantiello}, M., {et~al.} 2022{\natexlab{b}}, \aap, 665, A105

\bibitem[{{Laudato} \& {Salzano}(2023)}]{Laudato}
{Laudato}, E. \& {Salzano}, V. 2023, European Physical Journal C, 83, 402

\bibitem[{{Laureijs} {et~al.}(2011){Laureijs}, {Amiaux}, {Arduini}, {Augu{\`e}res}, {Brinchmann}, {Cole}, {Cropper}, {Dabin}, {Duvet}, {Ealet}, {Garilli}, {Gondoin}, {Guzzo}, {Hoar}, {Hoekstra}, {Holmes}, {Kitching}, {Maciaszek}, {Mellier}, {Pasian}, {Percival}, {Rhodes}, {Saavedra Criado}, {Sauvage}, {Scaramella}, {Valenziano}, {Warren}, {Bender}, {Castander}, {Cimatti}, {Le F{\`e}vre}, {Kurki-Suonio}, {Levi}, {Lilje}, {Meylan}, {Nichol}, {Pedersen}, {Popa}, {Rebolo Lopez}, {Rix}, {Rottgering}, {Zeilinger}, {Grupp}, {Hudelot}, {Massey}, {Meneghetti}, {Miller}, {Paltani}, {Paulin-Henriksson}, {Pires}, {Saxton}, {Schrabback}, {Seidel}, {Walsh}, {Aghanim}, {Amendola}, {Bartlett}, {Baccigalupi}, {Beaulieu}, {Benabed}, {Cuby}, {Elbaz}, {Fosalba}, {Gavazzi}, {Helmi}, {Hook}, {Irwin}, {Kneib}, {Kunz}, {Mannucci}, {Moscardini}, {Tao}, {Teyssier}, {Weller}, {Zamorani}, {Zapatero Osorio}, {Boulade}, {Foumond}, {Di Giorgio}, {Guttridge}, {James}, {Kemp}, {Martignac}, {Spencer}, {Walton}, {Bl{\"u}mchen}, {Bonoli},
  {Bortoletto}, {Cerna}, {Corcione}, {Fabron}, {Jahnke}, {Ligori}, {Madrid}, {Martin}, {Morgante}, {Pamplona}, {Prieto}, {Riva}, {Toledo}, {Trifoglio}, {Zerbi}, {Abdalla}, {Douspis}, {Grenet}, {Borgani}, {Bouwens}, {Courbin}, {Delouis}, {Dubath}, {Fontana}, {Frailis}, {Grazian}, {Koppenh{\"o}fer}, {Mansutti}, {Melchior}, {Mignoli}, {Mohr}, {Neissner}, {Noddle}, {Poncet}, {Scodeggio}, {Serrano}, {Shane}, {Starck}, {Surace}, {Taylor}, {Verdoes-Kleijn}, {Vuerli}, {Williams}, {Zacchei}, {Altieri}, {Escudero Sanz}, {Kohley}, {Oosterbroek}, {Astier}, {Bacon}, {Bardelli}, {Baugh}, {Bellagamba}, {Benoist}, {Bianchi}, {Biviano}, {Branchini}, {Carbone}, {Cardone}, {Clements}, {Colombi}, {Conselice}, {Cresci}, {Deacon}, {Dunlop}, {Fedeli}, {Fontanot}, {Franzetti}, {Giocoli}, {Garcia-Bellido}, {Gow}, {Heavens}, {Hewett}, {Heymans}, {Holland}, {Huang}, {Ilbert}, {Joachimi}, {Jennins}, {Kerins}, {Kiessling}, {Kirk}, {Kotak}, {Krause}, {Lahav}, {van Leeuwen}, {Lesgourgues}, {Lombardi}, {Magliocchetti}, {Maguire},
  {Majerotto}, {Maoli}, {Marulli}, {Maurogordato}, {McCracken}, {McLure}, {Melchiorri}, {Merson}, {Moresco}, {Nonino}, {Norberg}, {Peacock}, {Pello}, {Penny}, {Pettorino}, {Di Porto}, {Pozzetti}, {Quercellini}, {Radovich}, {Rassat}, {Roche}, {Ronayette}, {Rossetti}, {Sartoris}, {Schneider}, {Semboloni}, {Serjeant}, {Simpson}, {Skordis}, {Smadja}, {Smartt}, {Spano}, {Spiro}, {Sullivan}, {Tilquin}, {Trotta}, {Verde}, {Wang}, {Williger}, {Zhao}, {Zoubian}, \& {Zucca}}]{laureijs2011euclid}
{Laureijs}, R., {Amiaux}, J., {Arduini}, S., {et~al.} 2011, arXiv e-prints, arXiv:1110.3193

\bibitem[{{Lazar} {et~al.}(2024){Lazar}, {Kaviraj}, {Watkins}, {Martin}, {Bichang'a}, \& {Jackson}}]{Lazar_dwarf_morphology}
{Lazar}, I., {Kaviraj}, S., {Watkins}, A.~E., {et~al.} 2024, \mnras, 529, 499

\bibitem[{{Lee} {et~al.}(2020){Lee}, {Hodges-Kluck}, \& {Gallo}}]{Lee_UV_udgs}
{Lee}, C.~H., {Hodges-Kluck}, E., \& {Gallo}, E. 2020, \mnras, 497, 2759

\bibitem[{{Lim} {et~al.}(2020){Lim}, {C{\^o}t{\'e}}, {Peng}, {Ferrarese}, {Roediger}, {Durrell}, {Mihos}, {Wang}, {Gwyn}, {Cuillandre}, {Liu}, {S{\'a}nchez-Janssen}, {Toloba}, {Sales}, {Guhathakurta}, {Lan{\c{c}}on}, \& {Puzia}}]{Lim}
{Lim}, S., {C{\^o}t{\'e}}, P., {Peng}, E.~W., {et~al.} 2020, \apj, 899, 69

\bibitem[{{Magnier} {et~al.}(2013){Magnier}, {Schlafly}, {Finkbeiner}, {Juric}, {Tonry}, {Burgett}, {Chambers}, {Flewelling}, {Kaiser}, {Kudritzki}, {Morgan}, {Price}, {Sweeney}, \& {Stubbs}}]{Magnier_PanSTARS}
{Magnier}, E.~A., {Schlafly}, E., {Finkbeiner}, D., {et~al.} 2013, \apjs, 205, 20

\bibitem[{{Mancera Pi{\~n}a} {et~al.}(2019){Mancera Pi{\~n}a}, {Aguerri}, {Peletier}, {Venhola}, {Trager}, \& {Choque Challapa}}]{Mancera_udg}
{Mancera Pi{\~n}a}, P.~E., {Aguerri}, J.~A.~L., {Peletier}, R.~F., {et~al.} 2019, \mnras, 485, 1036

\bibitem[{{Mancera Pi{\~n}a} {et~al.}(2018){Mancera Pi{\~n}a}, {Peletier}, {Aguerri}, {Venhola}, {Trager}, \& {Choque Challapa}}]{Mancera}
{Mancera Pi{\~n}a}, P.~E., {Peletier}, R.~F., {Aguerri}, J.~A.~L., {et~al.} 2018, \mnras, 481, 4381

\bibitem[{{Marleau} {et~al.}(2021){Marleau}, {Habas}, {Poulain}, {Duc}, {M{\"u}ller}, {Lim}, {Durrell}, {S{\'a}nchez-Janssen}, {Paudel}, {Ahad}, {Chougule}, {B{\'\i}lek}, \& {Fensch}}]{Marleau}
{Marleau}, F.~R., {Habas}, R., {Poulain}, M., {et~al.} 2021, \aap, 654, A105

\bibitem[{{Martin} {et~al.}(2005){Martin}, {Fanson}, {Schiminovich}, {Morrissey}, {Friedman}, {Barlow}, {Conrow}, {Grange}, {Jelinsky}, {Milliard}, {Siegmund}, {Bianchi}, {Byun}, {Donas}, {Forster}, {Heckman}, {Lee}, {Madore}, {Malina}, {Neff}, {Rich}, {Small}, {Surber}, {Szalay}, {Welsh}, \& {Wyder}}]{Martin_galaex}
{Martin}, D.~C., {Fanson}, J., {Schiminovich}, D., {et~al.} 2005, \apjl, 619, L1

\bibitem[{{Martin} {et~al.}(2019){Martin}, {Kaviraj}, {Laigle}, {Devriendt}, {Jackson}, {Peirani}, {Dubois}, {Pichon}, \& {Slyz}}]{Martin}
{Martin}, G., {Kaviraj}, S., {Laigle}, C., {et~al.} 2019, \mnras, 485, 796

\bibitem[{{McGaugh}(1996)}]{McGaugh}
{McGaugh}, S.~S. 1996, \mnras, 280, 337

\bibitem[{{McGaugh} \& {Bothun}(1994)}]{McGaugh_bothun}
{McGaugh}, S.~S. \& {Bothun}, G.~D. 1994, \aj, 107, 530

\bibitem[{{Merritt} {et~al.}(2016){Merritt}, {van Dokkum}, {Danieli}, {Abraham}, {Zhang}, {Karachentsev}, \& {Makarova}}]{Merritt}
{Merritt}, A., {van Dokkum}, P., {Danieli}, S., {et~al.} 2016, \apj, 833, 168

\bibitem[{{Mihos} {et~al.}(2015){Mihos}, {Durrell}, {Ferrarese}, {Feldmeier}, {C{\^o}t{\'e}}, {Peng}, {Harding}, {Liu}, {Gwyn}, \& {Cuillandre}}]{Mihos}
{Mihos}, J.~C., {Durrell}, P.~R., {Ferrarese}, L., {et~al.} 2015, \apjl, 809, L21

\bibitem[{{Minchin} {et~al.}(2004){Minchin}, {Disney}, {Parker}, {Boyce}, {de Blok}, {Banks}, {Ekers}, {Freeman}, {Garcia}, {Gibson}, {Grossi}, {Haynes}, {Knezek}, {Lang}, {Malin}, {Price}, {Putman}, {Stewart}, \& {Wright}}]{Minchin}
{Minchin}, R.~F., {Disney}, M.~J., {Parker}, Q.~A., {et~al.} 2004, \mnras, 355, 1303

\bibitem[{{Miyazaki} {et~al.}(2018){Miyazaki}, {Komiyama}, {Kawanomoto}, {Doi}, {Furusawa}, {Hamana}, {Hayashi}, {Ikeda}, {Kamata}, {Karoji}, {Koike}, {Kurakami}, {Miyama}, {Morokuma}, {Nakata}, {Namikawa}, {Nakaya}, {Nariai}, {Obuchi}, {Oishi}, {Okada}, {Okura}, {Tait}, {Takata}, {Tanaka}, {Tanaka}, {Terai}, {Tomono}, {Uraguchi}, {Usuda}, {Utsumi}, {Yamada}, {Yamanoi}, {Aihara}, {Fujimori}, {Mineo}, {Miyatake}, {Oguri}, {Uchida}, {Tanaka}, {Yasuda}, {Takada}, {Murayama}, {Nishizawa}, {Sugiyama}, {Chiba}, {Futamase}, {Wang}, {Chen}, {Ho}, {Liaw}, {Chiu}, {Ho}, {Lai}, {Lee}, {Jeng}, {Iwamura}, {Armstrong}, {Bickerton}, {Bosch}, {Gunn}, {Lupton}, {Loomis}, {Price}, {Smith}, {Strauss}, {Turner}, {Suzuki}, {Miyazaki}, {Muramatsu}, {Yamamoto}, {Endo}, {Ezaki}, {Ito}, {Kawaguchi}, {Sofuku}, {Taniike}, {Akutsu}, {Dojo}, {Kasumi}, {Matsuda}, {Imoto}, {Miwa}, {Suzuki}, {Takeshi}, \& {Yokota}}]{HSC_cam3}
{Miyazaki}, S., {Komiyama}, Y., {Kawanomoto}, S., {et~al.} 2018, \pasj, 70, S1

\bibitem[{{Montes} {et~al.}(2024){Montes}, {Trujillo}, {Karunakaran}, {Infante-Sainz}, {Spekkens}, {Golini}, {Beasley}, {Cebri{\'a}n}, {Chamba}, {D'Onofrio}, {Kelvin}, \& {Rom{\'a}n}}]{Montes}
{Montes}, M., {Trujillo}, I., {Karunakaran}, A., {et~al.} 2024, \aap, 681, A15

\bibitem[{{Moore} {et~al.}(1999){Moore}, {Ghigna}, {Governato}, {Lake}, {Quinn}, {Stadel}, \& {Tozzi}}]{Moore_cosmology}
{Moore}, B., {Ghigna}, S., {Governato}, F., {et~al.} 1999, \apjl, 524, L19

\bibitem[{{Moore} {et~al.}(1996){Moore}, {Katz}, {Lake}, {Dressler}, \& {Oemler}}]{Moore1996}
{Moore}, B., {Katz}, N., {Lake}, G., {Dressler}, A., \& {Oemler}, A. 1996, \nat, 379, 613

\bibitem[{{Morganson} {et~al.}(2018){Morganson}, {Gruendl}, {Menanteau}, {Carrasco Kind}, {Chen}, {Daues}, {Drlica-Wagner}, {Friedel}, {Gower}, {Johnson}, {Johnson}, {Kessler}, {Paz-Chinch{\'o}n}, {Petravick}, {Pond}, {Yanny}, {Allam}, {Armstrong}, {Barkhouse}, {Bechtol}, {Benoit-L{\'e}vy}, {Bernstein}, {Bertin}, {Buckley-Geer}, {Covarrubias}, {Desai}, {Diehl}, {Goldstein}, {Gruen}, {Li}, {Lin}, {Marriner}, {Mohr}, {Neilsen}, {Ngeow}, {Paech}, {Rykoff}, {Sako}, {Sevilla-Noarbe}, {Sheldon}, {Sobreira}, {Tucker}, {Wester}, \& {DES Collaboration}}]{Morganson}
{Morganson}, E., {Gruendl}, R.~A., {Menanteau}, F., {et~al.} 2018, \pasp, 130, 074501

\bibitem[{{Morrissey} {et~al.}(2007){Morrissey}, {Conrow}, {Barlow}, {Small}, {Seibert}, {Wyder}, {Budav{\'a}ri}, {Arnouts}, {Friedman}, {Forster}, {Martin}, {Neff}, {Schiminovich}, {Bianchi}, {Donas}, {Heckman}, {Lee}, {Madore}, {Milliard}, {Rich}, {Szalay}, {Welsh}, \& {Yi}}]{Morrissey}
{Morrissey}, P., {Conrow}, T., {Barlow}, T.~A., {et~al.} 2007, \apjs, 173, 682

\bibitem[{{M{\"u}ller} {et~al.}(2018){M{\"u}ller}, {Jerjen}, \& {Binggeli}}]{muller}
{M{\"u}ller}, O., {Jerjen}, H., \& {Binggeli}, B. 2018, \aap, 615, A105

\bibitem[{{O'Neil} \& {Bothun}(2000)}]{Neil}
{O'Neil}, K. \& {Bothun}, G. 2000, \apj, 529, 811

\bibitem[{{Paudel} {et~al.}(2023){Paudel}, {Yoon}, {Yoo}, {Smith}, {Chhatkuli}, {Kumar Bachchan}, {Aryal}, {Adhikari}, {Adhikari}, {Sedain}, {Sheikh}, {Dhital}, {Giri}, \& {Baral}}]{Paudel_sanjaya}
{Paudel}, S., {Yoon}, S.-J., {Yoo}, J., {et~al.} 2023, \apjs, 265, 57

\bibitem[{{Pearson} {et~al.}(2022){Pearson}, {Suelves}, {Ho}, {Oi}, {Brough}, {Holwerda}, {Hopkins}, {Huang}, {Hwang}, {Kelvin}, {Kim}, {L{\'o}pez-S{\'a}nchez}, {Ma{\l}ek}, {Pearson}, {Poliszczuk}, {Pollo}, {Rodriguez-Gomez}, {Shim}, {Toba}, \& {Wang}}]{Pearson}
{Pearson}, W.~J., {Suelves}, L.~E., {Ho}, S.~C.~C., {et~al.} 2022, \aap, 661, A52

\bibitem[{{Peng} {et~al.}(2010){Peng}, {Ho}, {Impey}, \& {Rix}}]{GALFIT}
{Peng}, C.~Y., {Ho}, L.~C., {Impey}, C.~D., \& {Rix}, H.-W. 2010, \aj, 139, 2097

\bibitem[{{P{\'e}rez-Carrasco} {et~al.}(2019){P{\'e}rez-Carrasco}, {Cabrera-Vives}, {Martinez-Marin}, {Cerulo}, {Demarco}, {Protopapas}, {Godoy}, \& {Huertas-Company}}]{2019PASP..131j8002P}
{P{\'e}rez-Carrasco}, M., {Cabrera-Vives}, G., {Martinez-Marin}, M., {et~al.} 2019, \pasp, 131, 108002

\bibitem[{{Poulain} {et~al.}(2021){Poulain}, {Marleau}, {Habas}, {Duc}, {S{\'a}nchez-Janssen}, {Durrell}, {Paudel}, {Ahad}, {Chougule}, {M{\"u}ller}, {Lim}, {B{\'\i}lek}, \& {Fensch}}]{Poulain}
{Poulain}, M., {Marleau}, F.~R., {Habas}, R., {et~al.} 2021, \mnras, 506, 5494

\bibitem[{{Rines} {et~al.}(2003){Rines}, {Geller}, {Kurtz}, \& {Diaferio}}]{Rines_abell_194_z}
{Rines}, K., {Geller}, M.~J., {Kurtz}, M.~J., \& {Diaferio}, A. 2003, \aj, 126, 2152

\bibitem[{{Rom{\'a}n} \& {Trujillo}(2017)}]{Roman_2017}
{Rom{\'a}n}, J. \& {Trujillo}, I. 2017, \mnras, 468, 4039

\bibitem[{{Rom{\'a}n} {et~al.}(2020){Rom{\'a}n}, {Trujillo}, \& {Montes}}]{Roman_surface_brightness_depth}
{Rom{\'a}n}, J., {Trujillo}, I., \& {Montes}, M. 2020, \aap, 644, A42

\bibitem[{{Saburova} {et~al.}(2021){Saburova}, {Chilingarian}, {Kasparova}, {Sil'chenko}, {Grishin}, {Katkov}, \& {Uklein}}]{Saburova2021}
{Saburova}, A.~S., {Chilingarian}, I.~V., {Kasparova}, A.~V., {et~al.} 2021, \mnras, 503, 830

\bibitem[{{Saburova} {et~al.}(2023){Saburova}, {Chilingarian}, {Kulier}, {Galaz}, {Grishin}, {Kasparova}, {Toptun}, \& {Katkov}}]{Saburova2023}
{Saburova}, A.~S., {Chilingarian}, I.~V., {Kulier}, A., {et~al.} 2023, \mnras, 520, L85

\bibitem[{{Sales} {et~al.}(2020){Sales}, {Navarro}, {Pe{\~n}afiel}, {Peng}, {Lim}, \& {Hernquist}}]{Sales}
{Sales}, L.~V., {Navarro}, J.~F., {Pe{\~n}afiel}, L., {et~al.} 2020, \mnras, 494, 1848

\bibitem[{{Sandage} \& {Binggeli}(1984)}]{Sandage}
{Sandage}, A. \& {Binggeli}, B. 1984, \aj, 89, 919

\bibitem[{{Schlafly} \& {Finkbeiner}(2011)}]{Schlafly2011}
{Schlafly}, E.~F. \& {Finkbeiner}, D.~P. 2011, \apj, 737, 103

\bibitem[{{Schlegel} {et~al.}(1998){Schlegel}, {Finkbeiner}, \& {Davis}}]{Schlegel1998}
{Schlegel}, D.~J., {Finkbeiner}, D.~P., \& {Davis}, M. 1998, \apj, 500, 525

\bibitem[{{Singh} {et~al.}(2019){Singh}, {Zaritsky}, {Donnerstein}, \& {Spekkens}}]{Singh_UDG_nuv}
{Singh}, P.~R., {Zaritsky}, D., {Donnerstein}, R., \& {Spekkens}, K. 2019, \aj, 157, 212

\bibitem[{{Somalwar} {et~al.}(2020){Somalwar}, {Greene}, {Greco}, {Huang}, {Beaton}, {Goulding}, \& {Lancaster}}]{Somalwar}
{Somalwar}, J.~J., {Greene}, J.~E., {Greco}, J.~P., {et~al.} 2020, \apj, 902, 45

\bibitem[{{Sprayberry} {et~al.}(1995){Sprayberry}, {Impey}, {Bothun}, \& {Irwin}}]{Sprayberry1995}
{Sprayberry}, D., {Impey}, C.~D., {Bothun}, G.~D., \& {Irwin}, M.~J. 1995, \aj, 109, 558

\bibitem[{{Su} {et~al.}(2024){Su}, {Yi}, {Liang}, {Du}, {Liu}, {Kong}, {Bu}, \& {Wu}}]{Su_LSBGNET}
{Su}, H., {Yi}, Z., {Liang}, Z., {et~al.} 2024, \mnras, 528, 873

\bibitem[{{Tanoglidis} {et~al.}(2021{\natexlab{a}}){Tanoglidis}, {{\'C}iprijanovi{\'c}}, \& {Drlica-Wagner}}]{Tanoglidis2}
{Tanoglidis}, D., {{\'C}iprijanovi{\'c}}, A., \& {Drlica-Wagner}, A. 2021{\natexlab{a}}, Astronomy and Computing, 35, 100469

\bibitem[{{Tanoglidis} {et~al.}(2021{\natexlab{b}}){Tanoglidis}, {Drlica-Wagner}, {Wei}, {Li}, {S{\'a}nchez}, {Zhang}, {Peter}, {Feldmeier-Krause}, {Prat}, {Casey}, {Palmese}, {S{\'a}nchez}, {DeRose}, {Conselice}, {Gagnon}, {Abbott}, {Aguena}, {Allam}, {Avila}, {Bechtol}, {Bertin}, {Bhargava}, {Brooks}, {Burke}, {Rosell}, {Kind}, {Carretero}, {Chang}, {Costanzi}, {da Costa}, {De Vicente}, {Desai}, {Diehl}, {Doel}, {Eifler}, {Everett}, {Evrard}, {Flaugher}, {Frieman}, {Garc{\'\i}a-Bellido}, {Gerdes}, {Gruendl}, {Gschwend}, {Gutierrez}, {Hartley}, {Hollowood}, {Huterer}, {James}, {Krause}, {Kuehn}, {Kuropatkin}, {Maia}, {March}, {Marshall}, {Menanteau}, {Miquel}, {Ogando}, {Paz-Chinch{\'o}n}, {Romer}, {Roodman}, {Sanchez}, {Scarpine}, {Serrano}, {Sevilla-Noarbe}, {Smith}, {Suchyta}, {Tarle}, {Thomas}, {Tucker}, {Walker}, \& {DES Collaboration}}]{Tanoglidis1}
{Tanoglidis}, D., {Drlica-Wagner}, A., {Wei}, K., {et~al.} 2021{\natexlab{b}}, \apjs, 252, 18

\bibitem[{{Tempel} {et~al.}(2016){Tempel}, {Kipper}, {Tamm}, {Gramann}, {Einasto}, {Sepp}, \& {Tuvikene}}]{Tempel_cluster_memebership}
{Tempel}, E., {Kipper}, R., {Tamm}, A., {et~al.} 2016, \aap, 588, A14

\bibitem[{{Thuruthipilly} {et~al.}(2024{\natexlab{a}}){Thuruthipilly}, {Grespan}, \& {Zadro{\.z}ny}}]{Hareesh2}
{Thuruthipilly}, H., {Grespan}, M., \& {Zadro{\.z}ny}, A. 2024{\natexlab{a}}, in American Institute of Physics Conference Series, Vol. 3061, American Institute of Physics Conference Series, 040003

\bibitem[{{Thuruthipilly} {et~al.}(2024{\natexlab{b}}){Thuruthipilly}, {Junais}, {Pollo}, {Sureshkumar}, {Grespan}, {Sawant}, {Ma{\l}ek}, \& {Zadrozny}}]{haressh_lsbs}
{Thuruthipilly}, H., {Junais}, {Pollo}, A., {et~al.} 2024{\natexlab{b}}, \aap, 682, A4

\bibitem[{{Thuruthipilly} {et~al.}(2022){Thuruthipilly}, {Zadrozny}, {Pollo}, \& {Biesiada}}]{Hareesh}
{Thuruthipilly}, H., {Zadrozny}, A., {Pollo}, A., \& {Biesiada}, M. 2022, \aap, 664, A4

\bibitem[{{van der Burg} {et~al.}(2017){van der Burg}, {Hoekstra}, {Muzzin}, {Sif{\'o}n}, {Viola}, {Bremer}, {Brough}, {Driver}, {Erben}, {Heymans}, {Hildebrandt}, {Holwerda}, {Klaes}, {Kuijken}, {McGee}, {Nakajima}, {Napolitano}, {Norberg}, {Taylor}, \& {Valentijn}}]{vanderBurg2017}
{van der Burg}, R. F.~J., {Hoekstra}, H., {Muzzin}, A., {et~al.} 2017, \aap, 607, A79

\bibitem[{{van der Burg} {et~al.}(2016){van der Burg}, {Muzzin}, \& {Hoekstra}}]{Burg}
{van der Burg}, R. F.~J., {Muzzin}, A., \& {Hoekstra}, H. 2016, \aap, 590, A20

\bibitem[{{van Dokkum} {et~al.}(2015{\natexlab{a}}){van Dokkum}, {Abraham}, {Merritt}, {Zhang}, {Geha}, \& {Conroy}}]{Dokkum}
{van Dokkum}, P.~G., {Abraham}, R., {Merritt}, A., {et~al.} 2015{\natexlab{a}}, \apjl, 798, L45

\bibitem[{{van Dokkum} {et~al.}(2015{\natexlab{b}}){van Dokkum}, {Romanowsky}, {Abraham}, {Brodie}, {Conroy}, {Geha}, {Merritt}, {Villaume}, \& {Zhang}}]{vanDokkum}
{van Dokkum}, P.~G., {Romanowsky}, A.~J., {Abraham}, R., {et~al.} 2015{\natexlab{b}}, \apjl, 804, L26

\bibitem[{Vaswani {et~al.}(2017)Vaswani, Shazeer, Parmar, Uszkoreit, Jones, Gomez, Kaiser, \& Polosukhin}]{vaswani2017attention}
Vaswani, A., Shazeer, N., Parmar, N., {et~al.} 2017, in Advances in Neural Information Processing Systems 30: Annual Conference on Neural Information Processing Systems 2017, December 4-9, 2017, Long Beach, CA, {USA}, 5998--6008

\bibitem[{{Venhola} {et~al.}(2018){Venhola}, {Peletier}, {Laurikainen}, {Salo}, {Iodice}, {Mieske}, {Hilker}, {Wittmann}, {Lisker}, {Paolillo}, {Cantiello}, {Janz}, {Spavone}, {D'Abrusco}, {van de Ven}, {Napolitano}, {Verdoes Kleijn}, {Maddox}, {Capaccioli}, {Grado}, {Valentijn}, {Falc{\'o}n-Barroso}, \& {Limatola}}]{Venhola_dwarfs}
{Venhola}, A., {Peletier}, R., {Laurikainen}, E., {et~al.} 2018, \aap, 620, A165

\bibitem[{{Venhola} {et~al.}(2019){Venhola}, {Peletier}, {Laurikainen}, {Salo}, {Iodice}, {Mieske}, {Hilker}, {Wittmann}, {Paolillo}, {Cantiello}, {Janz}, {Spavone}, {D'Abrusco}, {van de Ven}, {Napolitano}, {Verdoes Kleijn}, {Capaccioli}, {Grado}, {Valentijn}, {Falc{\'o}n-Barroso}, \& {Limatola}}]{Venhola_optical_fornax}
{Venhola}, A., {Peletier}, R., {Laurikainen}, E., {et~al.} 2019, \aap, 625, A143

\bibitem[{{Venhola} {et~al.}(2022){Venhola}, {Peletier}, {Salo}, {Laurikainen}, {Janz}, {Haigh}, {Wilkinson}, {Iodice}, {Hilker}, {Mieske}, {Cantiello}, \& {Spavone}}]{Venhola_udgs}
{Venhola}, A., {Peletier}, R.~F., {Salo}, H., {et~al.} 2022, \aap, 662, A43

\bibitem[{{Wang} {et~al.}(2023){Wang}, {Peng}, {Liu}, {Mihos}, {C{\^o}t{\'e}}, {Ferrarese}, {Taylor}, {Blakeslee}, {Cuillandre}, {Duc}, {Guhathakurta}, {Gwyn}, {Ko}, {Lan{\c{c}}on}, {Lim}, {MacArthur}, {Puzia}, {Roediger}, {Sales}, {S{\'a}nchez-Janssen}, {Spengler}, {Toloba}, {Zhang}, \& {Zhu}}]{Wang_2023}
{Wang}, K., {Peng}, E.~W., {Liu}, C., {et~al.} 2023, \nat, 623, 296

\bibitem[{{Wei} {et~al.}(2020){Wei}, {Huerta}, {Whitmore}, {Lee}, {Hannon}, {Chandar}, {Dale}, {Larson}, {Thilker}, {Ubeda}, {Boquien}, {Chevance}, {Kruijssen}, {Schruba}, {Blanc}, \& {Congiu}}]{Wei}
{Wei}, W., {Huerta}, E.~A., {Whitmore}, B.~C., {et~al.} 2020, \mnras, 493, 3178

\bibitem[{{Xing} {et~al.}(2023){Xing}, {Yi}, {Liang}, {Su}, {Du}, {He}, {Liu}, {Kong}, {Bu}, \& {Wu}}]{Xing2023}
{Xing}, Y., {Yi}, Z., {Liang}, Z., {et~al.} 2023, \apjs, 269, 59

\bibitem[{{Yagi} {et~al.}(2016){Yagi}, {Koda}, {Komiyama}, \& {Yamanoi}}]{Yagi}
{Yagi}, M., {Koda}, J., {Komiyama}, Y., \& {Yamanoi}, H. 2016, \apjs, 225, 11

\bibitem[{{Yi} {et~al.}(2022){Yi}, {Li}, {Du}, {Liu}, {Liang}, {Xing}, {Pan}, {Bu}, {Kong}, \& {Wu}}]{Yi_2022}
{Yi}, Z., {Li}, J., {Du}, W., {et~al.} 2022, \mnras, 513, 3972

\bibitem[{Yosinski {et~al.}(2014)Yosinski, Clune, Bengio, \& Lipson}]{Yosinski}
Yosinski, J., Clune, J., Bengio, Y., \& Lipson, H. 2014, in Advances in Neural Information Processing Systems, ed. Z.~Ghahramani, M.~Welling, C.~Cortes, N.~Lawrence, \& K.~Weinberger, Vol.~27 (Curran Associates, Inc.)

\bibitem[{{Zaritsky} {et~al.}(2023){Zaritsky}, {Donnerstein}, {Dey}, {Karunakaran}, {Kadowaki}, {Khim}, {Spekkens}, \& {Zhang}}]{Zaritsky_2023}
{Zaritsky}, D., {Donnerstein}, R., {Dey}, A., {et~al.} 2023, \apjs, 267, 27

\bibitem[{{Zhong} {et~al.}(2008){Zhong}, {Liang}, {Liu}, {Hammer}, {Hu}, {Chen}, {Deng}, \& {Zhang}}]{Zhong_SDSS_lsbs}
{Zhong}, G.~H., {Liang}, Y.~C., {Liu}, F.~S., {et~al.} 2008, \mnras, 391, 986

\bibitem[{{Z{\"o}ller} {et~al.}(2024){Z{\"o}ller}, {Kluge}, {Staiger}, \& {Bender}}]{zoller_coma}
{Z{\"o}ller}, R., {Kluge}, M., {Staiger}, B., \& {Bender}, R. 2024, \apjs, 271, 52

\end{thebibliography}

\begin{appendix}
\section{Transformer models}
\subsection{Training}\label{training_on_des}
 All of the LSBG DETR and LSBG ViT models were trained with an initial learning rate of $\alpha= 10^{-4}$. We used the exponential linear unit (ELU) function as the activation function for all the layers in these models \citep{Clevert}. We initialise the weights of our model with the Xavier uniform initialiser \citep{Glorot}. All layers are trained with the ADAM optimiser using the default exponential decay rates \citep{kingma2017adam}. We have used the early stopping callback from {\tt Keras} \footnote{\url{https://keras.io/api/callbacks}} to monitor the validation loss of the model and stop training once the loss was converged. The models LSBG DETR 1, 2, 3, and 4 were trained for 110, 87, 128, and 168 epochs, respectively. Similarly, the LSBG ViT 1, 2, 3, and 4 were trained for 82, 57, 87 and 58 epochs, respectively. Our code for LSBG DETR  and LSBG ViT is publicly available \footnote{\url{https://github.com/hareesht23/}}.
 \subsection{Model performance on the testing set}\label{model_perfomance_on_DES}
We have developed four of each transformer models, LSBG DETR and LSBG ViT, each with different hyperparameters. Each model is a regression model, predicting the probability of an input being an LSBG. To improve overall performance, we create an ensemble for LSBG DETR and LSBG ViT by averaging the output probabilities from the four models of each type. We set the classification threshold at 0.5, meaning inputs with a predicted probability greater than or equal to 0.5 are classified as LSBGs. The performance of all the models is listed in Table \ref{table:perfomace_on_DES} along with the architecture, accuracy and  AUROC  for each model on the test dataset from DES.

\begin{table}[h]
\centering 
\addtolength{\tabcolsep}{-03.50pt}
\caption{Table comprising the architecture, accuracy, true positive rate (TPR), false positive rate (FPR) and AUROC of all the models in chronological order of creation.}
\begin{tabular}{c c c c c}\hline\hline
Model   name  &Accuracy (\%) &TPR & FPR& AUROC \\\hline
LSBG ViT 1 & 95.58 & 0.96 & 0.05 &  0.9908   \\
LSBG ViT 2 & 95.48 & 0.96 & 0.05 & 0.9906 \\
LSBG ViT 3 & 95.58 & 0.97 & 0.06 & 0.9906\\
LSBG ViT 4 & 95.14 & 0.96 & 0.05 &  0.9895 \\
LSBG ViT Ensemble & 95.62 & 0.96 & 0.05 & 0.9911 \\
LSBG DETR 1 & 95.68 & 0.96 & 0.04  & 0.9893  \\
LSBG DETR 2 & 95.36 &0.95  &0.04  &  0.9887  \\
LSBG DETR 3 & 95.48 & 0.96 & 0.05 &  0.9891   \\
LSBG DETR 4 & 95.54 & 0.97 & 0.06 & 0.9904 \\
LSBG DETR Ensemble & 95.62 & 0.96 & 0.05 &0.9903 \\
 \hline
\end{tabular}
\label{table:perfomace_on_DES}
\end{table}

The TPR and FPR reported in Table \ref{table:perfomace_on_DES} are computed using a threshold of 0.5 to distinguish between LSBGs and contaminants. 

\section{Sample of LSBGs}
 \begin{sidewaystable*}
\caption{Sample of LSBGs identified in this work.}
\centering
\addtolength{\tabcolsep}{-2.75 pt}
\begin{tabular}{lrrrrrrrrrrrrrrrrrr}
\hline
 NAME &  RA & DEC & $n$ & $q$ & PA &$log(\Sigma_{star})$&$log(M_{star})$ & $r_{\mathrm{eff},g}$ & \textit{g} & $\overline{\mu}_{\mathrm{eff},g}$ &$r_{\mathrm{eff},r}$ & \textit{r} & $\overline{\mu}_{\mathrm{eff},r}$ &\textit{NUV}& $\Delta$ \textit{NUV}&\textit{FUV}& $\Delta$ \textit{FUV} & DES \\
   &  (deg) & (deg) &  &  & (deg) &($M_{\odot} kpc^{-2}$)& ($M_{\odot}$) & ($\arcsec$) & (mag) &( $\frac{\text{mag}}{\text{arcsec}^2}$) & ($\arcsec$) & (mag) &($\frac{\text{mag}}{\text{arcsec}^2}$) & (mag) & (mag) & (mag) & (mag) &  \\
\hline
J01:22:31.599-01:37:29.491 & 20.63166 & -1.62486 & 0.98 & 0.44 &  39.5 &                    6.7 &               7.4 &          3.7 & 21.4 &                 25.3 &          3.8 & 20.7 &                 24.8 & >23.7 &            - & >23.4 &            - &    0 \\
J01:22:44.406-01:21:05.462 & 20.68503 & -1.35152 & 0.73 & 0.47 & -68.7 &                    6.7 &               7.3 &          3.4 & 20.7 &                 24.5 &          3.2 & 20.4 &                 24.1 &  22.4 &          0.1 &  22.9 &          0.2 &    0 \\
J01:22:57.659-01:03:57.566 & 20.74025 & -1.06599 & 1.09 & 0.74 & -35.6 &                    6.8 &               7.4 &          2.7 & 21.1 &                 24.9 &          2.5 & 20.6 &                 24.2 & >23.5 &            - & >23.5 &            - &    0 \\
J01:23:09.271-01:28:26.721 & 20.78863 & -1.47409 & 0.98 & 0.60 &  67.3 &                    6.4 &               7.3 &          3.8 & 21.6 &                 25.9 &          3.8 & 21.0 &                 25.3 & >24.0 &            - & >24.5 &            - &    1 \\
J01:23:22.943-01:17:38.620 & 20.84560 & -1.29406 & 0.69 & 0.65 & -59.3 &                    6.4 &               7.0 &          3.0 & 22.3 &                 26.3 &          2.9 & 21.7 &                 25.5 & >24.2 &            - & >23.9 &            - &    0 \\
J01:23:27.773-00:55:21.039 & 20.86572 & -0.92251 & 0.62 & 0.38 & -11.9 &                    6.8 &               7.1 &          2.8 & 21.2 &                 24.4 &          2.5 & 20.9 &                 23.9 &  23.1 &          0.3 &  22.6 &          0.2 &    0 \\
J01:23:31.637-01:23:25.319 & 20.88182 & -1.39037 & 1.93 & 0.94 &  -1.5 &                    7.7 &               8.1 &          3.0 & 20.6 &                 24.9 &          1.9 & 19.5 &                 22.8 & >23.7 &            - & >24.0 &            - &    0 \\
J01:23:44.851-01:37:27.984 & 20.93688 & -1.62444 & 0.89 & 0.89 & -65.9 &                    6.5 &               7.5 &          3.6 & 21.1 &                 25.8 &          3.6 & 20.5 &                 25.2 & >24.2 &            - & >24.1 &            - &    1 \\
J01:23:47.268-01:17:26.726 & 20.94695 & -1.29076 & 0.76 & 0.76 &  77.1 &                    6.4 &               7.3 &          3.9 & 21.3 &                 25.9 &          3.9 & 20.7 &                 25.4 & >24.2 &            - & >24.1 &            - &    1 \\
J01:23:59.643-01:24:43.713 & 20.99851 & -1.41214 & 0.88 & 0.73 &  50.3 &                    6.8 &               7.7 &          4.0 & 20.3 &                 24.9 &          3.8 & 19.7 &                 24.3 & >24.0 &            - & >24.0 &            - &    1 \\
J01:24:00.601-01:35:15.808 & 21.00250 & -1.58772 & 1.14 & 0.87 &  70.5 &                    7.0 &               7.8 &          3.4 & 20.1 &                 24.6 &          3.2 & 19.6 &                 23.9 & >24.0 &            - & >24.0 &            - &    1 \\
J01:24:02.742-01:49:30.931 & 21.01142 & -1.82526 & 0.87 & 0.97 &  74.5 &                    6.8 &               7.7 &          3.2 & 20.6 &                 25.1 &          3.2 & 20.0 &                 24.5 & >24.1 &            - & >24.1 &            - &    1 \\
J01:24:06.822-01:42:08.144 & 21.02843 & -1.70226 & 0.70 & 0.63 &  64.1 &                    6.4 &               7.4 &          4.5 & 21.3 &                 26.1 &          4.5 & 20.7 &                 25.5 & >24.1 &            - & >24.1 &            - &    0 \\
J01:24:09.443-01:34:51.826 & 21.03935 & -1.58106 & 1.16 & 0.72 & -30.8 &                    7.0 &               7.7 &          3.2 & 20.2 &                 24.4 &          2.9 & 19.7 &                 23.7 & >24.1 &            - & >24.2 &            - &    0 \\
J01:24:09.514-01:37:19.400 & 21.03964 & -1.62206 & 0.92 & 0.55 & -43.3 &                    6.7 &               7.4 &          3.5 & 21.2 &                 25.3 &          3.4 & 20.7 &                 24.6 & >23.9 &            - & >24.1 &            - &    1 \\
J01:24:10.378-01:24:03.853 & 21.04324 & -1.40107 & 0.92 & 0.50 &  59.1 &                    6.9 &               7.4 &          3.2 & 20.5 &                 24.2 &          3.0 & 20.2 &                 23.8 &  22.2 &          0.1 &  22.3 &          0.1 &    0 \\
J01:24:12.524-01:15:53.605 & 21.05218 & -1.26489 & 0.78 & 0.78 & -51.8 &                    6.9 &               7.6 &          3.1 & 20.4 &                 24.5 &          3.0 & 19.9 &                 24.0 & >24.1 &            - & >24.2 &            - &    1 \\
J01:24:16.553-01:24:54.715 & 21.06897 & -1.41520 & 0.99 & 0.89 & -38.6 &                    6.3 &               7.0 &          2.9 & 21.8 &                 26.0 &          2.6 & 21.4 &                 25.3 & >23.7 &            - & >23.9 &            - &    0 \\
J01:24:17.177-01:30:07.435 & 21.07157 & -1.50207 & 0.84 & 0.51 &  46.0 &                    6.9 &               7.5 &          3.2 & 21.1 &                 24.9 &          3.3 & 20.5 &                 24.3 & >23.9 &            - & >23.9 &            - &    0 \\
J01:24:19.742-01:36:32.984 & 21.08226 & -1.60916 & 0.59 & 0.85 &  57.1 &                    6.7 &               7.4 &          2.9 & 20.6 &                 24.7 &          2.9 & 20.3 &                 24.4 &  22.9 &          0.2 &  23.6 &          0.3 &    0 \\
J01:24:24.290-01:41:09.630 & 21.10121 & -1.68601 & 0.71 & 0.80 & -20.5 &                    6.2 &               7.1 &          3.6 & 21.7 &                 26.3 &          3.4 & 21.3 &                 25.7 & >24.1 &            - & >24.2 &            - &    0 \\
J01:24:24.399-01:13:30.857 & 21.10166 & -1.22524 & 0.94 & 0.57 & -26.9 &                    6.7 &               7.8 &          5.4 & 20.2 &                 25.2 &          5.3 & 19.6 &                 24.7 & >24.5 &            - & >24.4 &            - &    1 \\
J01:24:24.479-01:25:08.939 & 21.10200 & -1.41915 & 0.73 & 0.94 & -89.8 &                    6.4 &               7.1 &          3.0 & 21.6 &                 25.9 &          2.8 & 21.2 &                 25.3 & >24.2 &            - & >24.2 &            - &    0 \\
J01:24:26.226-01:24:45.783 & 21.10927 & -1.41272 & 1.13 & 0.42 & -24.5 &                    7.0 &               8.0 &          6.0 & 19.4 &                 24.3 &          5.5 & 19.0 &                 23.7 & >23.2 &            - & >24.1 &            - &    0 \\
J01:24:28.295-01:25:03.997 & 21.11790 & -1.41778 & 0.75 & 0.60 & -49.6 &                    6.6 &               7.5 &          4.0 & 21.0 &                 25.4 &          3.8 & 20.4 &                 24.8 & >24.2 &            - & >24.0 &            - &    1 \\
J01:24:35.281-01:31:10.590 & 21.14700 & -1.51961 & 1.00 & 0.60 & -88.6 &                    6.5 &               7.2 &          3.4 & 21.6 &                 25.7 &          3.1 & 21.1 &                 25.0 & >24.1 &            - & >24.0 &            - &    0 \\
J01:24:36.340-02:04:22.976 & 21.15142 & -2.07305 & 0.52 & 0.67 &  70.6 &                    6.4 &               7.7 &          6.5 & 20.1 &                 25.8 &          6.3 & 19.7 &                 25.2 & >24.1 &            - & >24.2 &            - &    0 \\
J01:24:40.652-01:38:11.829 & 21.16939 & -1.63662 & 0.68 & 0.78 & -77.6 &                    6.8 &               8.0 &          5.0 & 19.8 &                 25.0 &          5.0 & 19.2 &                 24.5 & >23.6 &            - & >23.8 &            - &    1 \\
J01:24:40.794-01:25:25.069 & 21.16998 & -1.42363 & 0.67 & 0.77 &  87.4 &                    6.8 &               7.3 &          2.6 & 21.2 &                 24.9 &          2.4 & 20.7 &                 24.3 & >24.3 &            - & >24.4 &            - &    0 \\
J01:24:41.234-01:43:58.390 & 21.17181 & -1.73289 & 0.50 & 0.87 & -86.9 &                    6.1 &               7.6 &          7.2 & 20.4 &                 26.5 &          7.0 & 19.9 &                 26.0 & >24.5 &            - & >24.5 &            - &    0 \\
J01:24:41.836-01:15:10.522 & 21.17432 & -1.25292 & 0.74 & 0.67 &  70.0 &                    6.8 &               7.4 &          3.1 & 21.0 &                 25.0 &          2.9 & 20.5 &                 24.4 & >24.1 &            - & >24.1 &            - &    0 \\
J01:24:43.896-01:30:26.899 & 21.18290 & -1.50747 & 1.04 & 0.92 &  62.3 &                    7.0 &               7.7 &          2.6 & 20.4 &                 24.4 &          2.4 & 19.9 &                 23.7 & >24.0 &            - & >24.0 &            - &    0 \\
J01:24:45.955-02:03:44.573 & 21.19148 & -2.06238 & 0.61 & 0.77 &   6.3 &                    6.8 &               7.6 &          3.5 & 19.9 &                 24.3 &          3.3 & 19.6 &                 24.0 &  21.6 &          0.1 &  21.8 &          0.1 &    1 \\
J01:24:46.134-00:50:40.281 & 21.19223 & -0.84452 & 0.71 & 0.85 & -32.2 &                    6.3 &               7.5 &          5.4 & 20.6 &                 26.1 &          5.3 & 20.1 &                 25.6 & >24.3 &            - & >24.3 &            - &    0 \\
J01:24:47.646-01:02:23.619 & 21.19852 & -1.03989 & 0.84 & 0.83 & -23.7 &                    6.9 &               8.0 &          4.3 & 19.7 &                 24.6 &          4.2 & 19.1 &                 24.0 & >22.5 &            - & >24.0 &            - &    1 \\
\hline
\end{tabular}\label{table:HSC_catalog}
\tablefoot{'NAME' (Column 1) is the unique object ID associated with each LSBG, while 'RA' (Column 2) and 'DEC' (Column 3) provide the sky coordinates of the LSBG. Columns 4, 5, 6, 7, and 8 represent the S\'ersic index, axis ratio, position angle, log of stellar mass surface density, and log of stellar mass of the LSBGs, respectively.
Columns 9 through 11 represent the half-light radius, magnitude and mean surface brightness within the half-light radius estimated using {\tt GALFIT} for \textit{g}-band. Similarly, Columns 12 through 14 represent the half-light radius, magnitude, and mean surface brightness within the half-light radius also estimated using {\tt GALFIT} for \textit{r}-band. Columns 15 and 16 provide the \textit{NUV} magnitude and its corresponding error, while columns 17 and 18 provide the \textit{FUV} magnitude and corresponding error. For both NUV and FUV, non-detections are indicated by the estimated lower limit on magnitude at the $3\sigma$ level. Finally, Column 19 labels the LSBG as 1 if it has been reported in \citet{Tanoglidis1} or \citet{haressh_lsbs}, and 0 if it has not been previously reported.}
 \end{sidewaystable*}

\end{appendix}
\end{document}